\documentclass[brazil,12pt,a4paper]{report}
\usepackage{graphicx}
\usepackage{latexsym}
\usepackage[portuges]{babel}

\newlength{\defbaselineskip}
\newcommand{\setlinespacing}[1]
           {\setlength{\baselineskip}{#1 \defbaselineskip}}

\newcommand{\lla}[1]{\left\{#1\right\}}
\newcommand{\co}[1]{\left[#1\right]}

\newcommand{\bq}{\begin{equation}}
\newcommand{\bqa}{\begin{eqnarray}}
\newcommand{\eq}{\end{equation}}
\newcommand{\eqa}{\end{eqnarray}}

\newcommand{\bqas}{\begin{eqnarray*}}
\newcommand{\eqas}{\end{eqnarray*}}

\newcommand{\abs}[1]{\vert #1\vert }

\newcommand{\cao}{\mbox{\c{c}\~ao}}
\newcommand{\coes}{\mbox{\c{c}\~oes}}
\newcommand{\cc}{\mbox{\c{c}}}
\newcommand{\ia}{\mbox{\'{\i}}}

\newcommand{\aspas}[1]{``#1''}

\begin{document}
\begin{titlepage}
\begin{center}
\vspace{0.0cm}

{\Large$\mathrm{Instituto~de~F\ia sica}$}\\ 
\vspace{0.5cm}
{\Large$\mathrm{Universidade~Federal~Fluminense}$}

\vspace{2cm}
\begin{center}
\huge{{\it Modelos Computacionais para estudar processos de 
Especia\cao~simp\'atrica}}
\end{center}
\vspace{0.5cm}
{\Large {\bf Karen Luz Burgoa Rosso}\\
\vspace{0.5cm}
{\normalsize Orientadora: Profa. Dra. Suzana Moss de Oliveira}}
\vspace{2.5cm}

\hspace{5cm}\normalsize{
Tese apresentada como requisito\\ \hspace{5cm}
para a obten\c{c}\~ao do\\ \hspace{5cm}
t\'{\i}tulo de Doutora~em F\'\i sica}\\
\end{center}
\vspace{1cm}
\begin{center}
{**************}\\
{\large{Niter\'oi\\ Mar\c{c}o de 2005}}\\
{*************************}
\end{center}
\end{titlepage}

\pagenumbering      {arabic}
\setcounter         {page}{1}
\setcounter         {chapter}{0}

\titlepage
\begin{flushright}
$~$\\
\vspace{10cm}
\Large{\it Dedicada \`a minha grande fam\'{\i}lia:\\
Rene, Miriam, Herald, Luxem, Nena, Billy,\\
Nelsy, Lesly, Dina, Jaime, Carlita, Hitaty e Jos\'e.}
\end{flushright}


\setlinespacing{1.5}
\titlepage
\addcontentsline{toc}{chapter}{Agradecimentos}
\chapter*{Agradecimentos}

\`A professora Suzana Moss de Oliveira pela orienta\cao,~incentivo, carinho 
e sobretudo, paci\^encia para comigo. Ela fez com que este per\'{\i}odo de 
doutorado fosse maravilhoso: muito obrigada, Suzana.

\`A professora Rosane Freira (PUC) pela ajuda na cadeira de Econof\ia sica.
Ao professor Thadeu Penna pela espl\^endida rede de computadores da UFF sem a 
qual eu n\~ao poderia ter feito minhas simula\coes.
Aos seguintes professores, que sempre estiveram dispostos a tirar minhas
d\'uvidas, Paulo Murilo, Thadeu Penna, Jorge de S\'a Martins,
J$\ddot{u}$rgen Stilck, Constantino 
Tsallis (CBPF), Mucio Continentino e Anna Chame.
 
Aos meus amigos e companheiros de Doutorado, Luciana Rios, Eliel Eleut\'erio, 
Andr\'e Schwartz, Cinthya Chianca, M\^onica Barcellos, Lu\ia s Fernando, Lisa Cordeiro, 
Adriana Soares, Luiz Alberto L. Wanderley,
Sohyoung Skime, Tony Dell, Toshinori Okuyama, Klauko Mota,
 Claudia Gomes, Wellington Gomes,
Scheilla M. Ramos, Aquino Lauri Espínola, Marcus Moldes, Marcos S\'ergio, Suzana 
Planas, Beatriz Maria Boechat, Luciano Fonseca, Felipe Dimer de Oliveira, 
\`Erica Renata, Daniel Grimm, Bruno de Siqueira, Armando Bustillos, Ant\^onio Delfino
 Claudette Elisea, Adriana Racco e Evandro de Mello.

Ao pessoal da Biblioteca de f\ia sica da UFF, Ana Maria de Andrade, Rita de C\'assia, 
L\'ucia Regina da Silva e K\'atia Maria da Silva Domas, pela ajuda que me prestaram
ao conseguir o material que eu precisei e tamb\'em pela organiza\cao~das 
festas de confraterniza\cao,~que para mim, foram muito importantes.

Agrade\cc o tamb\'em ao pessoal administrativo do Instituto da F\ia sica da UFF, 
ao pessoal de limpeza e aos guardas.
Muitas outras pessoas colaboraram de in\'umeras formas para que este trabalho
se realizasse: a elas, o meu muito obrigada.

\addcontentsline{toc}{chapter}{Resumo}
\chapter*{Resumo}

Foram feitas simula\coes~baseadas no modelo Penna para envelhecimento biol\'ogico, 
mas com o objetivo de estudar a especia\cao~simp\'atrica, isto \'e, a divis\~ao de 
uma esp\'ecie em duas ou mais popula\c{c}\~oes, isoladas reprodutivamente, mas sem 
que nenhuma barreira f\'{\i}sica as separe.
Para tal, introduzimos um novo tipo de competi\cao~entre os indiv\'{\i}duos, atrav\'es  
de um fator de Verhulst modificado. Esta nova competi\cao~depende de certas caracter\ia sticas 
fenot\ia picas de cada indiv\ia duo, que s\~ao representadas por pares de tiras de bits.
Estas tiras de bits s\~ao lidas em paralelo e n\~ao t\^em 
estrutura de idade. Desta forma, o genoma de cada indiv\'{\i}duo tem duas partes: uma 
estruturada por idade e relacionada ao aparecimento de doen\c{c}as heredit\'arias e 
outra, sem estrutura, ligada \`a competi\c{c}\~ao por alimento. 
Tamb\'em foi introduzida uma sele\cao~sexual
no modelo, representada por um par de tiras de bits n\~ao estruturadas independente. 

Nesta tese apresentamos tr\^es modelos diferentes;  
dois deles se utilizam, al\'em da competi\c{c}\~ao entre os indiv\'{\i}duos, 
de uma ecologia que muda abruptamente, para obter a especia\c{c}\~ao.  
Foram inspirados na especia\cao~dos tentilh\~oes 
de Darwin, uma fam\'{\i}lia de p\'assaros que habitam as Ilhas Gal\'apagos 
e nos cicl\ia deos, uma fam\'{i}lia de peixes que vivem nos Lagos da Nicar\'agua
e no Lago Vit\'oria, na \'Africa. O terceiro modelo n\~ao utiliza nenhuma mudan\c{c}a 
ecol\'ogica: a especia\c{c}\~ao \'e obtida dependendo apenas do grau   
de competi\cao~entre grupos de indiv\'{\i}duos com caracter\'{\i}sticas fenot\'{\i}picas 
semelhantes. 

No modelo para descrever o processo de especia\cao~dos tentilh\~oes 
utilizamos apenas um par de tiras de bits para descrever uma \'unica caracter\'{\i}stica 
fenot\'{\i}pica, 
que est\'a ligada tanto \`a sele\cao~ecol\'ogica como \`a sele\cao~sexual. Esta 
caracter\'{\i}stica equivale \`a morfologia dos bicos. 
Estudamos ent\~ao a mudan\cc a na distribui\cao~destas 
morfologias em fun\cao~das regras de acasalamento, dos valores dos par\^ametros do modelo  
e da posi\cao~da esp\'ecie na cadeia alimentar.  
Neste caso encontramos comportamentos qualitativos
equivalentes aos que v\^em sendo estudados nestes p\'assaros. 

Para simular os cicl\ia deos, utilizamos dois pares de tiras de bits
independentes; um deles representa o tipo de maxilar do peixe e est\'a 
diretamente ligado \`a competi\c{c}\~ao por alimento. O outro par representa a cor, 
que \'e o tra\c{c}o ligado \`a sele\cao~sexual. O objetivo principal foi o de 
analisar a estabilidade das distribui\coes~tanto dos maxilares como das cores 
destes peixes. Obtivemos 
a especia\cao~simp\'atrica com 
este modelo, embora n\~ao exatamente aquela observada nos cicl\ia deos. 
Verificamos ainda que se estabelece uma forte correla\c{c}\~ao entre os dois tra\c{c}os, 
sempre que a especia\c{c}\~ao ocorre. 

Com o modelo que estuda a especia\cao~simp\'atrica em fun\cao~do grau de 
competi\cao~numa ecologia fixa, encontramos uma transi\cao~de fase que possibilitar\'a   
o estudo da reversibilidade ou n\~ao deste processo de especia\cao.

\addcontentsline{toc}{chapter}{Abstract}
\chapter*{Abstract}
We perform simulations based on the Penna model for biological ageing, now with 
the purpose of studying sympatric speciation, that is, the division of a single species 
into two or more populations, reproductively isolated, but without any physical barrier 
separating them. For that we introduce a new kind of competition among the individuals, 
using a modified Verhulst factor. The new competition depends on some specific phenotypic 
characteristic of each individual, which is represented by a pair of bitstrings. These 
strings are read in parallel and have no age structure. In this way, each individual genome 
consists of two parts. The first one has an age-structure and is related to the appearance of 
inherited diseases; the second part is not structured and takes into account the competition 
for the available resources. We also introduce sexual selection into the model, making use 
of another non-structured and independent pair of bitstrings.     

In this thesis we present three different models; two of them use, besides the competition, a 
sudden change in the ecology to obtain speciation. They were motivated by the speciation 
process observed in the Darwin finches, a family of birds that inhabits the Galapagos Islands, 
and also by that observed in the cichlids, a family of fish that lives in the 
Nicaragua Lakes and in the Vitoria Lake, in Africa. The third model does not use any ecological 
change: sympatric speciation is obtained depending only on the strength of competition among 
individuals with similar phenotypic characteristics. 

In the model to describe the speciation process of the finches we use a single pair of 
non age-structured bitstrings to describe one phenotypic characteristic which is related 
to both competition and mating preference. This characteristic corresponds to the beak 
morphology. We study the changes in the distribution of the beaks according to 
the mating rules, to the values of the parameters and to the position of the species 
in a food-chain.
In this case we obtain qualitative results that are in agreement with field 
observations of these birds. To simulate the cichlids, we use two pairs of non-structured 
and independent bitstrings. One pair represents the fish type of jaw and is 
directly related to competition for food. The other pair represents the colour, which 
is the trait related to sexual selection. Our main purpose has been to analyze the 
stability of the distributions of both jaws and colours. In this case we obtain 
sympatric spectiation, although not exactly the one experimentally observed in the 
cichlids.    

Using the model without ecological changes, we study the probability of speciation as 
a function of the competition strength and find a phase transition, which may allow 
the study of the reversibility or not of the speciation process.



$~$\\



\def\contentsname{\'Indice}
\tableofcontents

\addcontentsline{toc}{chapter}{Introdu\cao}

\chapter*{Introdu\cao} 

O processo de especia\c{c}\~ao simp\'atrica, cuja defini\c{c}\~ao ser\'a 
vista mais adiante, est\'a baseado na teoria da 
evolu\c{c}\~ao de Darwin. Esta teoria tem uma matem\'atica pouco formalizada, n\~ao s\'o  
por tratar de sistemas adaptativos mas tamb\'em devido ao tipo de intera\c{c}\~ao local
entre os indiv\'{\i}duos, que inclui o cruzamento e a recombina\c{c}\~ao de seus genomas. 
Erwin Schr$\ddot{o}$edinger, no seu livro {\it What is life?} \cite{schro}, foi um 
dos primeiros cientistas a tentar interpretar e conjecturar a vida em termos dos 
conhecimentos da qu\'{\i}mica e da f\'{\i}sica. 

\begin{figure}[htb]
\begin{center}
\includegraphics[width=3.0cm,angle=0]{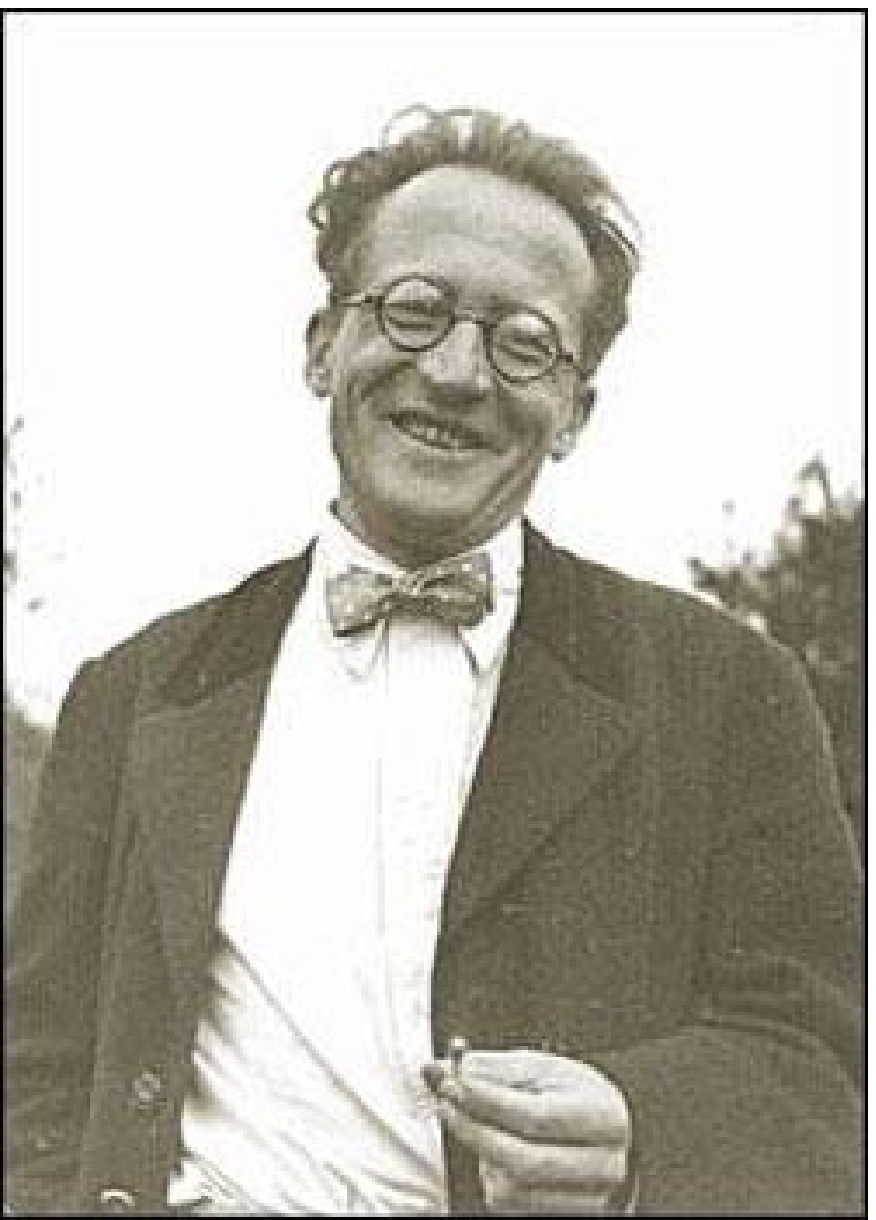}
\includegraphics[width=5.9cm,angle=0]{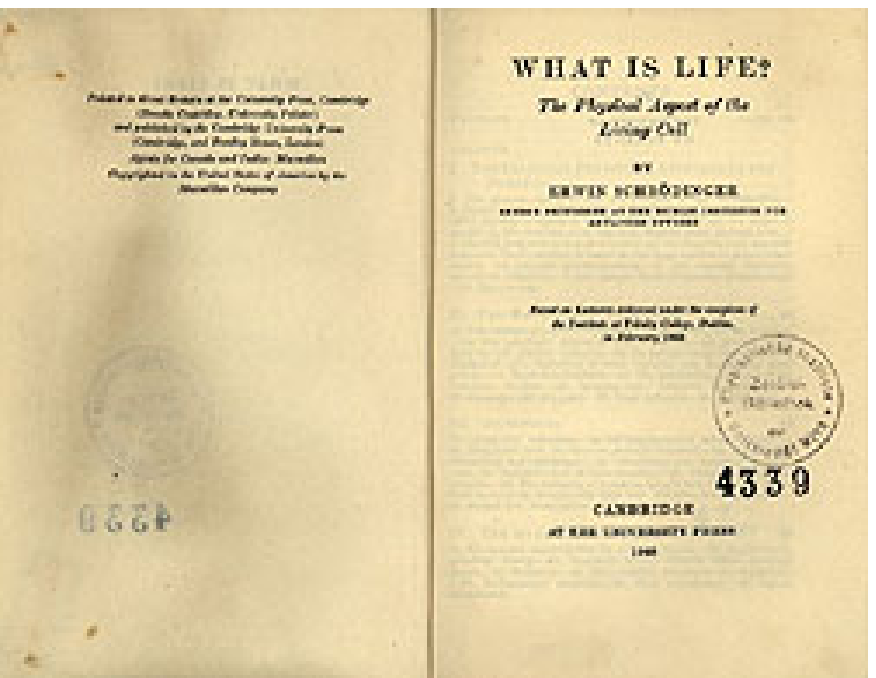}
\end{center}
\caption{\it Erwin Schr$\ddot{o}$edinger escreveu, em 1948, o primeiro livro dedicado a 
problemas de Biologia sob o ponto de vista f\'{\i}sico.}
\label{figsong}
\end{figure}

Numa palestra a respeito de diferentes 
aspectos das c\'elulas vivas, Schr$\ddot{o}$edinger explica \`a uma plat\'eia de leigos 
que o assunto era dif\'{\i}cil, embora a dedu\c{c}\~ao matem\'atica pouco fosse utilizada: 
``A raz\~ao disto n\~ao \'e que o assunto seja simples o bastante para poder ser explicado 
sem matem\'atica, mas sim porque \'e complexo demais para ser completamente access\'{\i}vel 
\`a matem\'atica.'' Atualmente um grande 
n\'umero de f\'{\i}sicos t\^em investido no assunto, principalmente utilizando  
computadores, os quais funcionam como um grande laborat\'orio no qual os passos da 
evolu\c{c}\~ao podem ser recriados e medidas estat\'{\i}sticas podem ser feitas. 
\addcontentsline{toc}{section}{Evolu\c{c}\~ao das esp\'ecies}
\section*{Evolu\c{c}\~ao das esp\'ecies}
O interesse pela evolu\c{c}\~ao org\^anica teve suas origens entre os Gregos e 
Romanos. As id\'eias de Emp\'edocles (cerca de 490-435 a.C.) \cite{librogene}, 
por exemplo, j\'a demonstravam s\'erios intentos de 
compreender as mudan\cc as biol\'ogicas. No entanto, at\'e meados 
do s\'eculo XIX muitos obst\'aculos se interpuseram no caminho de uma 
aprecia\c{c}\~ao cient\'{\i}fica racional da evolu\cao.~Um dos mais
importantes obst\'aculos foi a cren\c{c}a, amplamente aceita, da 
\aspas{imutabilidade das esp\'ecies}, que considerava todo tipo de formas
de vida como entidades est\'aticas, sem mudan\cc as no passado e 
sem possibilidades de mudan\cc a no futuro. 
A obten\c{c}\~ao sistem\'atica de provas e a constru\c{c}\~ao de uma teoria de  
evolu\c{c}\~ao das esp\'ecies por sele\c{c}\~ao natural  
foi conseguida por Charles Darwin, como resultado de 20 anos de trabalho \'arduo.  
Ela foi apresentada em 1857, ao mesmo tempo em que Alfred Russel Wallace tamb\'em o fazia, 
j\'a que chegara \`as mesmas conclus\~oes de Darwin.  
\begin{figure}[htb]
\begin{center}
\includegraphics[width=10cm,angle=0]{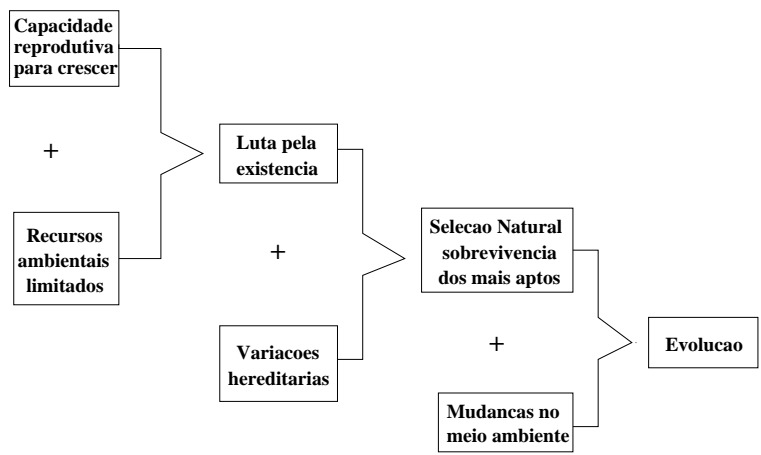}
\end{center}
\caption{\it Representa\c{c}\~ao esquem\'atica dos principais argumentos em 
favor da evolu\c{c}\~ao por sele\c{c}\~ao natural defendidos por Charles Darwin 
e Alfred Russell Wallace; tabela de Wallace.}
\label{figwallace}
\end{figure}
Ambos argumentaram que mesmo que existissem 
esp\'ecies com morfologias fixas,  
estas haviam aparecido por sele\c{c}\~ao natural entre os membros vari\'aveis 
de esp\'ecies anteriores. Um interessante esquema dos conceitos 
por eles desenvolvidos pode ser visto na figura \ref{figwallace}. 
Nessa \'epoca n\~ao se sabia explicar como as 
varia\c{c}\~oes podiam ser transmitidas aos descendentes, pois os estudos
desenvolvidos por Gregorio Mendel (1866) n\~ao tinham tido repercus\~ao. 
Somente em 1900, 34 anos depois, tr\^es bot\^anicos redescobriram
as mesmas leis b\'asicas da hereditariedade apontadas por Mendel \cite{biohoje}.
A adapata\c{c}\~ao e a sele\c{c}\~ao natural, embora de f\'acil comprens\~ao intuitiva,
s\~ao na verdade bastante complexas. Estud\'a-las significa compreender n\~ao apenas   
relacionamentos ecol\'ogicos intrincados, como tamb\'em a matem\'atica 
avan\cc ada envolvida na gen\'etica das popula\c{c}\~oes \cite{richard,theorygp}.

Atualmente entendemos que a 
evolu\c{c}\~ao tem como base a mudan\cc a na constitui\c{c}\~ao 
gen\'etica da popula\cao.~\'E sobre estas diferen\cc as gen\'eticas, originadas por muta\c{c}\~oes e 
recombina\cao,~que atua a sele\c{c}\~ao natural \cite{librogene}.
Os genes, tal 
como s\~ao entendidos hoje em dia, s\~ao se\c{c}\~oes de cadeias de DNA  
respons\'aveis pelo funcionamento e apar\^encia dos organismos. 
A distin\c{c}\~ao entre os efeitos dos genes num organismo e os genes propriamente ditos   
\'e considerada atrav\'es dos termos {\it fen\'otipo} e {\it gen\'otipo}, 
respectivamente (Futuyma 1979) \cite{robert}. O gen\'otipo cont\'em as informa\c{c}\~oes gen\'eticas, 
que ao serem ``compiladas'', d\~ao origem aos fen\'otipos. 

\addcontentsline{toc}{section}{Especia\cao}
\section*{Especia\cao}
\begin{figure}[htb]
\begin{center}
\includegraphics[width=2.9cm,angle=0]{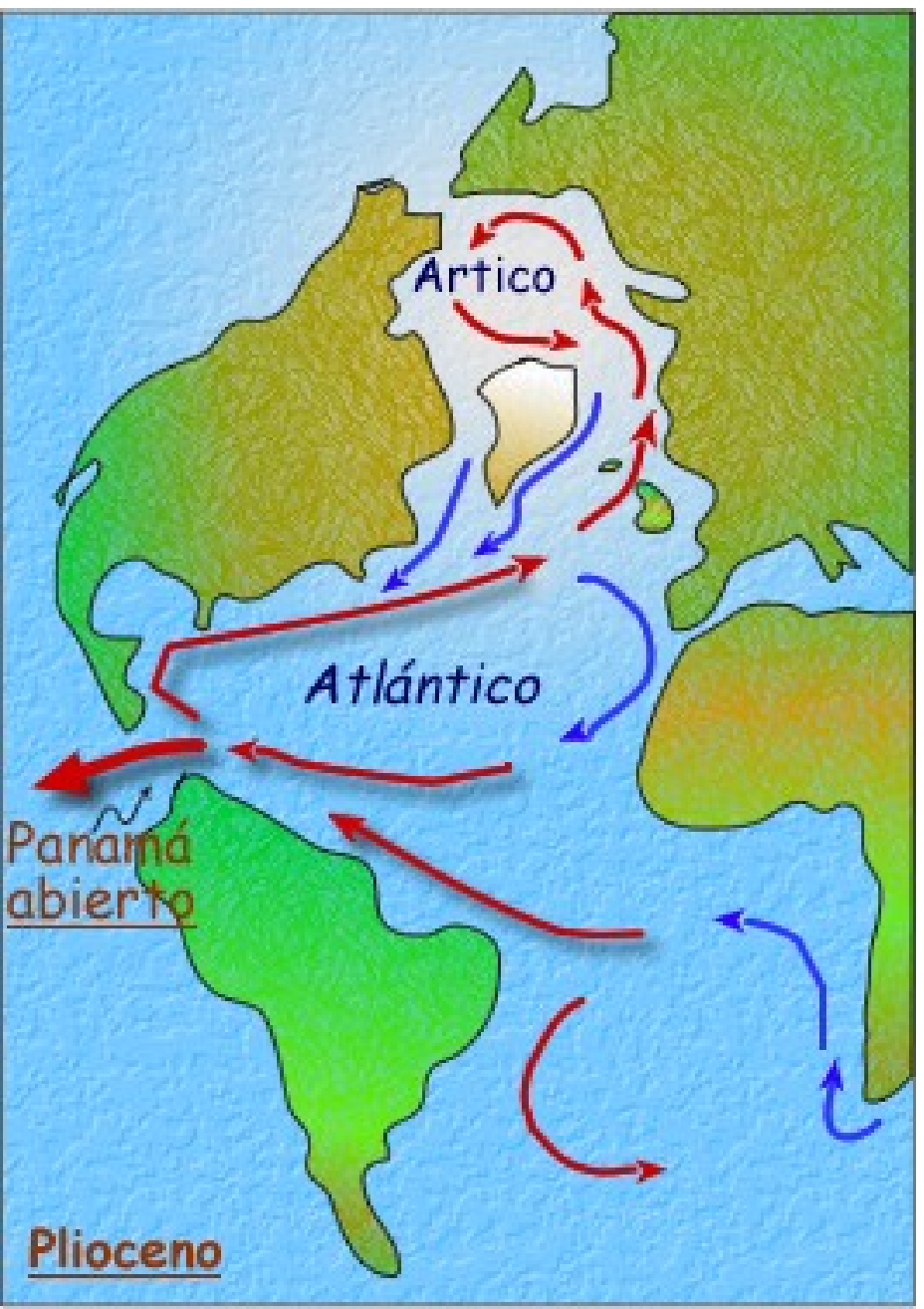}
\includegraphics[width=2.8cm,angle=0]{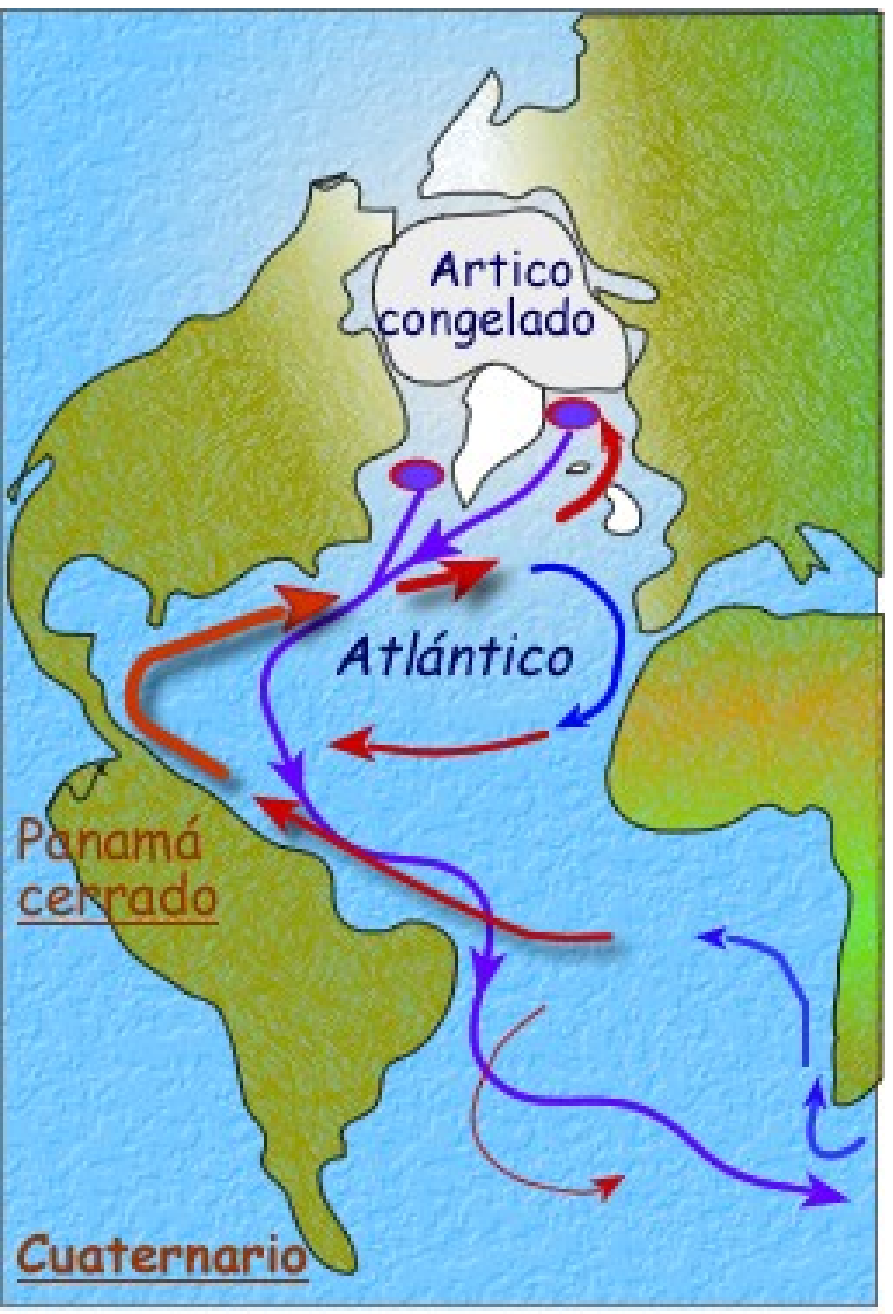}
\includegraphics[width=5.5cm,angle=0]{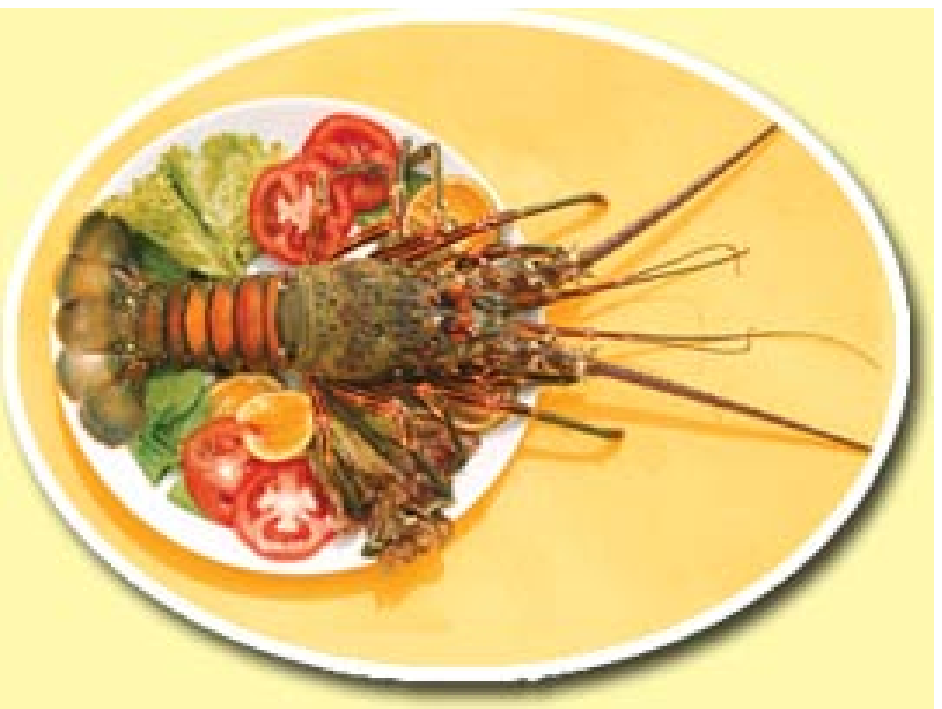}
\end{center}
\caption{\it O Istmo do Panam\'a isolou uma esp\'ecie  
do g\^enero de lagostas Alpheus, h\'a tr\^es milh\~oes de anos atr\'as, sendo 
um dos exemplos mais documentado de especia\c{c}\~ao alop\'atrica.}
\label{figlagostas}
\end{figure}
\underline{Especia\c{c}\~ao} \'e um processo pelo qual uma popula\c{c}\~ao original
se divide em duas ou mais, sem que haja um fluxo de genes entre elas
(sexualmente isoladas). Os dois tipos mais importantes 
de especia\c{c}\~ao s\~ao a alop\'atrica e a simp\'atrica. 
A especia\c{c}\~ao alop\'atrica basicamente consiste na 
separa\c{c}\~ao geogr\'afica da popula\c{c}\~ao em duas ou mais partes. Esta
separa\c{c}\~ao
pode ser devida \`a migra\c{c}\~ao de uma das partes ou ao aparecimento de
um obst\'aculo f\'{\i}sico que impe\c{c}a a troca de gens entre elas.  
A diferencia\c{c}\~ao destas novas 
esp\'ecies \'e bem conhecida, e deve-se a fatores estoc\'asticos (deriva
gen\'etica) e 
a processos seletivos de adapata\c{c}\~ao. 
O melhor exemplo documentado deste tipo de 
especia\c{c}\~ao foi aquele produzido na forma\c{c}\~ao do Istmo do Panam\'a,
h\'a tr\^es milh\~oes de anos 
atr\'as. O aparecimento desta barreira geogr\'afica separou as
popula\c{c}\~oes de organismos 
aqu\'aticos entre as \'aguas dos oceanos Pac\'{\i}fico e
Atl\^antico. Dentro do g\^enero 
{\it Alpheus} de lagostas, por exemplo, deu origem a 7 esp\'ecies distintas, com 
representantes de cada lado 
do Istmo. Estas esp\'ecies s\~ao consideradas g\^emeas, pois elas
diferem apenas ligeramente na morfologia, mas est\~ao isoladas geneticamente.

O segundo tipo de especia\c{c}\~ao \'e a chamada especia\c{c}\~ao 
simp\'atrica, que corresponde \`a divis\~ao da popula\c{c}\~ao dentro
de um mesmo espa\c{c}o geogr\'afico. Esta divis\~ao \'e produzida 
pela sele\c{c}\~ao natural, que atua atrav\'es da competi\c{c}\~ao pelas 
fontes de alimento em conjunto com a sele\c{c}\~ao sexual. A busca pelo
entendimento deste tipo de especia\c{c}\~ao tem gerado muitos modelos
te\'oricos e computacionais; no nosso caso, estes \'ultimos correspondem \`a ferramenta que 
temos utilizado para estudar este fascinante processo.

Os modelos computacionais usados nos cap\'{\i}tulos dois, tr\^es e quatro, t\^em uma  
origem comum, que est\'a descrita no primeiro cap\'{\i}tulo. 
Com o tipo de competi\c{c}\~ao que utilizamos observamos 
que existe um grau de competi\c{c}\~ao cr\'{\i}tico a partir do qual ocorre ou n\~ao a 
especia\cao.~Como ser\'a visto no cap\'{\i}tulo 4, para graus menores que o  
cr\'{\i}tico n\~ao \'e possivel achar especia\c{c}\~ao simp\'atrica
e para valores maiores, sempre \'e possivel encontr\'a-la. Nos  
cap\'{\i}tulos 2 e 3 trabalhamos exatamente com este grau  
cr\'{\i}tico de competi\cao, onde a especia\c{c}\~ao pode ou n\~ao ocorrer, devido \`as 
flutua\c{c}\~oes nas densidades populacionais. Afinal, \'e possivel encontrar 
especia\c{c}\~ao simp\'atrica na natureza, mas n\~ao o tempo todo. 
As simula\c{c}\~oes feitas foram inspiradas nos exemplos dados a seguir, 
observados experimentalmente. Maiores detalhes 
sobre estes processos de especia\c{c}\~ao ser\~ao vistos nos
cap\'{\i}tulos correspondentes.

\addcontentsline{toc}{section}{Exemplos de Especia\cao}
\subsection*{Exemplos de especia\cao}
Um dos mais famosos exemplos de especia\c{c}\~ao s\~ao os cl\'assicos
tentilh\~oes (finches) de Darwin, das ilhas 
Gal\'apagos, que constituem um grupo de 14 esp\'ecies com formas e
tamanhos de bicos diferentes.
O casal Peter Grant e Rosemary Grant, da Universidade de Princeton, EUA, 
come\cc ou em 1973 a estudar a popula\c{c}\~ao de duas esp\'ecies de uma pequena ilha, 
a Daphne Maior \cite{bgs214,sgs227}. Eles contaram as popula\c{c}\~oes e mediram 
certos tra\cc os do tentilh\~ao terrestre de bico m\'edio ({\it Geospiza fortis}) 
e do tentilh\~ao dos cactos ({\it Geospiza scandens}). 
Os tra\cc os b\'asicos eram o tamanho do corpo e o tamanho e a forma do bico. 
Eles notaram uma forte correla\c{c}\~ao dessas medidas com eventos naturais 
pelos quais a ilha passou. Ao fazer o estudo de longo prazo, foi 
poss\'{\i}vel medir o impacto nos tentilh\~oes dos momentos de seca 
intensa ou de chuvas torrenciais - como a provocada pelo fen\^omeno 
clim\'atico El Ni\~no, em 1983.
No caso do {\it G. fortis}, por exemplo, notaram que o tamanho m\'edio do 
bico aumentava nos anos de seca, ver figura \ref{figsong}, quando 
apenas sementes maiores e duras estavam dispon\'{\i}veis. 
Em tempos mais \'umidos, bicos menores eram mais comuns.
Eles tambem descobriram que estes p\'assaros acasalam atrav\'es 
do tipo de canto que os machos emitem para atrair as f\^emeas~\cite{gge50}. 

\begin{figure}[htbp]
\begin{center}
\includegraphics[width=7.5cm,angle=0]{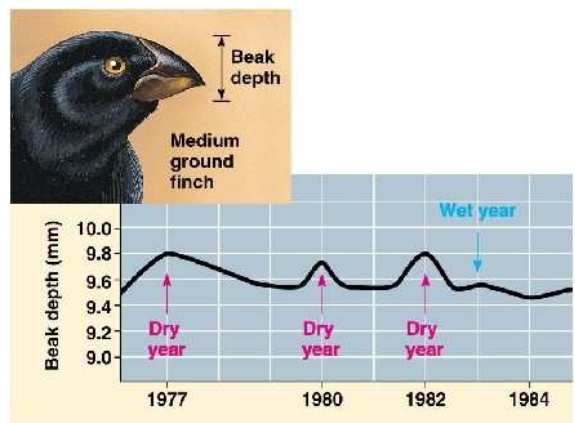}
\end{center}
\caption{\it Not\'avel mudan\cc a no tamanho dos bicos da esp\'ecie Geospiza em 
fun\c{c}\~ao das mudan\cc as clim\'aticas.}
\label{figsong}
\end{figure}

\begin{figure}[htbp]
\begin{center}
\includegraphics[width=13cm,angle=0]{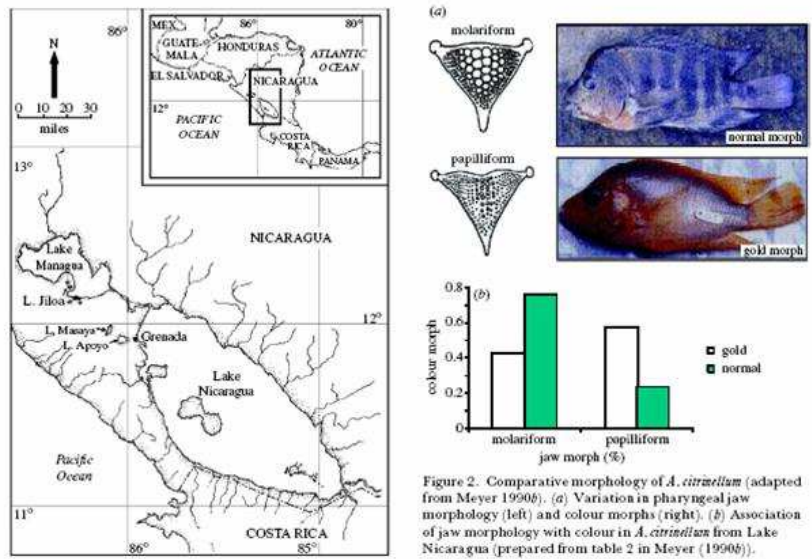}
\end{center}
\caption{\it Dados encontrados no trabalho de Wilson \cite{wilson}.
Lado esquerdo: Mapa da Nicar\'agua que mostra o lugar dos lagos 
onde foram colhidas as amostras da fam\'{\i}lia de peixes chamados cicl\'{\i}deos. 
Lado direito: A varia\c{c}\~ao da morfologia de cores e maxilares destes peixes,    
encontrados nos lagos da Nicar\'agua.}
\label{figpeixes}
\end{figure}

\begin{figure}[htbp]
\begin{center}
\includegraphics[width=6.5cm,angle=0]{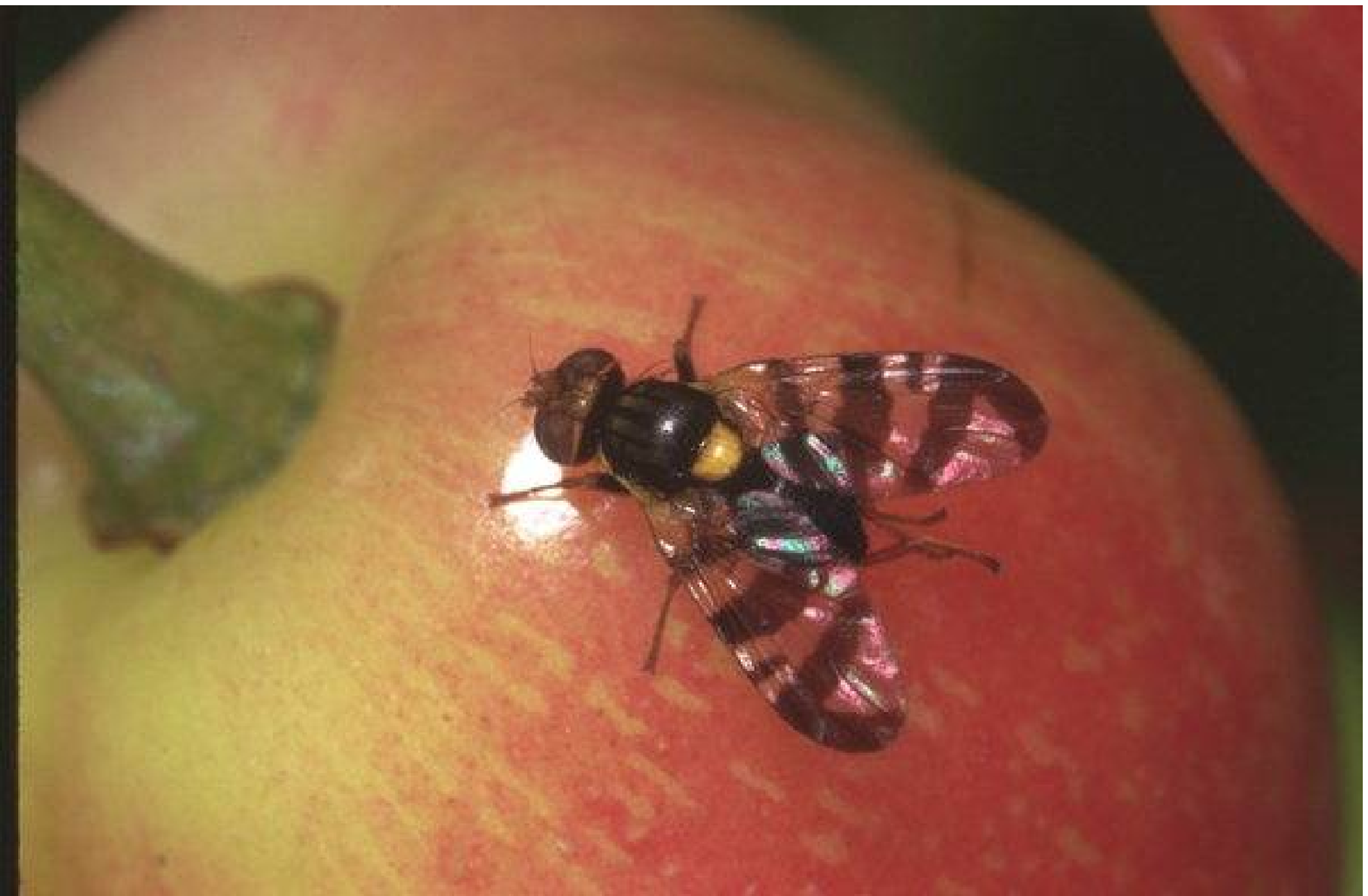}
\includegraphics[width=6.5cm,angle=0]{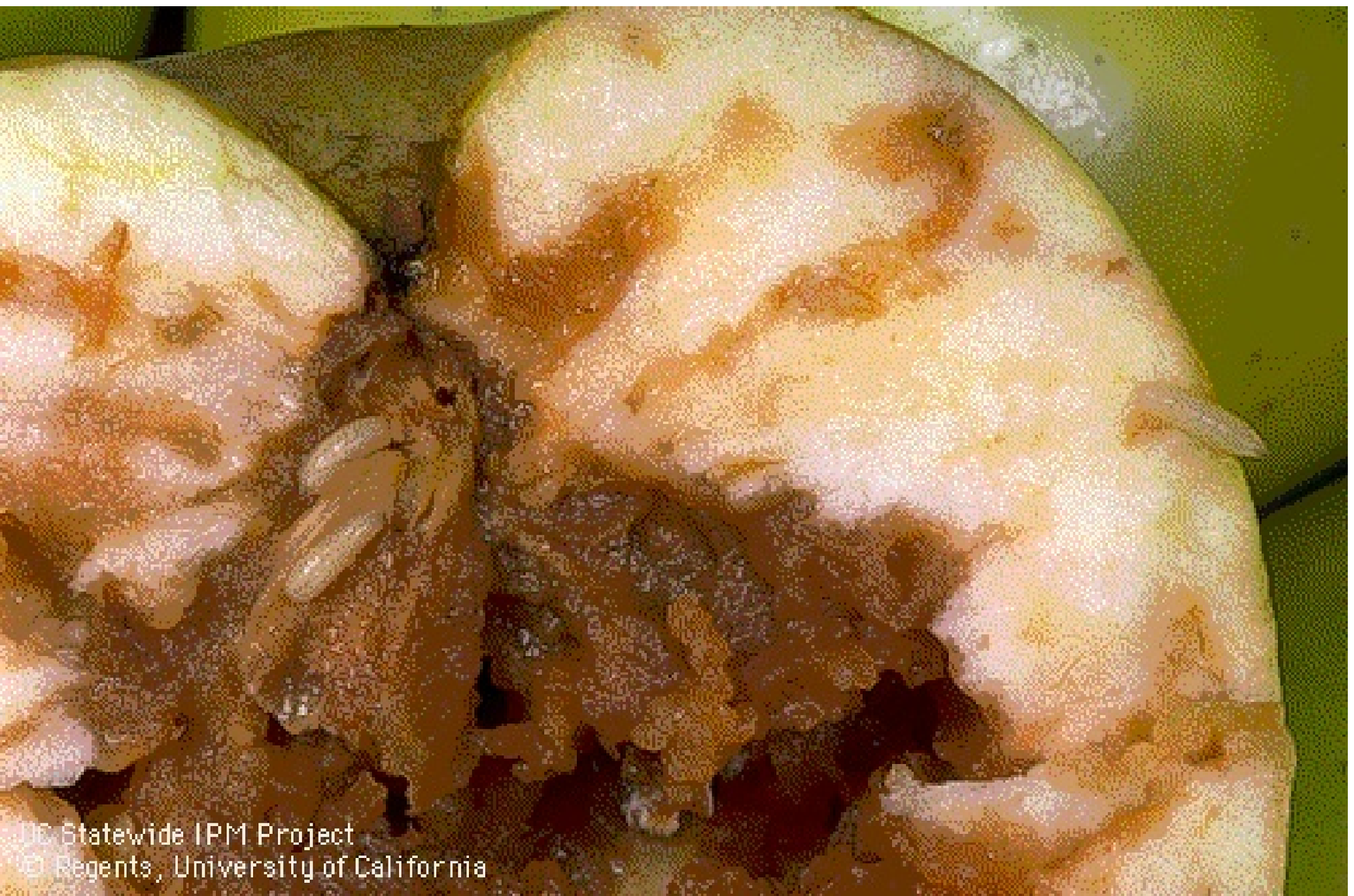}
\end{center}
\caption{\it A larva da mosca da ma\cc\~a \'e um outro exemplo
recente de processo de especia\c{c}\~ao simp\'atrica em curso. Lado 
esquerdo: A mosca da fruta Rhagoletis pomonella adulta. 
Lado direito: A larva da mosca da ma\c{c}\~a; dependendo da ra\cc a da mosca, 
a larva pode ser deixada na superf\'{\i}cie ou no centro da ma\c{c}\~a.}
\label{figmoscas}
\end{figure}

Como todo bom exemplo cl\'assico, os tentilh\~oes t\^em ajudado a entender os efeitos
da sele\c{c}\~ao ecol\'ogica, devida \`a competi\c{c}\~ao por comida, e da
sele\c{c}\~ao sexual, durante um processo de especia\cao. As simula\c{c}\~oes inspiradas
nestes p\'assaros podem ser vistas no cap\'{\i}tulo 2.

O cap\'{\i}tulo 3 corresponde \`a especia\c{c}\~ao da fam\'{\i}lia de peixes  
chamados cicl\'{\i}deos, figura \ref{figpeixes}. Existe uma forte evidencia, que 
o tipo de processo de especia\c{c}\~ao destes peixes 
\'e simp\'atrico \cite{editor}. Contudo, 
como ser\'a visto no cap\'{\i}tulo 3, as origens deste processo  
nos cicl\'{\i}deos t\^em dividido em dois grupos aqueles que estudam 
especia\c{c}\~ao simp\'atrica. O primeiro grupo 
acha que a especia\c{c}\~ao simp\'atrica \'e guiada pela 
diverg\^encia causada pela sele\c{c}\~ao natural devida \`a competi\c{c}\~ao pelas  
fontes de alimento. O outro grupo defende que este 
tipo de especia\c{c}\~ao \'e guiado pela diverg\^encia causada por sele\c{c}\~ao sexual. 

Finalmente o cap\'{\i}tulo 5 refere-se ao estudo 
da obten\c{c}\~ao da especia\c{c}\~ao simp\'atrica numa ecologia fixa, variando-se apenas o grau
de competi\c{c}\~ao entre os indiv\'{\i}duos. Encontramos uma transi\c{c}\~ao de fase 
cuja ordem, ainda n\~ao identificada, pode ser bastante relevante no que diz 
respeito \`a reversibilidade ou n\~ao do processo de especia\c{c}\~ao. 
Neste cap\'{\i}tulo a motiva\c{c}\~ao biol\'ogica veio da mosca da ma\c{c}\~a - ver 
figura \ref{figmoscas}.
O processo de especia\c{c}\~ao simp\'atrica em curso destas  
moscas parece ser revers\'{\i}vel, mas a bibliografia a respeito do mesmo ainda est\'a 
sendo acumulada, pois sua observa\c{c}\~ao \'e bastante recente.

A tese \'e conclu\'{\i}da com observa\c{c}\~oes gerais dos resultados das simula\c{c}\~oes apresentadas 
nos cap\'{\i}tulos 2,3 e 4.

\chapter{Modelos} 

Conforme mencionado na introdu\c{c}\~ao, os ingredientes principais para se obter a 
especia\c{c}\~ao simp\'atrica s\~ao a disputa por alimento e a sele\c{c}\~ao sexual. 
Ambos os ingredientes est\~ao relacionados a caracter\'{\i}sticas fenot\'{\i}picas 
externas do indiv\'{\i}duo, tais como tamanho e cor. 
Nossas simula\c{c}\~oes sobre especia\c{c}\~ao simp\'atrica baseam-se na introdu\c{c}\~ao de uma
nova parte ao \aspas{genoma cronol\'ogico} que representa o indiv\'{\i}duo no modelo 
Penna~\cite{penna}, a fim de atribuir ao mesmo um ``fen\'otipo''.   

O modelo Penna \'e baseado na teoria da sele\c{c}\~ao natural de Darwin para a evolu\c{c}\~ao das 
esp\'ecies e na teoria do ac\'umulo de muta\c{c}\~oes para explicar o envelhecimento biol\'ogico. 
Desde sua publica\c{c}\~ao em 1995, vem sendo utilizado com sucesso na compreens\~ao de  
muitos fen\^omenos evolucion\'arios observados na natureza, tais como a 
senesc\^encia catastr\'ofica do salm\~ao, a auto-organiza\c{c}\~ao da menopausa, as 
vantagens da reprodu\c{c}\~ao sexuada, etc... Para uma revis\~ao das diversas 
aplica\c{c}\~oes te\'oricas do modelo Penna veja \cite{spa257,livrosu} ; 
para poss\'{\i}veis 
aplica\c{c}\~oes pr\'aticas veja \cite{teseadriana} e para solu\c{c}\~oes 
anal\'{\i}ticas veja \cite{asppa253,prlpenna1}. 

Embora o estado final  do modelo Penna, calculado analiticamente \cite{asppa253}, 
seja a morte da popula\c{c}\~ao toda, 
o tempo que este estado final leva para ser atingido cresce exponencialmente 
com o tamanho da popula\c{c}\~ao. Assim, os indiv\'{\i}duos vivem num 
estado quase-estacionario de dura\c{c}\~ao quase infinita, onde a popula\c{c}\~ao 
apresenta uma distribui\c{c}\~ao de indiv\'{\i}duos por idade que se mant\'em constante 
durante todas as simula\c{c}\~oes, por mais longas que estas sejam (bilh\~oes de passos). 
Esta propriedade do modelo certamente facilita o estudo da din\^amica de popula\c{c}\~oes em  
processo de especia\cao.

\section{O modelo Penna}

Na vers\~ao sexuada do modelo Penna o genoma de cada indiv\'{\i}duo \'e representado 
por duas tiras de
32 bits cada, que s\~ao lidas em paralelo. Elas cont\^em a informa\c{c}\~ao de quando 
os sintomas de uma dada doen\cc a heredit\'aria v\~ao aparecer, sendo por isto chamadas  
de \aspas{genoma cronol\'ogico}. Cada uma das tiras cont\'em a heran\c{c}a gen\'etica de um 
dos pais, sendo as doen\c{c}as representadas por bits 1. Se um dado indiv\'{\i}duo possui dois 
bits iguais a 1, por exemplo na terceira posi\c{c}\~ao de ambas as tiras (homozigoto), isto 
indica que aquele indiv\'{\i}duo vai come\c{c}ar a sofrer dos sintomas de uma doen\c{c}a 
no terceiro per\'{\i}odo de sua vida. Assim, cada indiv\'{\i}duo pode viver no m\'aximo por 32  
per\'{\i}odos. Se o indiv\'{\i}duo for heterozigoto numa dada 
posi\c{c}\~ao (bit 1 numa tira e zero na outra), \^ele s\'o ficar\'a doente se naquela 
posi\c{c}\~ao o bit 1 for dominante. No in\'{\i}cio da simula\c{c}\~ao define-se 
quantas posi\c{c}\~oes 
ser\~ao dominantes e sorteia-se aleatoriamente quais ser\~ao elas. Estas posi\c{c}\~oes 
s\~ao as mesmas para todos os genomas e s\~ao mantidas fixas durante todo o processo de 
evolu\c{c}\~ao da popula\c{c}\~ao. Um passo computacional significa ler mais um bit do 
genoma de todos os indiv\'{\i}duos. Se em qualquer passo, o n\'umero de doen\c{c}as acumuladas 
num dado genoma atinge o limite $L$, aquele indiv\'{\i}duo morre.   

\begin{figure}
 \includegraphics[width=8cm,angle=0]{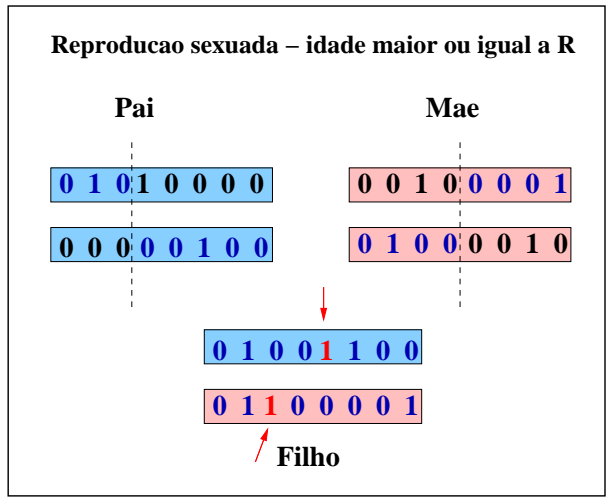}
 \includegraphics[width=8cm,angle=0]{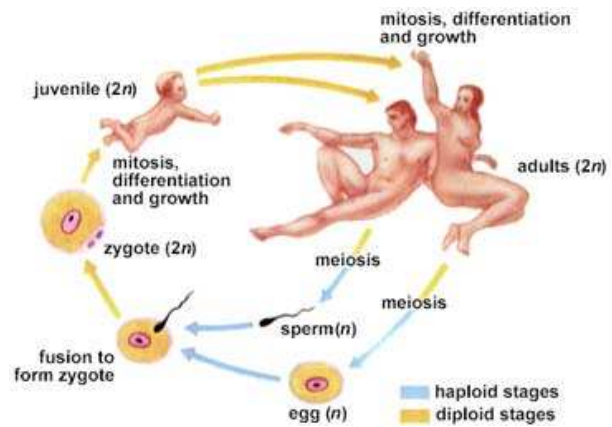}
\caption{\it Forma\c{c}\~ao dos gametas feminino e masculino que d\~ao origem 
ao genoma do filho. 
O processo de cruzamento e recombina\c{c}\~ao (lado esquerdo) \'e feito da mesma forma
tanto no genoma do pai quanto no da m\~ae. As setas indicam onde muta\c{c}\~oes nocivas  
aleat\'orias ocorreram. 
O desenho (lado direito) mostra
que uns dos elementos principais para a continuidade da vida \'e o processo
de divis\~ao celular chamado de meiose, que ocorre nas c\'elulas germinativas e
\'e o \'unico onde acontece um cruzamento entre dois cromossomos hom\'ologos.}
\label{figcrossing}
\end{figure}

A f\^emea que sobrevive at\'e a idade m\'{\i}nima de reprodu\c{c}\~ao $R$ passa a acasalar 
aleatoriamente e a cada passo, com um macho de idade tamb\'em $\ge R$, gerando $NF$ 
filhos a cada vez.   
O genoma do filho \'e construido atrav\'es do cruzamento e da recombina\c{c}\~ao
das tiras dos pais, conforme esquematizado na figura \ref{figcrossing}. Primeiro 
corta-se o genoma da m\~ae numa posi\c{c}\~ao aleat\'oria e une-se dois peda\c{c}os 
complementares para formar o gameta feminino (uma \'unica tira de bits). A seguir, 
introduz-se aleatoriamente $M$ muta\c{c}\~oes nocivas, \footnote{As muta\c{c}\~oes normalmente 
ocorrem durante a
duplica\c{c}\~ao  do DNA e devem-se \`a perda ou adi\c{c}\~ao de uma base, ou \`a
substitui\c{c}\~ao  de uma base por outra. A maioria das muta\c{c}\~oes s\~ao neutras, mas as 
muta\c{c}\~oes ruins s\~ao 100 
vezes mais prov\'aveis que as boas (corretoras) \cite{pamilo}, sendo que estas  
\'ultimas em geral d\~ao origem a uma nova esp\'ecie.} 
cujas posi\c{c}\~oes est\~ao indicadas por setas na figura \ref{figcrossing}.
 Neste caso, se o bit escolhido para mutar no gameta feminino vale 0, 
\^ele se torna 1 no genoma do filho, mas se o bit escolhido j\'a for 1, \^ele permanece 
inalterado no filho (n\~ao ocorre muta\c{c}\~ao). O mesmo processo de cruzamento e  
recombina\c{c}\~ao com taxa de muta\c{c}\~ao $M$ ocorre no genoma do pai, gerando o 
gameta masculino. O genoma do filho \'e o resultado da uni\~ao dos dois gametas e 
o sexo \'e aleatoriamente escolhido, com $50 \%$ de chance para cada um.

Mesmo permitindo apenas muta\c{c}\~oes ruins, a popula\c{c}\~ao que se obt\'em com 
a din\^amica at\'e aqui descrita cresce exponencialmente. Para evitar tal comportamento 
explosivo, a vers\~ao original do modelo Penna adota o chamado fator log\'{\i}stico de 
Verhulst, dado por $V(t)~=~P(t)/MAX$,
onde $P(t)$ \'e a popula\c{c}\~ao  total no tempo $t$ e $MAX$ \'e a capacidade de 
sustenta\c{c}\~ao do ambiente. Assim, em cada passo e para cada indiv\'{\i}duo 
gera-se um n\'umero 
aleat\'orio entre 0 e 1, cujo valor \'e comparado com o valor de $V$. Se o n\'umero 
gerado for menor que $V$, o indiv\'{\i}duo morre, independente da idade ou do genoma. 
Este termo tamb\'em representa uma competi\c{c}\~ao aleat\'oria por espa\c{c}o e comida, 
que depende do n\'umero m\'aximo, $MAX$, de indiv\'iduos que o ambiente pode sustentar.   
Exatamente por ser um termo totalmente aleat\'orio, o fator de Verhulst vem sendo criticado 
na literatura \cite{jsmctb119}. A introdu\c{c}\~ao de um fen\'otipo no modelo 
Penna evita a utiliza\c{c}\~ao de tal termo, como veremos a seguir.

\section{Modelos com fen\'otipo.}

\begin{figure}
 \includegraphics[width=8cm,angle=0]{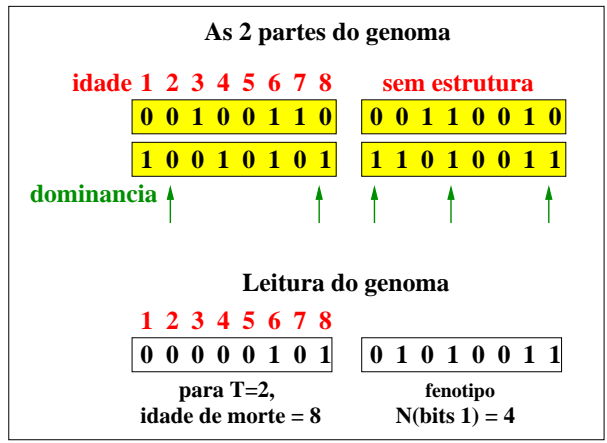}
 \includegraphics[width=8cm,angle=0]{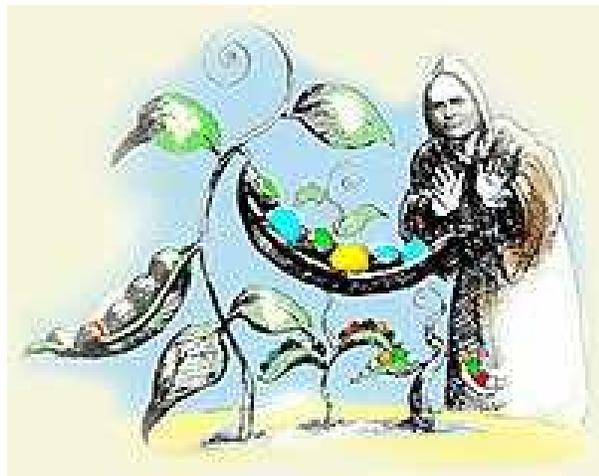}
\caption{\it Representa\c{c}\~ao das partes cronol\'ogica e sem estrutura de idade 
dos genomas. As setas indicam as posi\c{c}\~oes onde os bits 1 s\~ao dominantes e a leitura 
do genoma \'e feita de acordo com tais posi\c{c}\~oes. Para a parte estruturada por idade, 
tanto o n\'umero de bits 1 quanto suas respectivas posi\c{c}\~oes s\~ao relevantes; j\'a 
para a parte n\~ao estruturada, apenas o n\'umero efetivo de bits 1 importa.     
O desenho do lado direito mostra uma caricatura do pai da Gen\'etica
Moderna, Gregor Mendel, que introduziu os conceitos hoje conhecidos como
domin\^ancia e recessividade nos estudos estat\'{\i}sticos que realizou com ervilhas em 1865.}
\label{figparalelo}
\end{figure}

Para simular uma caracter\'{\i}stica fenot\'{\i}pica do indiv\'{\i}duo adotamos o mesmo 
procedimento introduzido por Medeiros, S\'a Martins, S. Moss de Oliveira e 
Medeiros \cite{medeiros}, isto \'e, acrescentamos mais 
um par de tiras de bits ao genoma cronol\'ogico do modelo Penna. Este novo par n\~ao apresenta 
estrutura de idade, mas suas tiras tamb\'em s\~ao lidas em paralelo. No momento da 
reprodu\c{c}\~ao este par sofre o mesmo processo de cruzamento e recombina\c{c}\~ao 
indicado na figura \ref{figcrossing} e existem tamb\'em posi\c{c}\~oes dominantes e recessivas 
para os bits 1. Este n\'umero de posi\c{c}\~oes dominantes \'e dominado $D$, para parte com estrutura 
e $DF$ para parte sem estrutura.
A \'unica diferen\cc a est\'a nas muta\c{c}\~oes, pois para 
este novo par elas podem se dar nas duas dire\c{c}\~oes ($0 \leftrightarrow 1$), com uma 
taxa $MF$ por bit (dos 32 bits escolhe-se s\'o 1 bit).  
O n\'umero total $k$ ($\in\co{0,32}$) de bits iguais a 1 \'e contado nestas duas tiras, 
levando-se em conta as posi\c{c}\~oes dominantes.  
Este n\'umero \'e ent\~ao relacionado a algum tipo de
carater\'{\i}stica fenot\'{\i}pica do ind\'{\i}viduo, conhecida na biologia 
como {\it tra\cc o}. Supondo que tal caracter\'{\i}stica seja, por exemplo, o tamanho do 
animal ou o tamanho da cauda de um p\'assaro, muta\c{c}\~oes para um lado ou para o outro 
apenas aumentam ou diminuem tal tamanho 
nas pr\'oximas gera\c{c}\~oes, o que em princ\'{\i}pio n\~ao pode ser considerado bom nem 
ruim. O car\'ater seletivo destas muta\c{c}\~oes s\'o ser\'a definido quando for estabelecida 
a ecologia \`a qual a popula\c{c}\~ao estar\'a sujeita e/ou a prefer\^encia sexual por 
determinado fen\'otipo na hora do acasalamento.  
A figura \ref{figparalelo} ilustra as duas partes do genoma de cada indiv\'{\i}duo e como 
\'e feita a leitura de cada parte. 

Ao nos referirmos apenas \`a parte n\~ao estruturada do modelo como fen\'otipo, estamos 
cometendo um abuso de linguagem. Como mencionado na introdu\c{c}\~ao, qualquer tradu\c{c}\~ao 
ou compila\c{c}\~ao do gen\'otipo corresponde a um fen\'otipo. Assim, a leitura da parte 
estruturada por idade do modelo Penna e representada por uma tira de bits na figura 
\ref{figparalelo}, lado esquerdo, tamb\'em corresponde a um fen\'otipo. Contudo, para 
distinguir com facilidade a parte estruturada da n\~ao estruturada, chamaremos de fen\'otipo 
apenas a tradu\c{c}\~ao da parte n\~ao estruturada do genoma. 

Na natureza existe uma enorme variedade de mecanismos
relacionados \`a capacidade do indiv\'{\i}duo com um dado fen\'otipo de produzir 
uma descend\^encia apta a sobreviver. Um destes mecanismos \'e o da sele\c{c}\~ao natural 
na competi\c{c}\~ao por diferentes fontes de alimento, que atua no indiv\'{\i}duo de acordo com  
seu tra\cc o ecol\'ogico, isto \'e, com sua caracter\'istica fenot\'{\i}pica relevante para   
a obten\c{c}\~ao de alimentos. Outro mecanismo \'e o da sele\c{c}\~ao sexual, que atua no 
indiv\'i{\i}duo de acordo com alguma caracter\'{\i}stica fenot\'{\i}pica relativa     
\`a prefer\^encia sexual no momento do acasalamento - o tra\c{c}o sexual. A sele\c{c}\~ao 
sexual \'e um ingrediente fundamental para se obter o isolamento reprodutivo na 
especia\c{c}\~ao simp\'atrica. Para as esp\'ecies cujo tra\c{c}o  
ecol\'ogico \'e o mesmo que o sexual, n\'os faremos uso de apenas um par de tiras de bits 
n\~ao estruturadas por idade. 
No caso em que a sele\c{c}\~ao natural devido \`a ecologia atua num tra\c{c}o que \'e 
diferente daquele utilizado para definir a prefer\^encia sexual, ser\~ao usados dois 
pares independentes de tiras de bits n\~ao estruturadas. 

\subsection{O tra\cc o ecol\'ogico} 

A caracter\'{\i}stica fenot\'{\i}pica de cada indiv\'{\i}duo ligada \`a competi\c{c}\~ao   
pelo alimento dispon\'{\i}vel fica ent\~ao definida pelo n\'umero 
$k$ de bits 1 efetivos no segundo par de tiras de bits de seu genoma ($0 \le k \le 32$).  
Em nossas simula\c{c}\~oes dividimos a popula\c{c}\~ao em tr\^es grupos, 
de acordo com os valores de $k$. O grupo $P_1$ consiste dos indiv\'{\i}duos com  
$0 \le k < n_1$; 
o grupo do meio, $P_m$, corresponde aos fen\'otipos intermedi\'arios onde   
$n_1 \le k \le n_2$ e o terceiro grupo, $P_2$, corresponde aos fen\'otipos com  
$n_2 < k \le 32$. 
Se supusermos que o valor de $k$ est\'a ligado ao tamanho dos indiv\'{\i}duos, os grupos 
$P_1$ e $P_2$ correspondem aos tamanhos extremos, sendo os indiv\'iduos do grupo $P_1$  
bem pequenos (com poucos bits 1 efetivos) e os do grupo $P_2$ bem grandes (ricos em bits 1).  

O car\'ater seletivo deste tra\c{c}o \'e dado pelo seguinte fator de Verhulst 
modificado:
\[
V_{grupo}(k,t)~=~\frac{\mbox{N\'umero de indiv\'{\i}duos com fen\'otipos que competem 
pela mesma fonte de comida}}
{\mbox{Capacidade de sustenta\c{c}\~ao da fonte de alimento dispon\'{\i}vel F(k,t)}}. 
\]

A seguinte competi\cc ao intra-espec\'{\i}fica \'e mantida fixa em todas as simula\coes, 
mas a capacidade de sustenta\c{c}\~ao da fonte de alimento dispon\'{\i}vel, $F(k,t)$, pode 
ou n\~ao variar no tempo. Com 
\^este novo fator de Verhulst as mortes por disputa de comida deixam de ser 
totalmente aleat\'orias e passam a depender dos fen\'otipos e da fonte de alimento 
dispon\'{\i}vel.   

A forma expl\'icita da competi\c{c}\~ao inter-espec\'{\i}fica, entre os indiv\'iduos de cada grupo 
\'e dada por \cite{toshitony}:
\begin{equation}\label{novo}
V(k,t)=\left\{
\begin{array}{cc} V_1(k,t) & se~0\le k<n_1, \\
V_m(k,t) & se~n_1\le k\le n_2, \\
V_2(k,t) & se~n_2< k\le 32.\end{array}\right.
\end{equation}
com a competi\c{c}\~ao intra-espec\'{\i}fica igual a:
\begin{eqnarray}
V_1(k,t)&=&\frac{P_1(k,t)+P_m(k,t)}{F(k,t)}, \nonumber\\
V_m(k,t)&=&\frac{P_m(k,t)+X~*~\co{P_1(k,t)+P_2(k,t)}}{F(k,t)}, \label{eqcompe}\\
V_2(k,t)&=&\frac{P_2(k,t)+P_m(k,t)}{F(k,t)}.\nonumber
\end{eqnarray}
onde $P_1(k,t)$ \'e a popula\c{c}\~ao com fen\'otipos em um dos extremos, $P_m(k,t)$ \'e a 
popula\c{c}\~ao do meio e 
$P_2(k,t)$ \'e a popula\c{c}\~ao com os fen\'otipos no outro extremo. O par\^ametro $X$ regula 
com que fra\c{c}\~ao das popula\c{c}\~oes dos extremos os indiv\'iduos de fen\'otipos 
intermedi\'arios ir\~ao competir. J\'a os indiv\'{\i}duos do grupo $P_1$ competem entre si 
e com todos aqueles de fen\'otipos intermedi\'arios, mas n\~ao competem com os fen\'otipos 
do outro extremo; o mesmo ocorre com os indiv\'{\i}duos do grupo $P_2$, que tamb\'em n\~ao 
competem com os do grupo $P_1$.  Em cada passo e para cada indiv\'{\i}duo com fen\'otipo $k$
gera-se um n\'umero aleat\'orio entre 0 e 1, cujo valor \'e comparado com o valor de $V(k,t)$. 
Se o n\'umero gerado for menor que $V(k,t)$, o indiv\'{\i}duo morre.
 
Este tipo de competi\c{c}\~ao j\'a \'e bastante adotado na 
literatura biol\'ogica da especia\c{c}\~ao simp\'atrica \cite{librogene}, porque  
produz um aumento na diferencia\c{c}\~ao gen\'etica entre 
grupos com fen\'otipos extremos, abrindo o caminho para o isolamento reprodutivo 
nestes grupos quando a sele\c{c}\~ao sexual \'e introduzida.  

\subsection{O tra\cc o sexual} 
A sele\c{c}\~ao sexual \'e definida atrav\'es do    
acasalamento n\~ao aleat\'orio. Adicionamos ao genoma de cada f\^emea mais um  
gene, representado por um \'unico bit, que determina se a f\^emea \'e seletiva ou n\~ao 
no momento da reprodu\c{c}\~ao. Se este bit for igual a 1, 
ela escolhe o parceiro de acordo com alguma prefer\^encia relacionada ao tra\c{c}o sexual; 
caso contr\'ario, ela escolhe aleatoriamente um macho para acasalar. 
Este gene \'e herdado da m\~ae pela filha com uma probabilidade de muta\c{c}\~ao (em ambas as 
dire\c{c}\~oes) igual a $MS$. Isto significa que as filhas de f\^emeas seletivas podem 
ou n\~ao ser seletivas e vice-versa. As tr\^es regras de sele\c{c}\~ao sexual que 
utilizaremos para obter o isolamento reprodutivo s\~ao:
\begin{itemize}
\item {\it Regra com forte dire\cao}~: As f\^emeas seletivas com fen\'otipos extremos 
escolhem entre um n\'umero $NM$ de machos, aquele com o fen\'otipo mais  
extremo do mesmo grupo que o dela. Supondo que o tra\c{c}o sexual seja tamb\'em o 
tamanho do animal, por exemplo, 
uma f\^emea grande tenta acasalar com o maior dos machos do conjunto $NM$, ao passo que 
uma f\^emea pequena escolhe o menor dos machos do conjunto.  
As f\^emeas seletivas com fen\'otipos intermedi\'arios   
atuam como as f\^emeas seletivas de um dos dois conjuntos de fen\'otipos extremos, 
com 50$\%$ de probabilidade para cada conjunto. 
\item {\it Regra com dire\c{c}\~ao}: F\^emeas seletivas com fen\'otipos extremos 
acasalam com o primeiro macho sorteado aleatoriamente que perten\c{c}a 
ao mesmo grupo fenot\'{\i}pico que ela. 
As f\^emeas seletivas com fen\'otipos intermedi\'arios se comportam como aquelas 
seletivas de um dos conjuntos extremos, definido a priori. 
\item {\it Regra da diferen\cc a}~: A f\^emea seletiva com fen\'otipo $k_F$ 
escolhe, dentre um n\'umero $NM$ de machos, aquele cujo fen\'otipo $k_M$ apresenta  
a menor diferen\c{c}a para o seu pr\'oprio $k_F$, isto \'e, 
\[
\mbox{m\'{\i}nimo}\lla{\abs{k_F-k_{M_1}},\abs{k_F-k_{M_2}},\cdots,\abs{k_F-k_{M_{NM}}}}
\]   
\end{itemize} 

Se o tra\c{c}o sexual for diferente do ecol\'ogico, acrescentamos mais um par de tiras 
de bits n\~ao estruturadas por idade ao genoma do indiv\'{\i}duo, cujo processo 
de cruzamento, recombina\c{c}\~ao e muta\c{c}\~ao \'e o mesmo do tra\c{c}o ecol\'ogico. 

A nomenclatura utilizada at\'e aqui para descrever os modelos ser\'a mantida nos 
cap\'{\i}tulos seguintes. 

Em todas as simula\c{c}\~oes a {\it popula\c{c}\~ao inicial}, de 3000 f\^emeas e 3000 machos, tem a 
parte cronol\'ogica do genoma constituida apenas de zeros, isto \'e, todos os indiv\'{\i}duos  
s\~ao beb\^es livres de doen\c{c}as heredit\'arias. 
A parte sem estrutura de idade, que 
define as caracter\'isticas externas dos indiv\'{\i}duos ligadas \`a ecologia e \`a 
prefer\^encia sexual \'e constituida de zeros 
e uns, aleatoriamente distribuidos. Para garantir que nossas medidas fossem tomadas em 
popula\c{c}\~oes j\'a equilibradas geneticamente, acompanh\'avamos sempre a distribui\c{c}\~ao 
de idades dos indiv\'{\i}duos e s\'o comput\'avamos os dados muitos passos depois da  
mesma ter se tornado constante.

\chapter{Modelos com um tra\cc o fenot\'{\i}pico} 
Com o tipo de din\^anima descrita no cap\'{\i}tulo anterior e inspirados num processo de 
especia\c{c}\~ao  
observado nos tentilh\~oes das ilhas Gal\'apagos, n\'os estudamos os poss\'{\i}veis 
mecanismos de especia\c{c}\~ao simp\'atrica em cadeias alimentares, utilizando     
popula\c{c}\~oes com um \'unico tra\cc o fenot\'{\i}pico e   
vivendo numa ecologia onde a fonte de alimento muda com o tempo.

Primeiro analisamos detalhadamente o processo de especia\c{c}\~ao simp\'atrica
de uma popula\c{c}\~ao isolada de herb\'{\i}voros, que s\~ao diretamente afetados  
pelas mudan\c{c}as morfol\'ogicas que ocorrem na fonte de alimento, causadas, por 
exemplo, por oscila\c{c}\~oes sazonais no regime de chuvas. 
Tentando ficar o mais perto poss\'{\i}vel da realidade e
levando em conta que na natureza existem predadores que comem herb\'{\i}voros, os quais 
por sua vez comem plantas, n\'os comparamos os processo de especia\c{c}\~ao simp\'atrica 
em cadeias alimentares de duas e de tr\^es esp\'ecies.
 
\section{Motiva\c{c}\~ao biol\'ogica}
Existem atualmente v\'arios trabalhos 
experimentais e te\'oricos sendo desenvolvidos acerca da evolu\c{c}\~ao dos tentilh\~oes, 
tamb\'em conhecidos como p\'assaros de Darwin. 
A significativa bibliografia das observa\c{c}\~oes  feitas nas ilhas
Gal\'apagos conta com estat\'{\i}sticas do tamanho do corpo, morfologia do bico e 
s\'eries temporais dos sons destes p\'assaros (ver figura~\ref{figsong}).
As medidas mostram que a diversifica\c{c}\~ao na morfologia dos 
bicos tem formado padr\~oes de evolu\c{c}\~ao vocal que indicam uma correla\c{c}\~ao entre o tamanho 
do bico e o tipo de canto~\cite{pn409,pnbio54}. 
Nesses trabalhos tamb\'em \'e mencionado que a restri\c{c}\~ao f\'{\i}sica que a morfologia do 
bico causa na emiss\~ao do tipo de som dos tentilh\~oes pode ter levado a popula\c{c}\~ao 
original a um isolamento 
reprodutivo e consequentemente, \`a uma r\'apida especia\c{c}\~ao \cite{psr207}. 

Nas simula\c{c}\~oes a seguir modelamos a especia\c{c}\~ao simp\'atrica em popula\c{c}\~oes  
onde existe uma rela\c{c}\~ao direta entre 
\aspas{o tamanho do bico} e \aspas{o som emitido} pelos \aspas{p\'assaros}. 
Isto \'e, popula\c{c}\~oes cujo tra\cc o ecol\'ogico \'e o mesmo que o sexual 
(tamanho do bico) e portanto necessitamos apenas de um par de tiras de bits n\~ao 
estruturadas por idade.   

\section{Especia\c{c}\~ao para herb\'{\i}voros}

Inicialmente estudaremos uma popula\c{c}\~ao de p\'assaros herb\'{\i}voros que se 
alimentam de plantas ou sementes cuja morfologia varia no tempo. 
Primeiro analisaremos a competi\c{c}\~ao intra-espec\'{\i}fica 
pelas fontes de alimento.  
A seguir introduziremos a sele\c{c}\~ao sexual que, em conjunto com a disputa 
por alimento, levar\'a a popula\c{c}\~ao a especiar. Isto \'e, a popula\c{c}\~ao inicial  
de fen\'otipo intermedi\'ario se dividir\'a em duas popula\c{c}\~oes  com 
fen\'otipos extremos diferentes e sexualmente isoladas. 

\subsection{Competi\c{c}\~ao sem sele\c{c}\~ao sexual}
Utilizaremos uma ecologia na qual a fonte de alimento
muda de uma fonte que \'e abundante para todos os fen\'otipos para outra
na qual os fen\'otipos intermedi\'arios s\~ao desfavorecidos em rela\c{c}\~ao aos 
extremos. A capacidade de sustenta\c{c}\~ao desta fonte varia no tempo da seguinte 
forma: 
\begin{equation}\label{figdplantas}
F(k,t)=2\times10^5\times\left\{\begin{array}{ll}
\mbox{1}& \mbox{ ~~~~~$\forall$ $0\le t\le 4\times10^3$}, \\ & \\
\mbox{0.1+\large{$\frac{\abs{16-k}}{20}$}} 
& \mbox{ ~~~~~$\forall$ $4\times10^3\le t\le 3\times10^4$,}\end{array}\right.
\end{equation}
onde $k$ \'e o fen\'otipo da popula\c{c}\~ao de herb\'{\i}voros 
e $t$ \'e o passo da simula\cao.

Os fatores de Verhulst aos quais os indiv\'{\i}duos est\~ao sujeitos s\~ao dados pela 
equa\c{c}\~ao \ref{eqcompe}.  
Para $t \le 4 \times 10^3$ a capacidade de sustenta\c{c}\~ao do ambiente independe  
dos fen\'otipos. Corresponde, por exemplo, a uma \'epoca de chuvas onde existe uma 
grande variedade de tamanhos de plantas ou sementes. A partir do passo $4 \times 10^3$ a 
capacidade de sustenta\c{c}\~ao passa a depender dos fen\'otipos, ou melhor, 
cada p\'assaro passa a ter maior ou menor facilidade de arrumar comida, dependendo de seu 
valor de $k$. Este per\'{\i}odo corresponderia \`as \'epocas de seca, que ocorrem devido a 
fen\^omenos clim\'aticos como el Ni\~no, quando a variabilidade de alimento diminui 
sensivelmente.

\subsection{Resultados}

As tabelas \ref{tabpennap} e \ref{tabpolyp} mostram 
os valores dos par\^ametros do modelo utilizados nas simula\coes.
\bigskip

\begin{table}[htbp]
\begin{center}
{\footnotesize
\begin{tabular}{|l||c||c|}
\hline
Descri\c{c}\~ao das grandezas para (macho e f\^emea) & Nome & Valor\\
\hline\hline
Muta\c{c}\~ao (na tira de bits) & M & 1 (bit de 32 bits)*\\
\hline
Domin\^ancia & D &  1 \\ \hline
Limite de doen\c{c}as & L & 1 \\ \hline
N\'umero de filhos& NF  & 10 \\ \hline
Idade m\'{\i}nima de reprodu\c{c}\~ao & $R$ &9 \\ \hline
\hline
\end{tabular}
\caption{\it Par\^ametros relativos \`a parte estruturada por idade.}
\label{tabpennap}
}
\end{center}
\end{table}
\begin{table}[htbp]
\begin{center}
{\footnotesize
\begin{tabular}{|l||c||c|}
\hline
Grandezas (para machos e f\^emeas) & Nome & Valor\\
\hline\hline
Domin\^ancia & DF & 16* \\ \hline
Muta\c{c}\~ao do fen\'otipo num bit  & MF & 0.8 \\ \hline
Grau de competi\c{c}\~ao   & X & 0.5* \\ \hline
Fen\'otipos intermedi\'arios & $k$ &    16*  \\ \hline
Fen\'otipos extremos & $k$ & $<16$ \, e \, $>16$*  \\ \hline
\hline
\end{tabular}
\caption{\it Par\^ametros relativos \`a parte n\~ao estruturada por idade. 
S\~ao considerados fen\'otipos intermedi\'arios apenas aqueles com $k=16$. 
*Grandezas que n\~ao mudam neste cap\'{\i}tulo.}
\label{tabpolyp}
}
\end{center}
\end{table}

A distribui\c{c}\~ao de fen\'otipos da popula\c{c}\~ao no primeiro per\'{\i}odo 
de tempo corresponde a uma gaussiana centrada em $k=16$ 
(ver figura \ref{figdoscila}, quadrados).
Este comportamento unimodal \'e est\'atico e deve-se ao n\'umero de 
posi\c{c}\~oes  dominantes do fen\'otipo ($DF=16$) e ao grau de competi\c{c}\~ao ($X=0.5$). 
Neste intervalo de tempo n\~ao existe uma sele\c{c}\~ao do fen\'otipo em fun\c{c}\~ao da  
fonte de alimento. 
\bigskip

A distribui\c{c}\~ao de fen\'otipos gerada no segundo per\'{\i}odo de tempo pode    
apresentar-se de duas formas diferentes, para um mesmo conjunto de par\^ametros 
(tabela~\ref{tabpolyp}), dependendo apenas da semente aleat\'oria inicial: 

\begin{figure}[htbp]
\begin{center}
\includegraphics[width=6.5cm,angle=270]{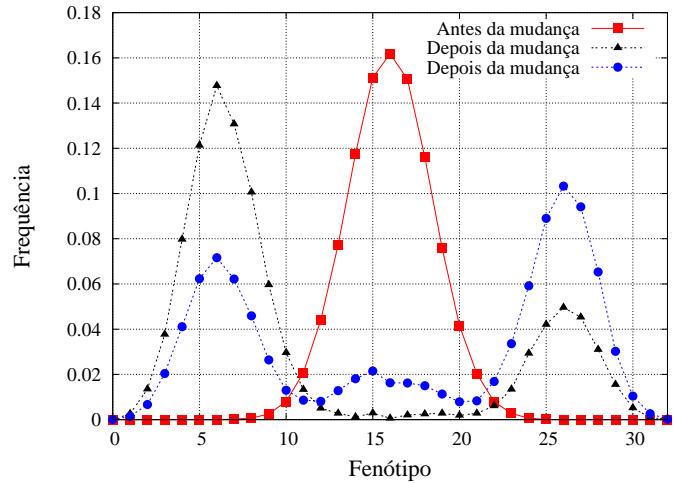}
\end{center}
\caption{\it Distribui\c{c}\~ao de fen\'otipos da popula\c{c}\~ao. No primero 
per\'{\i}odo de tempo, antes da ecologia mudar, a distribui\c{c}\~ao \'e unimodal (quadrados); 
no segundo per\'{\i}odo de tempo ($t > 40\times10^3$) aparece um estado 
estacion\'ario no qual a distribui\c{c}\~ao oscila entre aquela representada pelos c\'{\i}rculos 
e aquela representada pelos tri\^angulos. 
As m\'edias dos fen\'otipos extremos permanecem centradas em 6 e 26, respectivamente.}
\label{figdoscila}
\end{figure} 
\begin{figure}[htbp]
\begin{center}
\includegraphics[width=6cm,height=8cm,angle=270]{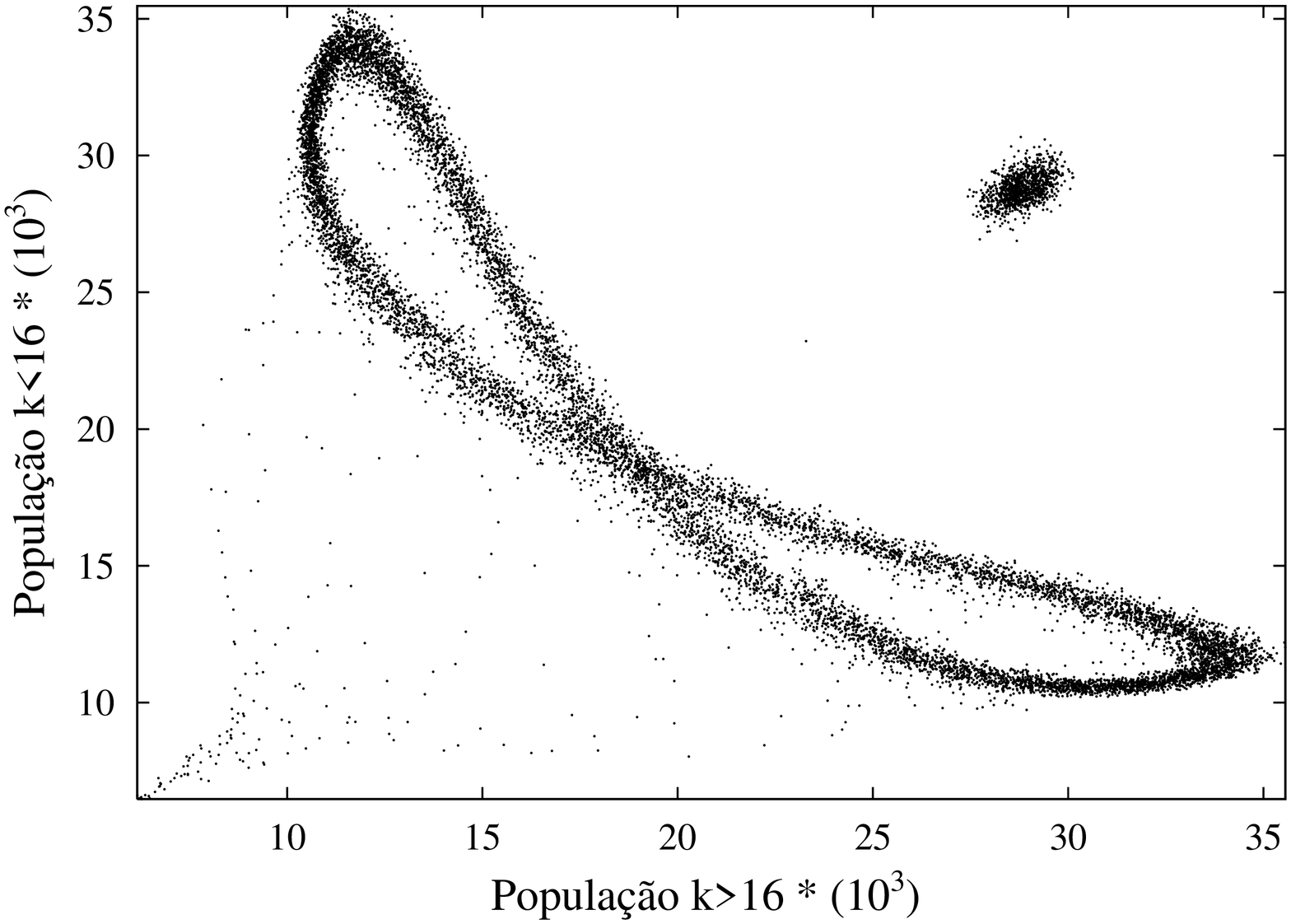}
\includegraphics[width=6cm,height=8cm,angle=270]{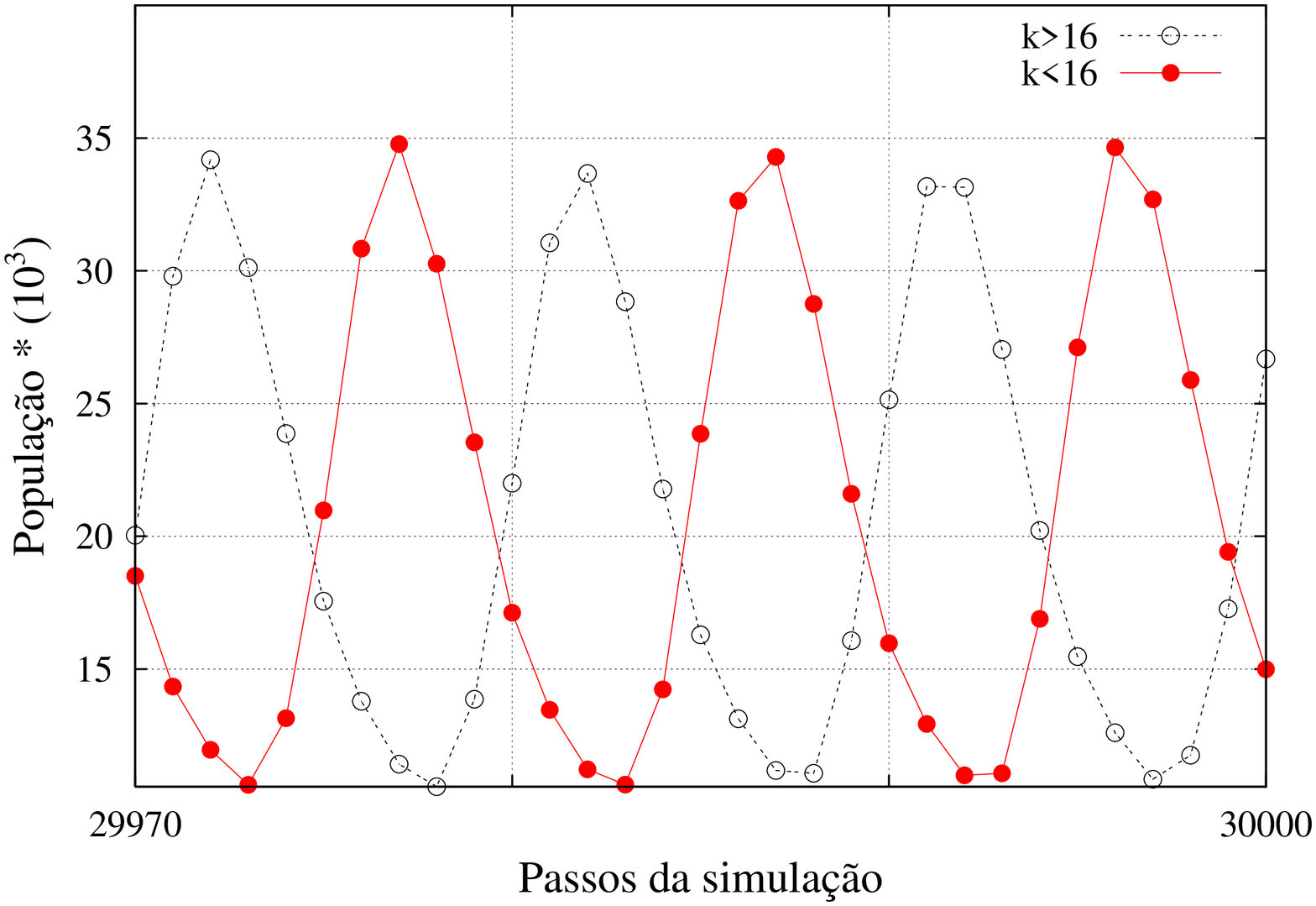}
\end{center}
\caption{\it Din\^amica da popula\c{c}\~ao de herb\'{\i}voros no caso do
estado estacion\'ario oscilat\'orio da figura anterior. Lado esquerdo: pode-se observar o 
comportamento dos tamanhos das popula\c{c}\~oes  com fen\'otipos extremos, \aspas{pequenos} versus  
\aspas{grandes}, durante toda a simula\c{c}\~ao. O atrator semelhante a um borr\~ao  
nesta figura corresponde aos tamanhos das popula\c{c}\~oes dos extremos antes da ecologia 
mudar. 
Lado direito: pode-se observar como o tamanho de cada popula\c{c}\~ao dos extremos varia com o 
tempo, depois da mudan\c{c}a na ecologia.}
\label{figpoposcila}
\end{figure}

\noindent i) Ela pode permanecer em um estado estacion\'ario 
oscilat\'orio entre duas {\it distribui\c{c}\~oes bimodais},    
representadas, respectivamente, pelos c\'{\i}rculos e tri\^angulos da figura \ref{figdoscila}. 
Esta variabilidade gen\'etica \'e chamada de {\it polimorfismo}, termo que foi introduzido 
por E.B. Ford \cite{librogene} em 1940 e est\'a associado \`a variabilidade gen\'etica 
mediante sele\cao. Indica que a popula\c{c}\~ao de  
p\'assaros como um todo carrega tanto os genes para tamanhos de bicos grandes quanto para 
pequenos. No caso da capacidade de alimento variar novamente, voltando \`a distribui\c{c}\~ao 
inicial, o mesmo ocorrer\'a com a distribui\c{c}\~ao de bicos (voltar\'a a ser unimodal), 
j\'a que nada impede o acasalamento entre bicos extremos oposto, gerando filhotes 
de bico m\'edio. Observe que a exist\^encia de um polimorfismo \'e uma condi\c{c}\~ao 
necess\'aria para que uma especia\c{c}\~ao venha a ocorrer, caso alguma sele\c{c}\~ao sexual 
seja introduzida para provocar o isolamento reprodutivo entre fen\'otipos extremos opostos.
\bigskip

\begin{figure}[ht]
\begin{center}
\includegraphics[width=5.5cm,angle=270]{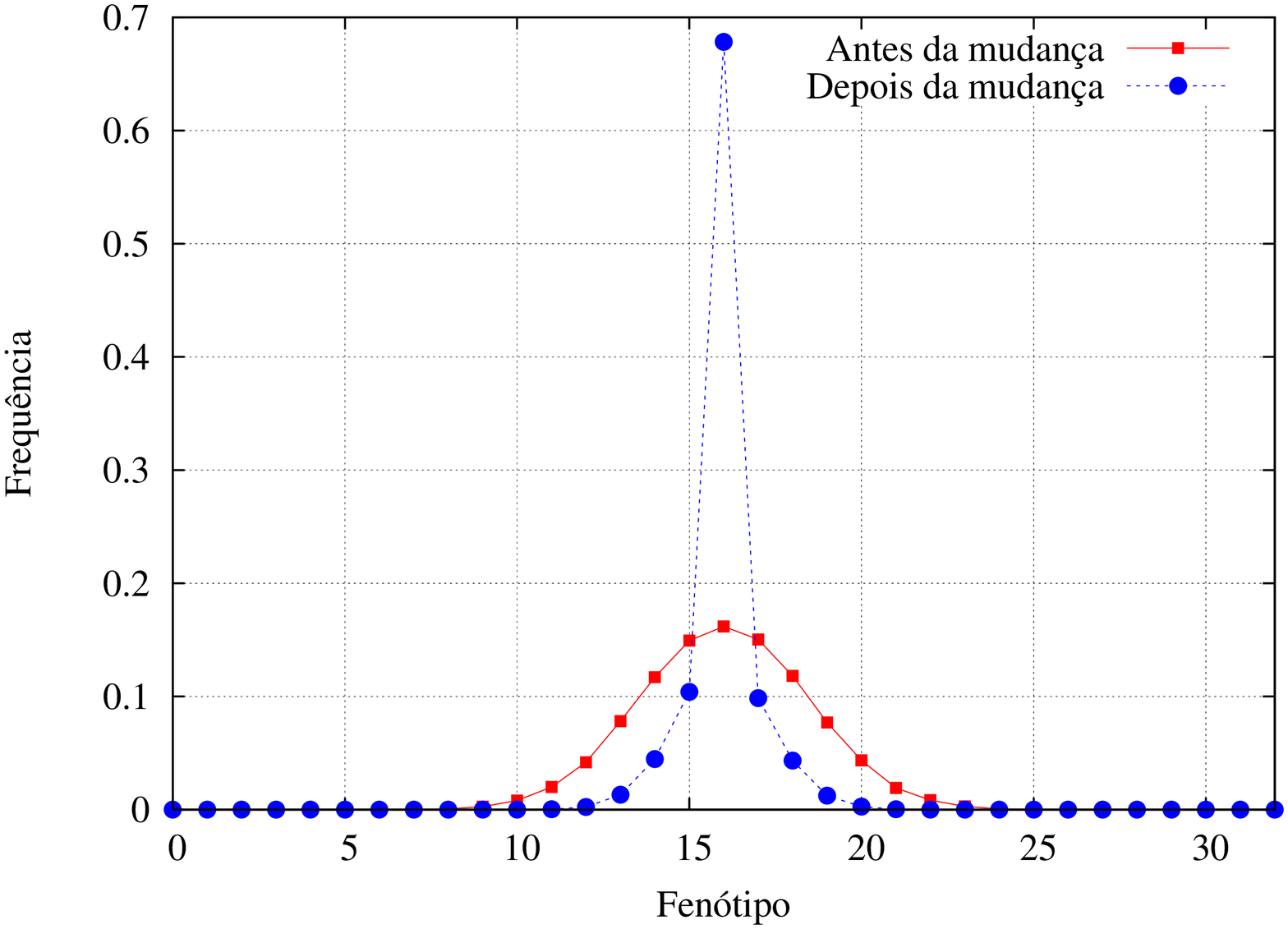}
\includegraphics[width=6cm,height=8cm,angle=270]{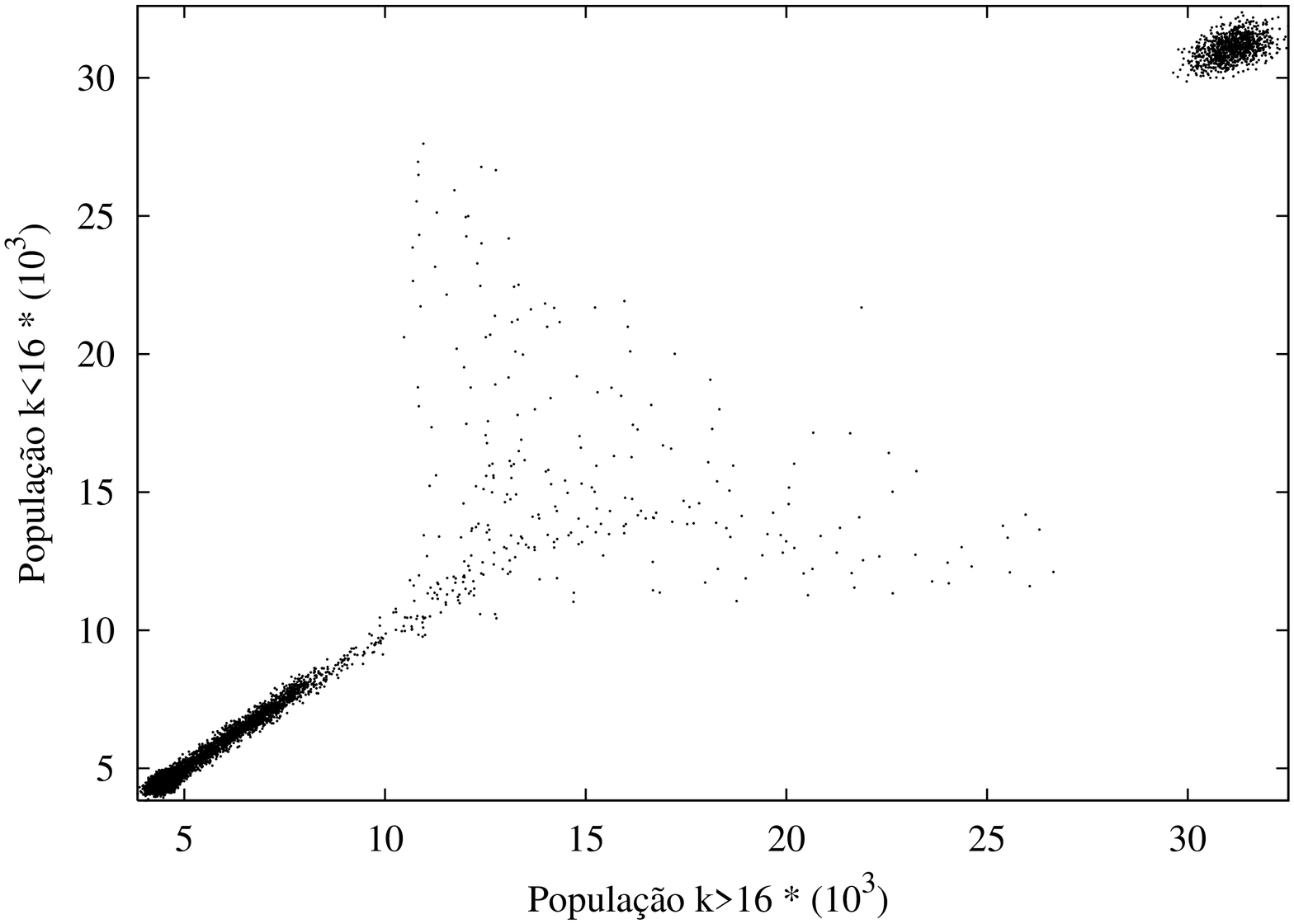}
\end{center}
\caption{\it Distribui\c{c}\~ao de fen\'otipos no caso estacion\'ario
sem variabilidade. Lado esquerdo: distribui\c{c}\~oes antes e depois da ecologia 
mudar. Lado direito: tamanhos das popula\c{c}\~oes com fen\'otipos extremos durante toda 
a simula\c{c}\~ao.}
\label{figdpopsem}
\end{figure}

Em ambas as distribui\c{c}\~oes bimodais da figura \ref{figdoscila} as popula\c{c}\~oes  com 
fen\'otipos extremos s\~ao maiores do que a 
popula\c{c}\~ao de fen\'otipo intermedi\'ario ($k=16$). Contudo, ora a popula\c{c}\~ao de 
bicos grandes \'e maior que a de bicos pequenos, ora \'e o contr\'ario. 
Este comportamento oscilat\'orio entre as 
popula\c{c}\~oes  dos extremos tem per\'{\i}odo igual a 10 passos de simula\c{c}\~ao e pode ser visto na 
figura~(\ref{figpoposcila}). Observe que este per\'{\i}odo corresponde tamb\'em ao n\'umero de 
intervalos de tempo (passos) desde o nascimento (idade zero) at\'e a idade 
m\'{\i}nima de reprodu\c{c}\~ao da popula\c{c}\~ao, $R=9$. 
\bigskip

\noindent ii) O polimorfismo pode tamb\'em desaparecer: a distribui\c{c}\~ao de 
fen\'otipos evolui para um estado estacion\'ario sem variabilidade,
onde a distribui\c{c}\~ao \'e tamb\'em unimodal mas de largura menor que a inicial 
(ver figura~\ref{figdpopsem}). 
Neste caso a popula\c{c}\~ao dos extremos permanece constante e pequena, 
e a popula\c{c}\~ao do meio domina, apesar da escassez da fonte de alimento para estes fen\'otipos.
\bigskip

\begin{figure}[htbp]
\begin{center}
\includegraphics[width=6.5cm,angle=270]{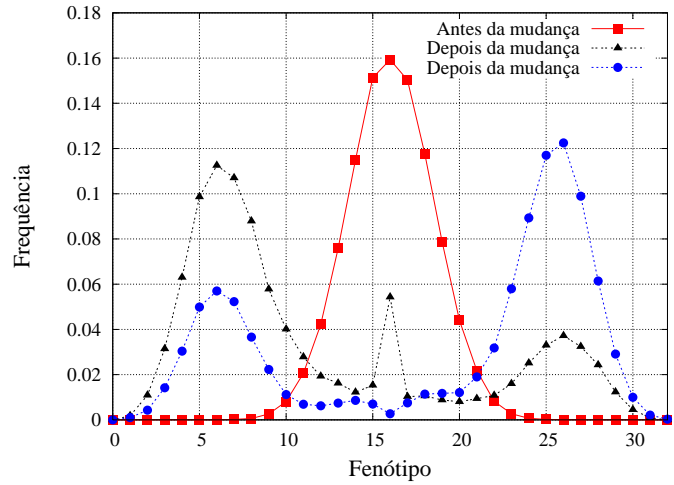}
\end{center}
\caption{\it Polimorfismo inst\'avel no qual a distribui\c{c}\~ao de fen\'otipos 
oscila entre a curva preta (tri\^angulos) e a curva azul (c\'{\i}rculos), para 
finalmente se tornar unimodal e est\'avel como na figura \ref{figdpopsem}}.  
\label{figdvaria}
\end{figure} 
\begin{figure}[htbp]
\begin{center}
\includegraphics[width=6cm,height=8cm,angle=270]{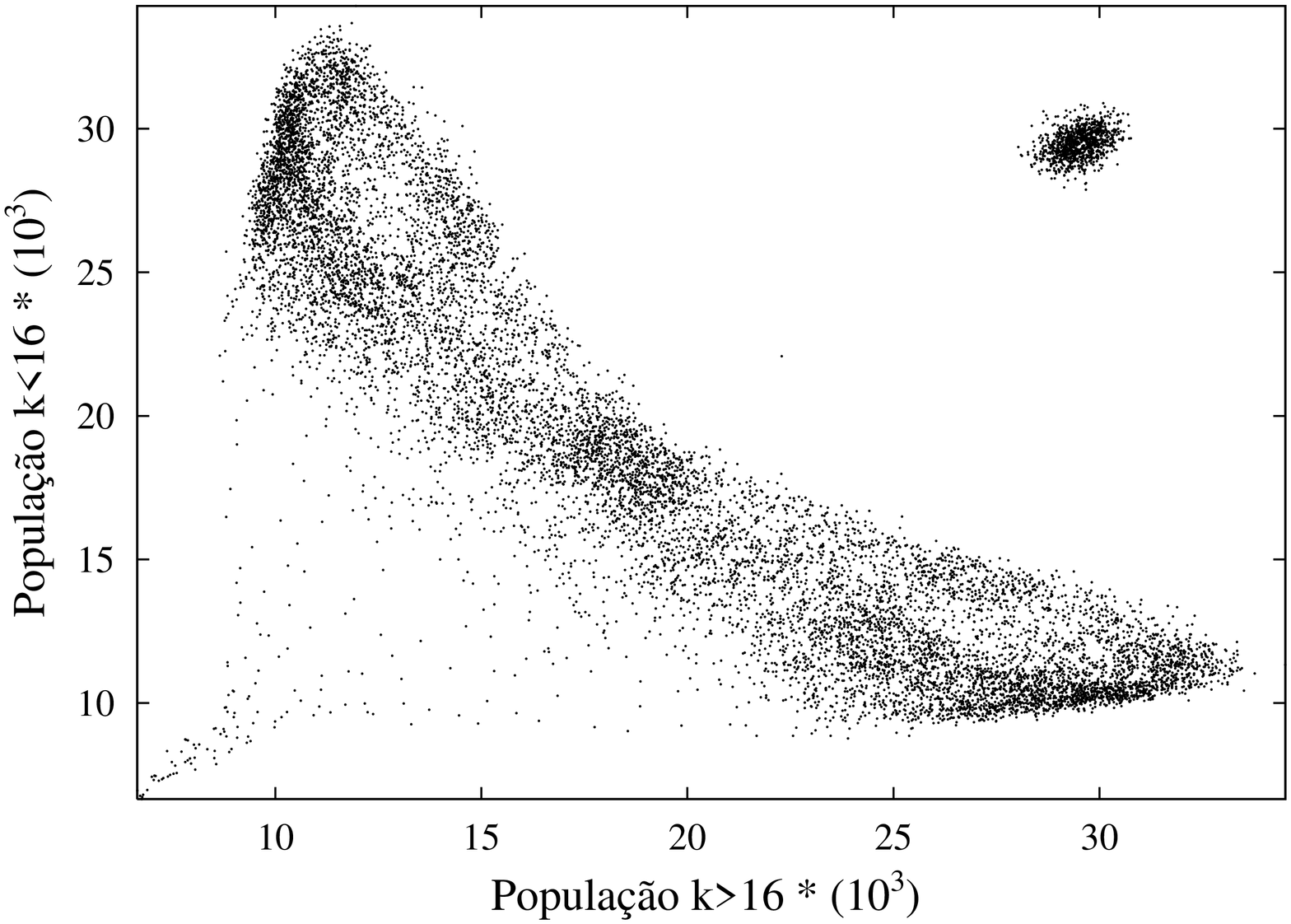}
\includegraphics[width=6cm,height=8cm,angle=270]{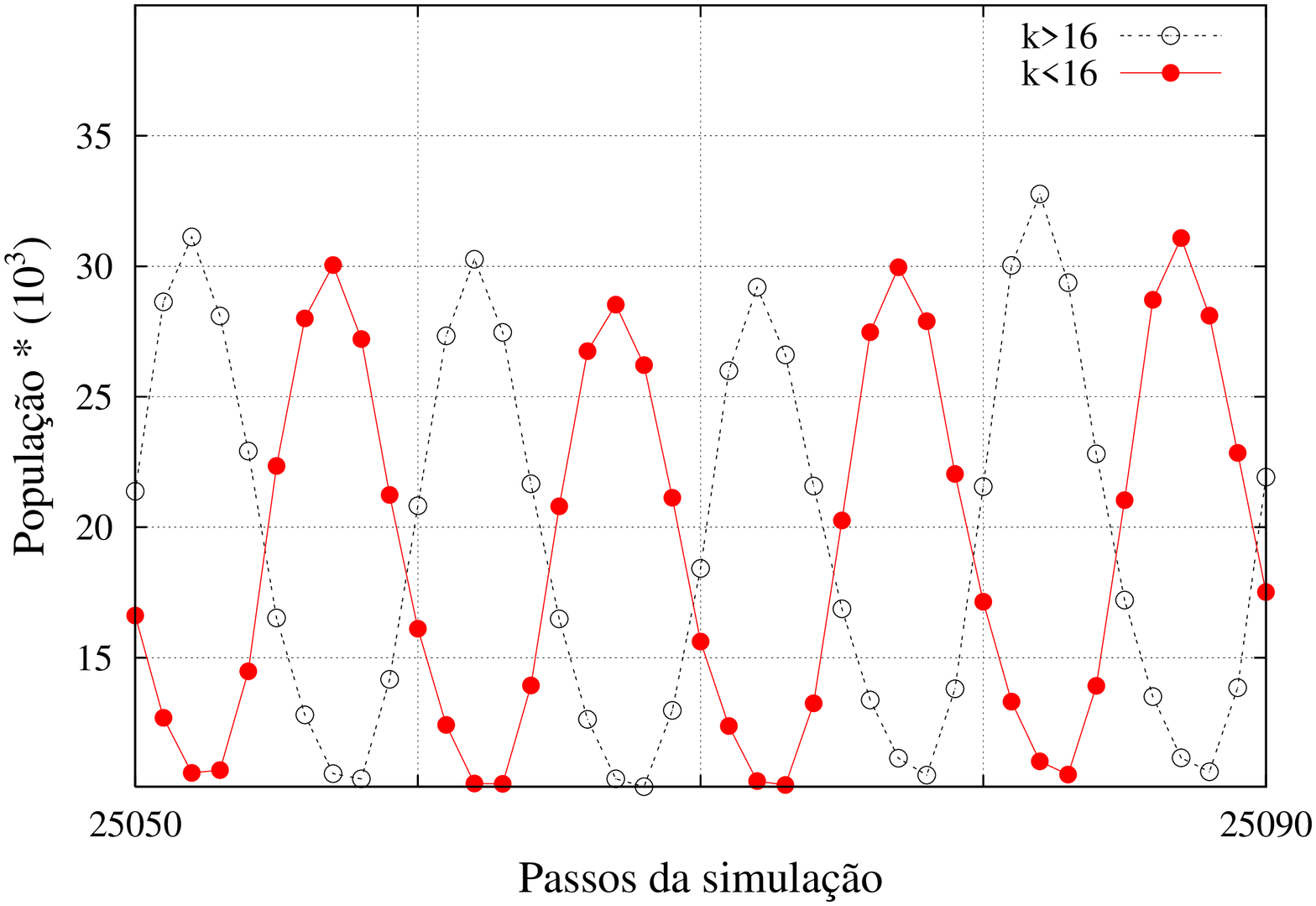}
\end{center}
\caption{\it Lado direito: Tamanhos das popula\c{c}\~oes de fen\'otipos  
extremos no estado em que o polimorfismo \'e meta-est\'avel, medidos durante in\'umeros 
passos ap\'os a ecologia mudar. 
Lado esquerdo: oscila\c{c}\~oes nos tamanhos das popula\c{c}\~oes dos extremos, cujas 
amplitudes variam no tempo.} 
\label{figpopvaria}
\end{figure}

As distribui\c{c}\~oes finais em (i) ou (ii), {\it antes de serem atingidas}, passam    
por um estado oscilat\'orio {\it inst\'avel} no qual a popula\c{c}\~ao de  
fen\'otipo intermedi\'ario chega a ter um tamanho compar\'avel aos das 
popula\c{c}\~oes  com fen\'otipos extremos 
(ver escala da figura~\ref{figdvaria}, tri\^angulos). O tempo que a distribui\c{c}\~ao passa 
neste estado inst\'avel pode, com uma pequena probabilidade, ser bastante longo, permitindo  
que se observe as mesmas oscila\c{c}\~oes nos tamanhos das popula\c{c}\~oes dos extremos 
presentes no caso (i), mas cujas amplitudes variam no tempo, como mostra o lado direito da 
figura \ref{figpopvaria}.
\bigskip

Nas simula\c{c}\~oes  pode-se favorecer o aparecimento do polimorfismo
oscilat\'orio est\'avel mudando-se 
o valor da taxa de muta\c{c}\~ao fenot\'{\i}pica $MF$ (veja tabela \ref{tabpolyresul}). 
Para valores pequenos desta muta\cao,~por exemplo para $MF=0.01$, de 10 
simula\c{c}\~oes  feitas com sementes aleat\'orias diferentes obtivemos 10 estados 
estacion\'arios deste tipo. Nestes casos a popula\c{c}\~ao final \'e grande, j\'a que 
as m\'edias dos fen\'otipos se concentram em valores pr\'oximos de 0 e 32, para os quais 
a fonte de comida \'e m\'axima. 

\begin{table}[htbp] 
\begin{center}
\begin{tabular}{lcl} 
 \hline 
 \hline 
 $MF$ & M\'edia dos fen\'otipos extremos & Estabilidade \\ \hline 
 1.0  & 7 e 25   & 7 oscilat\'orios e 3 unimodais \\
0.8   & 6 e 26   & 2 oscilat\'orios, 7 unimodais e 1 inst\'avel\\ 
0.1   & 1 e 31   & 10 oscilat\'orios\\
0.01  & 0 e 32   & 10 oscilat\'orios\\
\hline
\hline
 \end{tabular} 
 \caption{\it Resultados de 10 simula\c{c}\~oes  para diferentes taxas de muta\c{c}\~ao do fen\'otipo.} 
 \label{tabpolyresul} 
 \end{center}
 \end{table}

Uma outra forma de se favorecer o aparecimento de um polimorfismo est\'avel \'e aumentar 
o grau de competi\c{c}\~ao para o fen\'otipo do meio, por exemplo trocando $X=0.5$ por $X=1.0$. 
Neste caso os fen\'otipos intermedi\'arios ficam de tal forma prejudicados que impedem  
que a distribui\c{c}\~ao unimodal sem variabilidade (fig. \ref{figdpopsem}) seja atingida.

\subsection{Sele\c{c}\~ao sexual e Especia\cao}
Como mencionado no cap\'{\i}tulo anterior, a sele\c{c}\~ao sexual \'e definida pela 
f\^emea, pois \'e ela que escolhe o macho. As f\^emeas com \aspas{gene seletivo} 
igual a 1 decidem, dentre $NM$ machos, o melhor parceiro para acasalar;  
as f\^emeas com \aspas{gene seletivo} igual a 0 acasalam aleatoriamente. 
\'E importante lembrar que as f\^emeas herdam este gene  
da m\~ae com uma taxa de muta\c{c}\~ao revers\'{\i}vel, $MS$.

\begin{figure}[htbp]
 \includegraphics[width=5.5cm,angle=270]{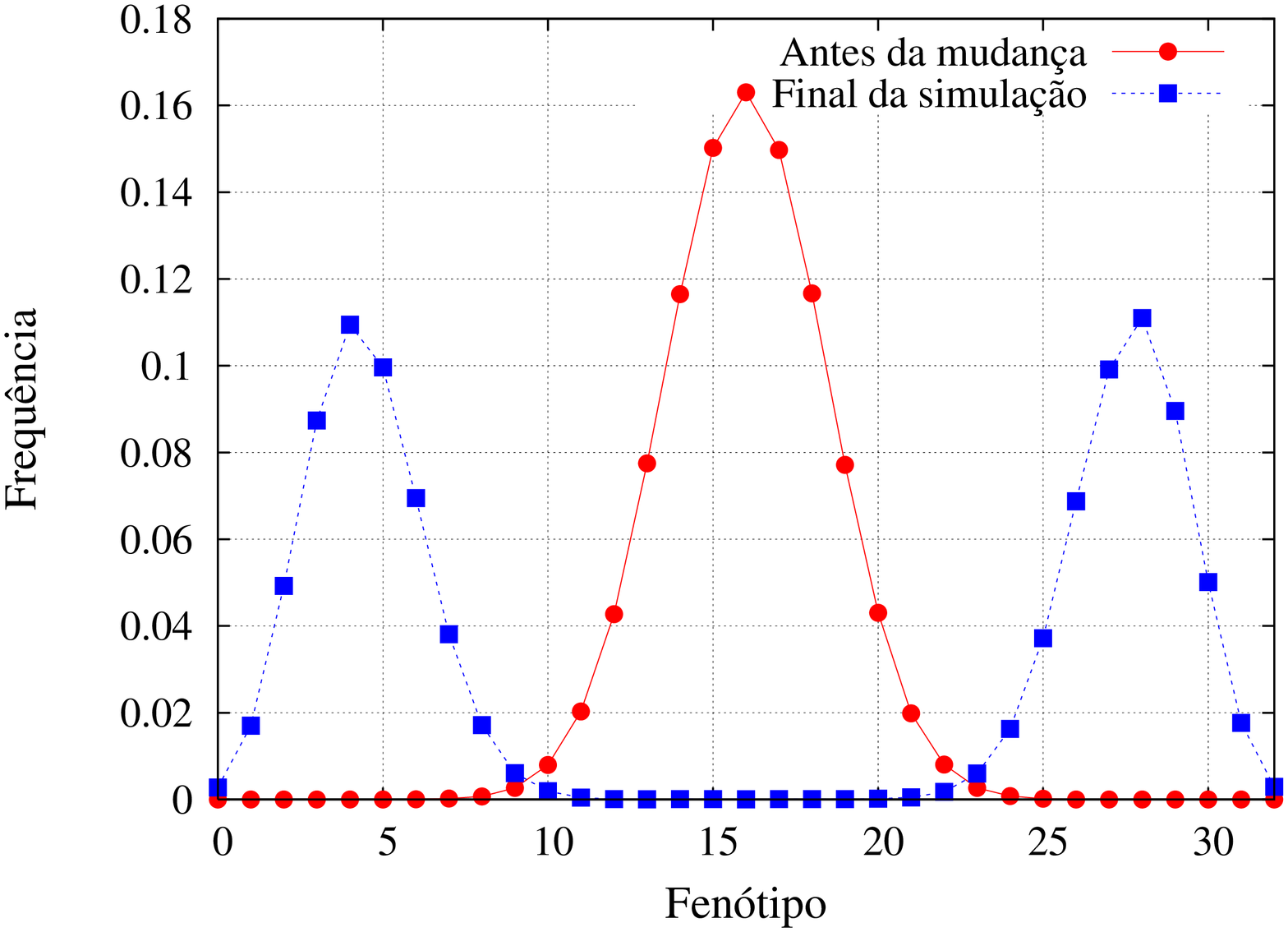}
 \includegraphics[width=5.5cm,angle=270]{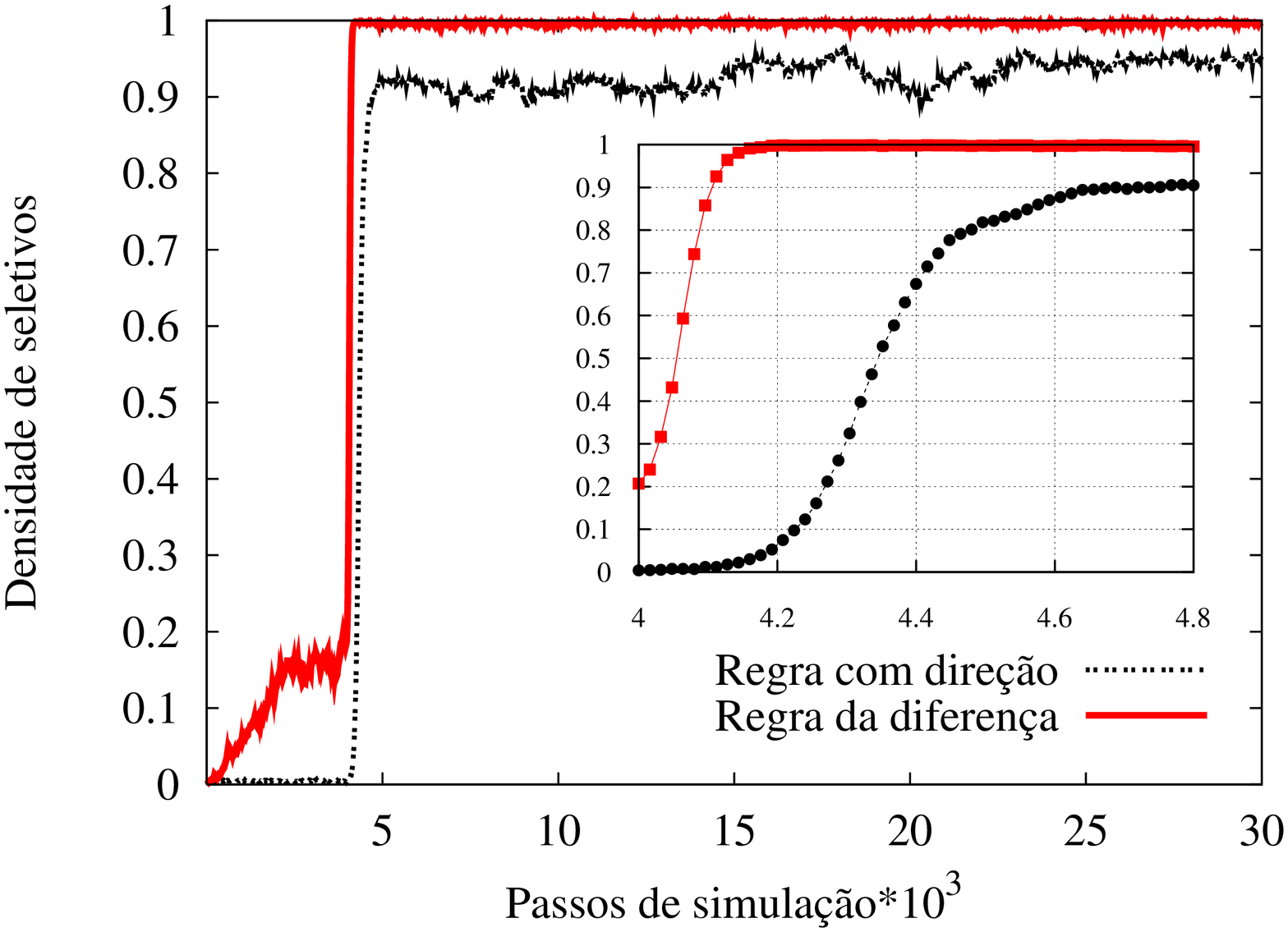}
\caption{\it Lado esquerdo: Distribui\c{c}\~ao de fen\'otipos da popula\c{c}\~ao com  
sele\c{c}\~ao ecol\'ogica e sexual. Lado direito: 
densidade de f\^emeas seletivas para duas regras diferentes de acasalamento.
O detalhe (``inset'') mostra a diferen\c{c}a de velocidades na obten\c{c}\~ao do isolamento 
reprodutivo, dependendo da regra.}
\label{figdbp}
\end{figure}

Iniciamos as simula\c{c}\~oes  com todas as f\^emeas n\~ao seletivas (bit de seletividade 
igual a zero para todas).
Para qualquer valor de $MF$, $MS=0.001$, $NM=6$ e com grau de 
competi\c{c}\~ao $X=0.5$, obt\'em-se uma distribui\c{c}\~ao est\'avel,  
com as duas popula\c{c}\~oes  de fen\'otipos extremos isoladas reprodutivamente:  
pode-se observar na figura~\ref{figdbp} que a popula\c{c}\~ao de fen\'otipo intermedi\'ario 
desaparece completamente (lado esquerdo), j\'a que $95 \%$ da popula\c{c}\~ao de f\^emeas 
se torna seletiva (lado direito).   
Esta estabilidade \'e obtida tanto com a regra com dire\c{c}\~ao como com a regra da 
diferen\c{c}a para o acasalamento, descritas no final do cap\'{\i}tulo anterior.  
Note que o n\'umero de f\^emeas 
seletivas aumenta rapidamente (entre 20 e 60 gera\c{c}\~oes), como pode ser visto pelo 
detalhe da figura~\ref{figdbp}, lado direito.

Comparando-se as figuras \ref{figdoscila} e \ref{figdbp} pode-se concluir que a 
popula\c{c}\~ao de fen\'otipo intermedi\'ario da figura \ref{figdoscila} era formada apenas 
por beb\^es resultantes do acasalamento entre fen\'otipos extremos diferentes. Voltaremos 
a este ponto na \'ultima se\c{c}\~ao. 
Aumentando-se o valor da taxa de muta\c{c}\~ao $MS$, a densidade de seletivos diminui,  
mas os valores m\'edios da distribui\c{c}\~ao n\~ao mudam e nem tampouco sua estabilidade. 

Ressaltamos ainda que tamb\'em realizamos simula\c{c}\~oes onde os fen\'otipos intermedi\'arios 
correspondiam a uma faixa de valores, e n\~ao apenas a $k=16$. Desde que estes fen\'otipos 
estivessem dentro do intervalo $n_1=10 \le k \le n_2=22$, os resultados obtidos foram 
qualitativamente os mesmos. Contudo, para $n_1 < 10$ e $n_2 > 22$ a probabilidade de se 
obter a especia\c{c}\~ao diminui, t\~ao mais quanto mais pr\'oximos os valores de 
$n_1$ e $n_2$ estiverem dos extremos.

\section{Especia\c{c}\~ao para predadores}
No processo de especia\c{c}\~ao anterior a esp\'ecie herb\'{\i}vora sofre o efeito 
direto da mudan\cc a na ecologia, isto \'e, da mudan\c{c}a na distribui\c{c}\~ao de plantas. 
Na natureza existem tamb\'em esp\'ecies que sofrem o efeito indireto de uma  
sele\c{c}\~ao ecol\'ogica, como por exemplo, as esp\'ecies predadoras que se alimentam do  
herb\'{\i}voro que se alimenta destas plantas. 
Nesta se\c{c}\~ao n\'os estudaremos o processo de especia\c{c}\~ao simp\'atrica neste tipo de 
cadeia alimentar, onde por exemplo, p\'assaros (predadores) se alimentam 
de insetos (herb\'{\i}voros), os quais se alimentam de plantas.    

Os insetos s\~ao representados por seus c\'odigos gen\'eticos  
mas n\~ao sofrem sele\c{c}\~ao sexual;  
alimentam-se das plantas cuja capacidade de sustenta\c{c}\~ao, $F(k,t)$, \'e dada pela 
equa\c{c}\~ao (\ref{figdplantas}).
A din\^amica desta popula\c{c}\~ao j\'a foi analisada na se\c{c}\~ao anterior e as 
poss\'{\i}veis distribui\c{c}\~oes  de fen\'otipos destes herv\'{\i}boros, $\rho_h(k,t)$, 
s\~ao as mesmas apresentadas nas  
figuras \ref{figdoscila}, \ref{figdpopsem} e \ref{figdvaria}. A tabela \ref{tabpassainse} 
mostra os par\^ametros do modelo utilizados nas simula\c{c}\~oes desta se\c{c}\~ao. Observe 
que os par\^ametros relativos aos herb\'{\i}voros nesta tabela s\~ao os mesmos utilizados 
na se\c{c}\~ao anterior (tabela \ref{tabpolyp}).

A capacidade de sustenta\c{c}\~ao da fonte de alimento dos predadores, que 
tamb\'em s\~ao representados por seus c\'odigos gen\'eticos mas que sofrem sele\c{c}\~ao 
sexual, \'e dada por:
\begin{equation}\label{eqrelacao}
F_p(k,t)=2\times10^6\times\rho_h(k,t),
\end{equation}
onde $k$ \'e o fen\'otipo dos herb\'{\i}voros e $t$, 
o n\'umero de passos da simula\cao.~Esta equa\c{c}\~ao liga diretamente o comportamento da 
popula\c{c}\~ao de predadores ao da popula\c{c}\~ao de herb\'{\i}voros. 
Desta forma, tanto a competi\c{c}\~ao por alimento entre os herb\'{\i}voros como aquela  
entre os predadores \'e dada pela equa\c{c}\~ao \ref{eqcompe}, mudando apenas a capacidade 
de sustenta\c{c}\~ao da fonte de alimento de cada esp\'ecie.   

\subsection{Resultados}

\begin{table}[ht]
\begin{center}
{\footnotesize
\begin{tabular}{|l||l||l|}
\hline
Grandeza & Predador& Herb\'{\i}voro\\
\hline\hline
 $M$ & 1 &  1\\ \hline
 $D$ & 10 & 1\\ \hline  
 $L$ & 10 & 1\\ \hline
 $R$ &10 & 9\\ \hline 
 $NF$  & 2&  10\\ \hline \hline
 $DF$ & 16 & 16\\ \hline
 $MF$ & 0.5 & 0.8\\ \hline
 $MS$ &  0.001 &  0.0\\ \hline
 $X$ & 0.5 & 0.5 \\ \hline
\end{tabular}
\caption{\it Par\^ametros do modelo para as duas esp\'ecies, relativos tanto aos    
genomas cronol\'ogicos (parte superior da tabela) quanto aos genomas sem estrutura 
de idade (parte inferior da tabela). Os fen\'otipos intermedi\'arios correspondem  
\`aqueles com $k=16$, tanto para herb\'{\i}voros como para predadores.} 
 \label{tabpassainse} 
 }
\end{center}
\end{table}

\begin{figure}[htbp]
\begin{center}
\includegraphics[width=6.5cm,angle=270]{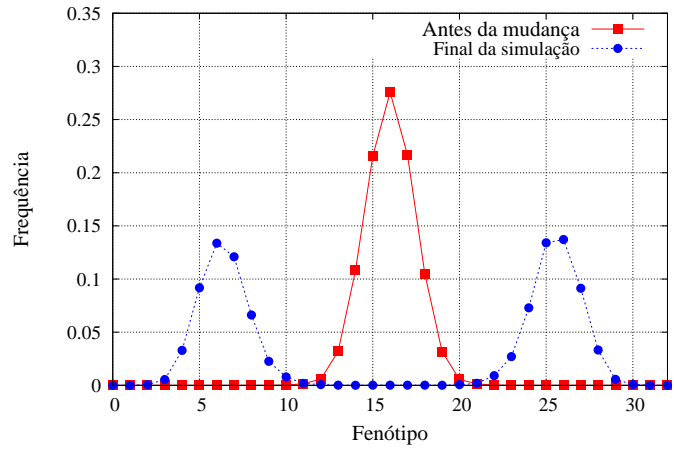}
\end{center}
\caption{\it Distribui\c{c}\~ao de fen\'otipos da popula\c{c}\~ao de 
predadores, que est\'a sendo afetada pela mudan\c{c}a s\'ubita na fonte de aliment\c{c}\~ao 
dos herb\'{\i}voros.} 
\label{figdbinsecto}
\end{figure}

A distribui\c{c}\~ao de fen\'otipos da popula\c{c}\~ao de predadores 
pode ser vista na figura \ref{figdbinsecto}, antes e depois da distribui\c{c}\~ao de 
plantas mudar. Note que apesar da taxa de muta\c{c}\~ao dos fen\'otipos dos  
predadores ($MF=0.5$) ser diferente daquela dos herb\'{\i}voros ($MF=0.8$), as 
distribui\c{c}\~oes  dos fen\'otipos extremos dos predadores est\~ao centradas em 6 e 26, 
isto \'e, nas mesmas posi\c{c}\~oes que as popula\c{c}\~oes  extremas de herb\'{\i}voros 
(veja lado esquerdo da figura~\ref{figdoscila}).

\begin{figure}
\begin{center}
\includegraphics[width=6cm,height=8cm,angle=270]{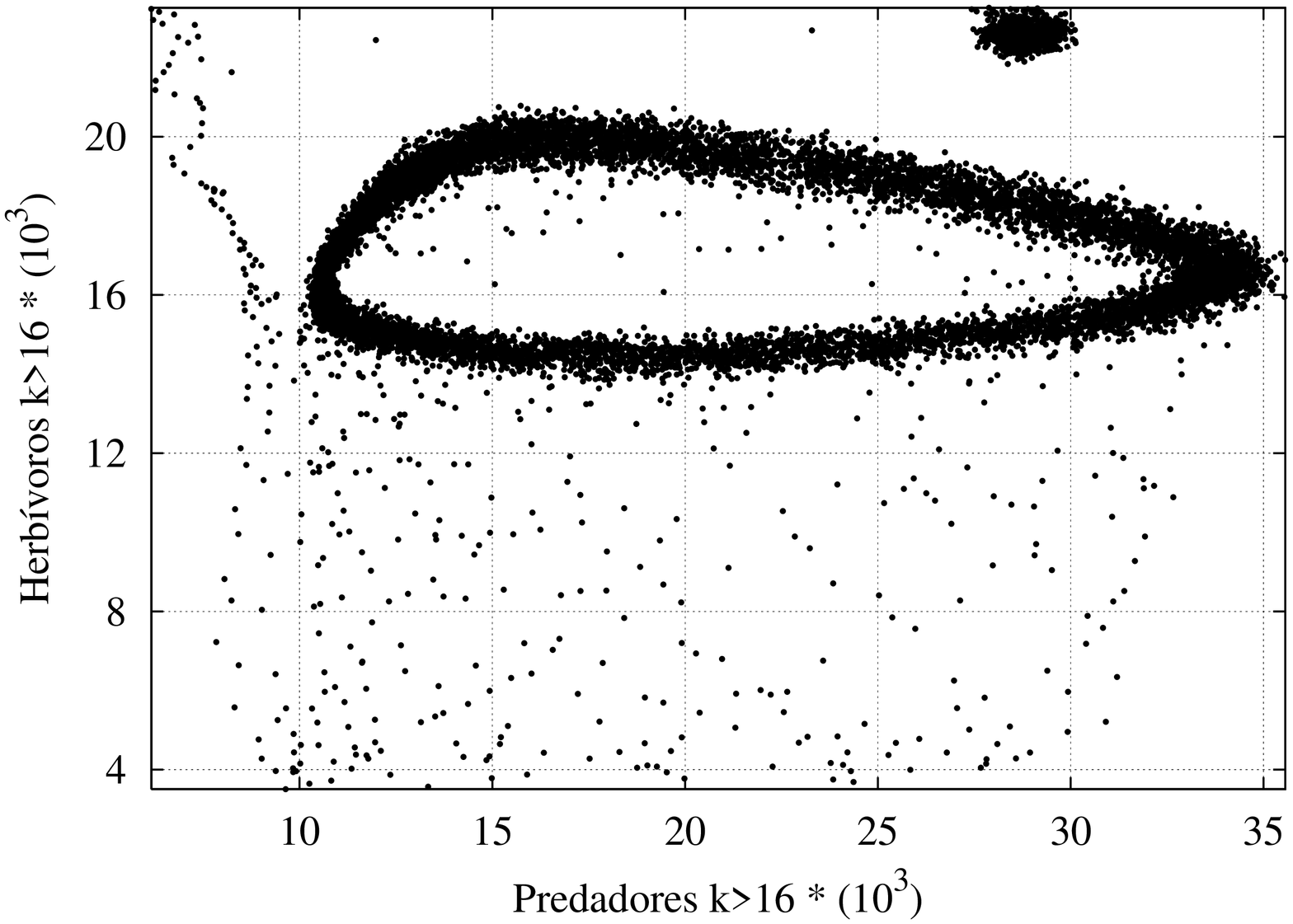}
\includegraphics[width=6cm,height=8cm,angle=270]{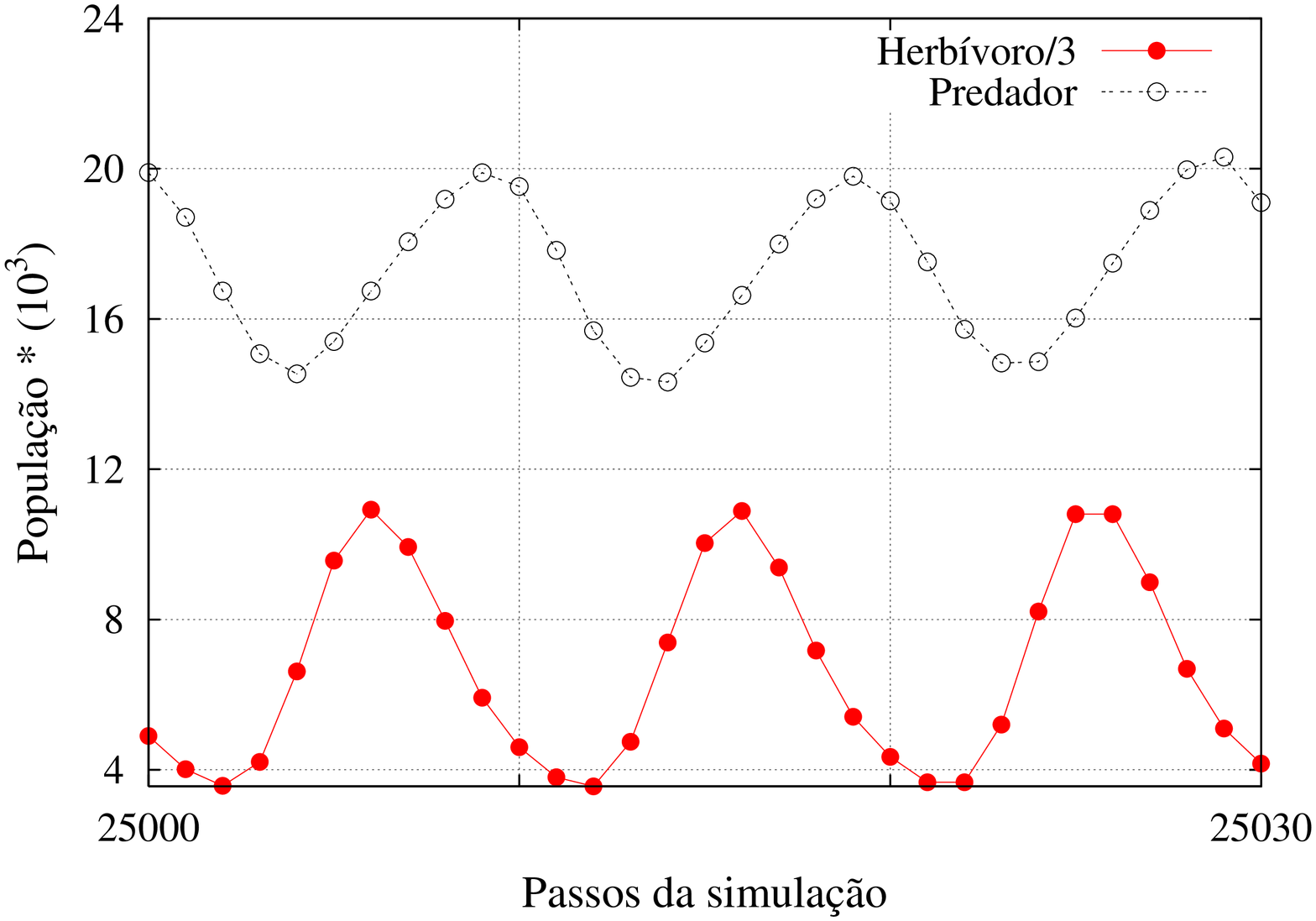}
\end{center}
\caption{\it Lado esquerdo: Popula\c{c}\~ao de herb\'{\i}voros com fen\'otipos  
extremos grandes em fun\c{c}\~ao da popula\c{c}\~ao de predadores tamb\'em com fen\'otipos 
extremos grandes, durante toda uma simula\c{c}\~ao onde ocorreu a especia\c{c}\~ao dos 
predadores.  
Lado direito: oscila\c{c}\~oes nas popula\c{c}\~oes de fen\'otipos extremos, tanto de 
herb\'{\i}voros quanto de predadores.}  
\label{figpopbird}
\end{figure}
\begin{figure}
\begin{center}
\includegraphics[width=5.5cm,angle=270]{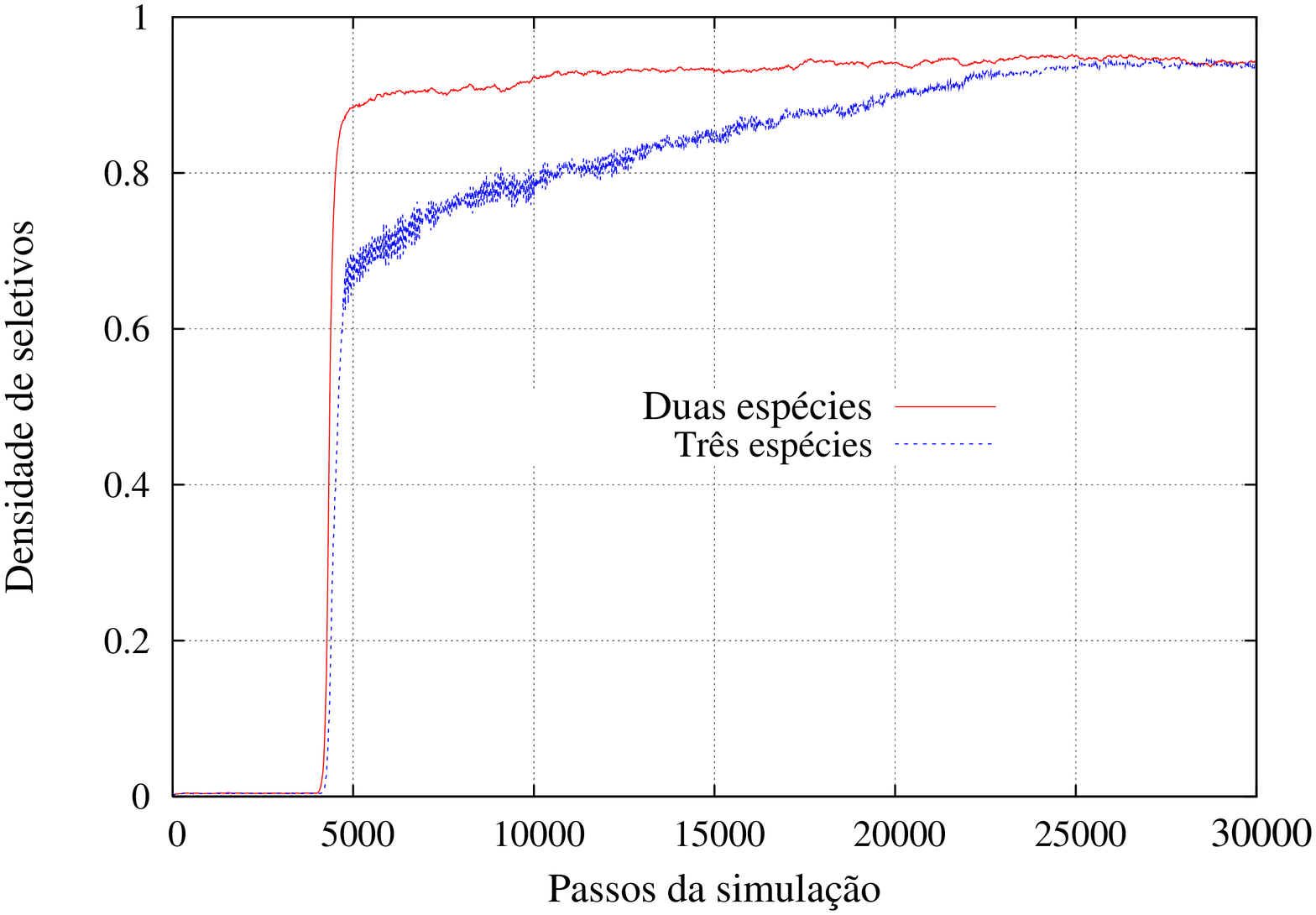}
\end{center}
\caption{\it \aspas{Velocidades} de especia\cao em cadeias alimentares  
de duas (curva superior, vermelha) e de t\^res esp\'ecies (curva inferior, 
azul).}
\label{figveloc}
\end{figure}

A especia\c{c}\~ao simp\'atrica dos predadores s\'o ocorre quando o  
o polimorfismo oscilat\'orio dos herb\'{\i}voros \'e atingido, mas a rec\'{\i}proca n\~ao \'e 
verdadeira, isto \'e, nem sempre o plimorfismo dos herb\'{\i}voros leva \`a especia\c{c}\~ao 
dos predadores. Quando esta especia\c{c}\~ao ocorre, a popula\c{c}\~ao de predadores com fen\'otipos 
extremos apresenta oscila\c{c}\~oes  com per\'{\i}odo igual ao 
das oscila\c{c}\~oes da popula\c{c}\~ao de herb\'{\i}voros tamb\'em com fen\'otipos extremos
\footnote{Lembre-se que o   
per\'{\i}odo das oscila\c{c}\~oes da popula\c{c}\~ao de herb\'{\i}voros depende da idade 
m\'{\i}nima de reprodu\c{c}\~ao dos mesmos - ver figura \ref{figpoposcila}.}, 
independentemente de qual seja a idade m\'{\i}nima de reprodu\c{c}\~ao dos predadores
(ver tabela~\ref{tabpassainse}).

Como pode-se ver no lado direito da figura \ref{figpopbird},  
existe uma defasagem de 3 passos entre as oscila\c{c}\~oes da popula\c{c}\~ao de 
herb\'{\i}voros com fen\'otipos extremos e aquelas da popula\c{c}\~ao de predadores com 
fen\'otipos extremos, defasagem esta que n\~ao muda 
com nehum dos par\^ametros da tabela \ref{tabpassainse}.

No caso em que a muta\c{c}\~ao no fen\'otipo dos herb\'ivoros vale $MF=1.0$ e a muta\c{c}\~ao no  
fen\'otipo dos predadores vale $MF=0.01$, obt\'em-se casos onde a popula\c{c}\~ao de herb\'{\i}voros 
apresenta um estado estacion\'ario com polimorfismo oscilat\'orio, 
mas a popula\c{c}\~ao de predadores n\~ao especia, isto \'e, apresenta s\'o uma 
carater\'{\i}stica fenot\'{\i}pica, que tanto pode ser a de um dos extremos ($k < 16$) como 
a do outro ($k > 16$). 

A figura \ref{figveloc} mostra que numa cadeia alimentar de tr\^es esp\'ecies (plantas, 
herb\'{\i}vores e predadores) a 
especia\c{c}\~ao \'e mais lenta do que numa cadeia alimentar de duas esp\'ecies (plantas e 
herb\'{\i}voros), como era de se esperar.

\section{Perspectivas}

\begin{figure}[ht]
\begin{center}
\includegraphics[width=4.5cm,angle=270]{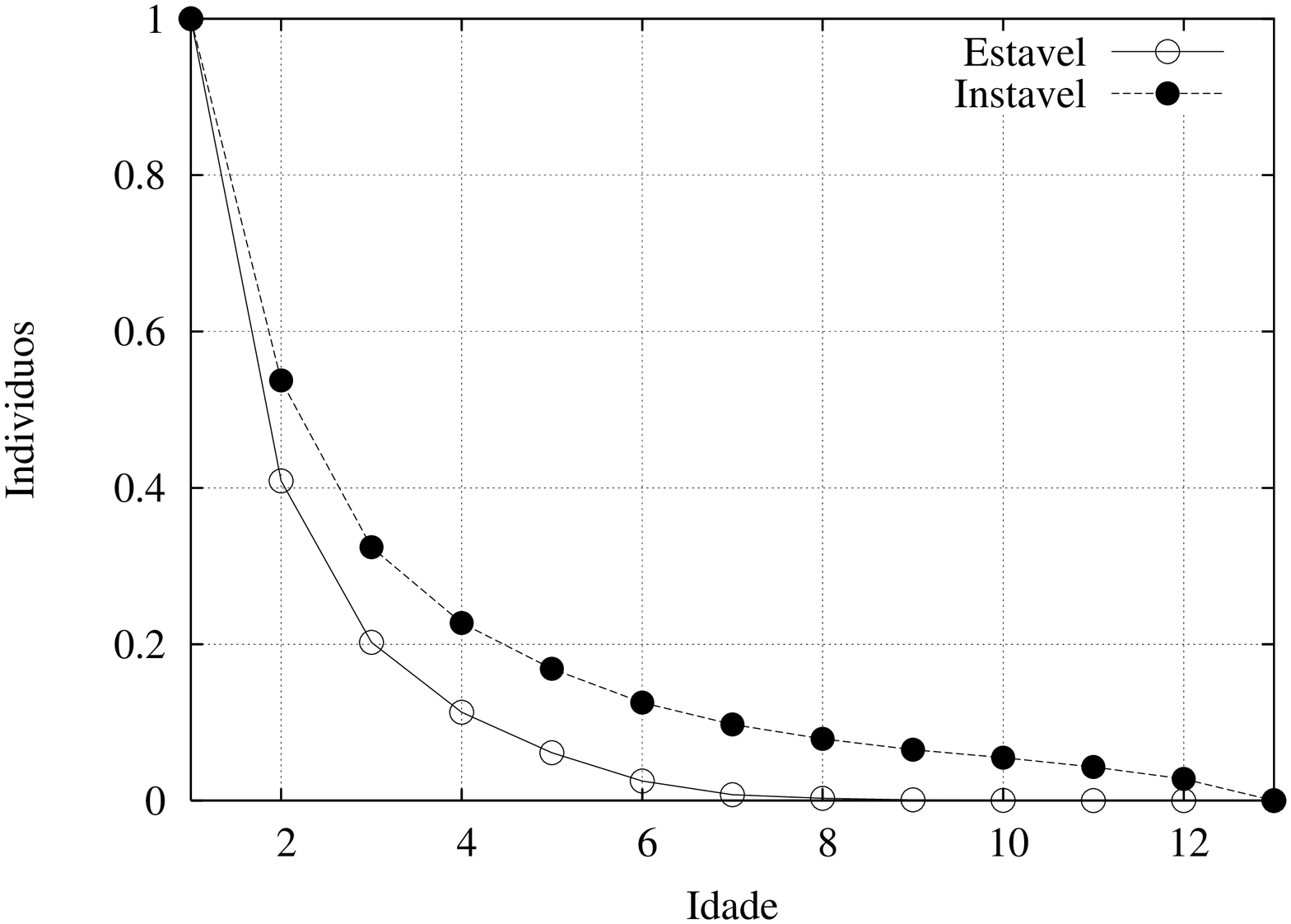}
\includegraphics[width=4.5cm,angle=270]{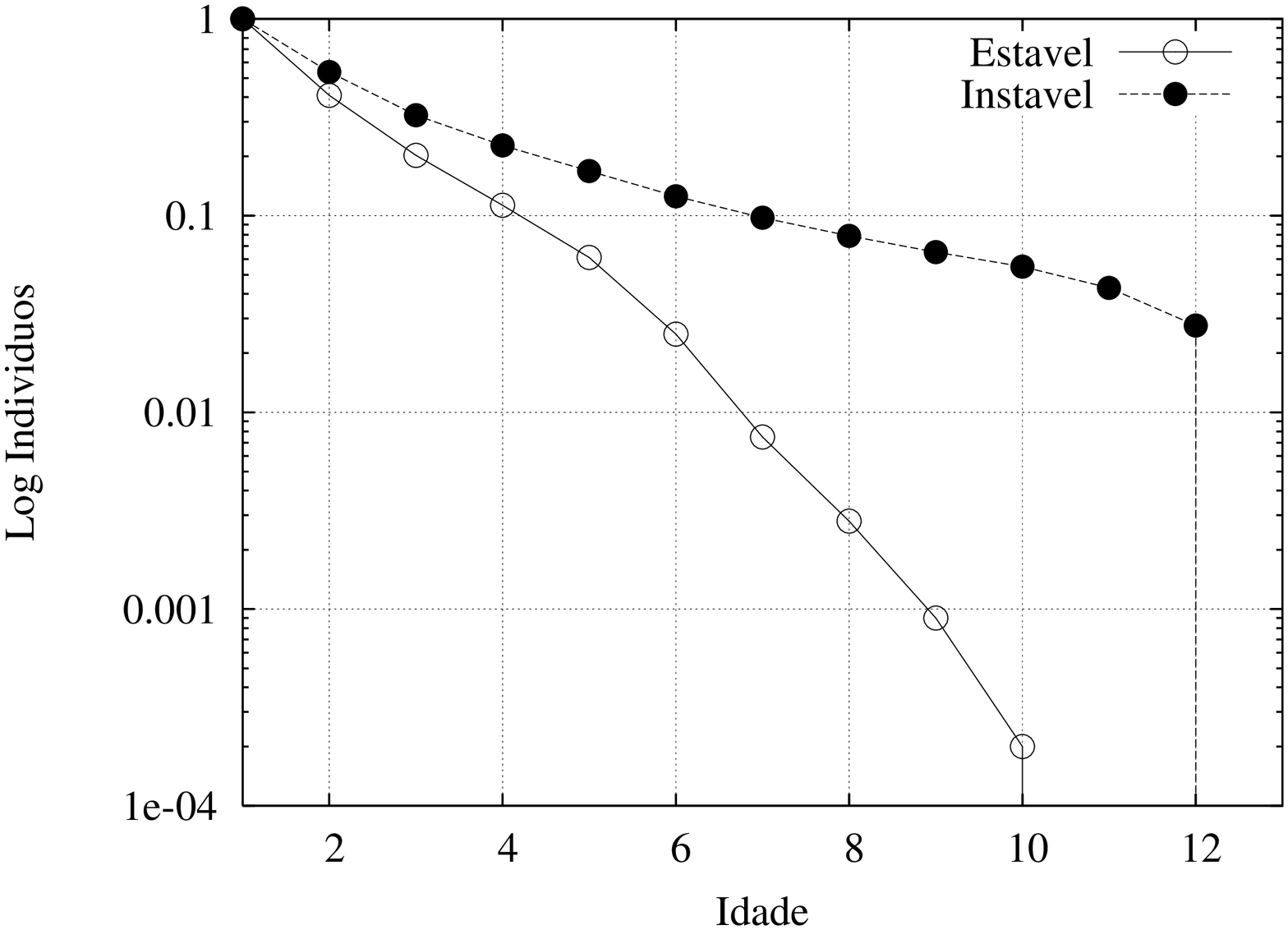}
\end{center}
\caption{\it Lado esquerdo: Distrubui\c{c}\~oes m\'edias por idade das   
popula\c{c}\~oes de fen\'otipo intermedi\'ario, medidas nos \'ultimos 
250 passos e normalizadas (divididas pelos seus respectivos valores 
m\'aximos). Pode-se observar que a longevidade dos h\'{\i}bridos \'e bem maior no caso 
inst\'avel do que no est\'avel, em $R=9$. Lado direito: as mesmas distribui\c{c}\~oes em escala 
semi-logar\'{\i}tmica mostram mais claramente a diferen\c{c}a entre os dois casos.}
\label{figsobrev}
\end{figure}

O polimorfismo oscilat\'orio inst\'avel 
(figuras~\ref{figdvaria} e~\ref{figpopvaria}), 
dif\'{\i}cil de ser observado, apresenta carater\'{\i}sticas de um estado meta-est\'avel 
(tempo de relaxa\c{c}\~ao grande). 
Al\'em disso, mant\'em os tr\^es tipos de fen\'otipos   
(dois extremos e um intermedi\'ario) com popula\c{c}\~oes de tamanhos compar\'aveis, pois 
neste caso a popula\c{c}\~ao intermedi\'aria n\~ao \'e composta apenas por beb\^es: apresenta  
a mesma curva de mortalidade das popula\c{c}\~oes  com fen\'otipos extremos. A figura 
\ref{figsobrev} compara as distribui\c{c}\~oes por idade das popula\c{c}\~oes 
{\it intermedi\'arias}, nos dois casos em que o polimorfismo aparece, o est\'avel e o 
inst\'avel. Esta distribui\c{c}\~ao tem a mesma forma daquela usualmente obtida com o modelo 
Penna e mostra que no caso est\'avel, de fato a quase totalidade dos h\'{\i}bridos 
s\~ao beb\^es resultantes do acasalamento entre fen\'otipos extremos, beb\^es estes que 
n\~ao chgar\~ao vivos at\'e a idade m\'{\i}nima de reprodu\c{c}\~ao, $R=9$. J\'a no caso 
inst\'avel, os h\'{\i}bridos conseguem sobreviver o suficiente para reproduzir durante 
alguns per\'{\i}odos.  

Seria interessante compreender que par\^ametros interferem no tempo de vida deste estado 
meta-est\'avel e verificar se existe alguma faixa de par\^ametros para a qual 
este polimorfismo se torna est\'avel.   

Os fato de que as oscila\c{c}\~oes  na popula\c{c}\~ao de predadores (cadeia alimentar 
de tr\^es esp\'ecies) tenham per\'{\i}odo igual ao das oscila\c{c}\~oes na popula\c{c}\~ao de 
herb\'{\i}voros, parece ser devido \`a rela\c{c}\~ao direta entre estas duas esp\'ecies, 
dada pela equa\c{c}\~ao (\ref{eqrelacao}). Contudo, n\~ao conseguimos identificar o(s) 
par\^ametro(s) respons\'avel(eis) pela diferen\cc a de fase de 3 passos entre estas 
oscila\c{c}\~oes. Para entender melhor este comportamento   
gostar\'{\i}amos de trabalhar numa  
cadeia alimentar de 3 esp\'ecies, na qual os predadores se alimentar\~ao tanto dos 
herb\'{\i}voros quanto das plantas cuja distribui\c{c}\~ao muda repentinamente, estando, 
desta forma, submetidos a uma sele\c{c}\~ao ecol\'ogica direta e a outra, indireta.

%
%


\chapter{Modelo com dois tra\cc os fenot\'{\i}picos} 

Apresentaremos agora um modelo no qual os indiv\'{\i}duos t\^em um 
tra\cc o relacionado \`a sele\c{c}\~ao ecol\'ogica e um outro tra\cc o,
independente, relativo \`a sele\c{c}\~ao sexual. A 
inspira\c{c}\~ao biol\'ogica surgiu a partir das fam\'{\i}lias de cicl\'{\i}deos  
encontradas na Nicar\'agua e na \'Africa e que v\^em atraindo grande interesse,  
em fun\c{c}\~ao dos dados estat\'{\i}sticos coletados acerca destas fam\'{\i}lias.

Atrav\'es de simula\c{c}\~oes  primeiro estudamos a 
distribui\c{c}\~ao dos fen\'otipos ligados \`a sele\c{c}\~ao sexual, 
numa ecologia onde a capacidade de sustenta\c{c}\~ao do ambiente 
independe do tra\cc o ecol\'ogico. O objetivo \'e o de compreender melhor  
as consequ\^encias da sele\c{c}\~ao sexual, independentemente da ecologia. A seguir, 
introduzimos a ecologia dependente do tra\c{c}o ecol\'ogico a fim de estudar o processo  
de especia\c{c}\~ao simp\'atrica nestes peixes.  
Analisamos os resultados do processo para diferentes valores das taxas de 
muta\cao,~com o objetivo de comparar qualitativamente os resultados do modelo com aqueles 
encontrados experimentalmente nos cicl\'{\i}deos da Nicar\'agua.

\section{Motiva\c{c}\~ao Biol\'ogica}
Os cicl\'{\i}deos parecem especiar rapidamente, 
pois an\'alises das 300 esp\'ecies encontradas no lago Vit\'oria, no leste da \'Africa, 
indicam a exist\^encia de um mesmo ancestral comum, que teria vivido h\'a apenas 12.400 
anos atr\'as \cite{mkn408}.
Os estudos gen\'eticos da esp\'ecie {\it Amphilophus citrinellum}
da Nicar\'agua~\cite{wilson} revelaram que estes peixes apresentam distintos 
tipos de maxilares, que s\~ao a chave da sua adapta\c{c}\~ao nas diversas 
ecologias existentes. 
Um tipo de maxilar \'e adequado \`a uma dieta de carac\'ois (molariform) 
e um outro tipo \'e adequado \`a ingest\~ao de presas moles (papilliform). 

Esperava-se que a especia\c{c}\~ao simp\'atrica dos cicl\'{\i}deos  
envolvesse grande diverg\^encia no tra\cc o ecol\'ogico e que houvesse uma forte 
correla\c{c}\~ao entre este tra\c{c}o e a cor dos peixes, que \'e o tra\cc o ligado 
ao acasalamento. Contudo, n\~ao \'e assim. 
As estat\'{\i}sticas mostram que existe uma forte diferencia\c{c}\~ao na 
cor (entre os tipos \aspas{gold} - dourado e \aspas{normal} - listrado de preto e branco), 
mas que n\~ao 
existe grande diferencia\c{c}\~ao na morfologia dos maxilares, isto \'e, dentre os peixes de uma 
dada cor os dois tipos de maxilares s\~ao encontrados, ao inv\'es de apenas um tipo de maxilar 
para cada cor. De fato, a correla\c{c}\~ao entre os 
dois tra\c{c}os mostrou-se pequena ($R=0.48$). Voltaremos a estas observa\c{c}\~oes 
no final deste cap\'{\i}tulo.

Dois importantes modelos te\'oricos (do estilo \aspas{campo m\'edio}) foram desenvolvidos 
concomitantemente \cite{kk,dd}, 
para explicar a especia\c{c}\~ao nos cicl\'{\i}deos. Ambos obt\^em 
a especia\c{c}\~ao envolvendo dois tra\cc os independentes, agregada 
a uma forte correla\c{c}\~ao entre estes 
dois tra\cc os \cite{kk}. Isto \'e, no caso dos cicl\'ideos, os modelos prev\^em que 
a quase totalidade dos peixes dourados se alimentaria de carac\'ois e os listrados, 
de presas moles (ou vice-versa). No trabalho \cite{kk} os autores admitem uma mudan\cc a 
brusca tanto na ecologia {\it como na competi\c{c}\~ao}, para obter a 
especia\c{c}\~ao simp\'atrica. A competi\c{c}\~ao s\'o se torna intra-espec\'{\i}fica quando 
a ecologia muda: antes disto, todos disputam igualmente com todos pela fonte de alimento. 
Esta mesma estrat\'egia foi utilizada com sucesso nas simula\c{c}\~oes apresentadas  
em \cite{medeiros} e tamb\'em nas nossas apresentadas em \cite{bjp,cise}, 
na obten\c{c}\~ao da especia\c{c}\~ao.  
Contudo, no modelo adotado em \cite{dd} e considerado mais realista \cite{turelli},  
n\~ao existem tais mudan\cc as:   
tanto a ecologia quanto a competi\c{c}\~ao por alimento s\~ao mantidas fixas e, ainda assim, a 
especia\c{c}\~ao \'e obtida. 

\begin{figure}[htbp]
\begin{center}
 \includegraphics[width=6.5cm,angle=270]{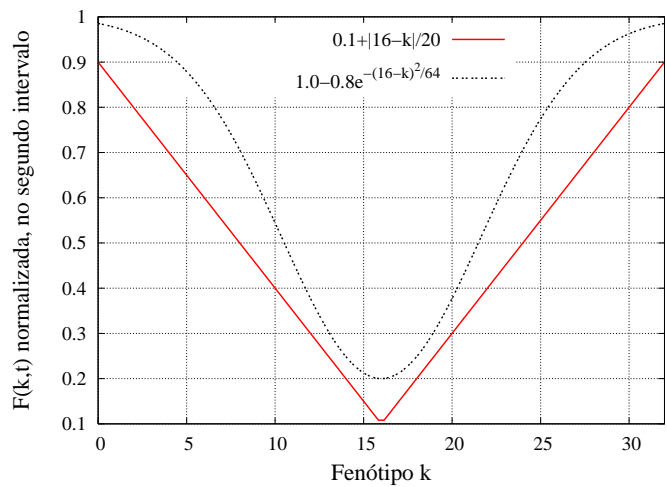}
\end{center}
\caption{\it Compara\c{c}\~ao das capacidades de sustenta\c{c}\~ao do ambiente, no
segundo intervalo de tempo ($t>4\times10^3$), utilizadas no cap\'{\i}tulo
anterior (linha solida) e neste cap\'{\i}tulo (linha pontilhada).}
\label{figfuncaoFe}
\end{figure}

Como visto nos cap\'{\i}tulos anteriores, nos modelos que apresentamos nesta tese a 
competi\c{c}\~ao \'e a mesma desde o in\'{\i}cio da simula\c{c}\~ao e 
dada pela equa\c{c}\~ao \ref{eqcompe}, com $X=0,5$. Contudo, a ecologia 
muda, de uma onde a abund\^ancia de alimento \'e a mesma para qualquer fen\'otipo 
para outra que desfavorece fortemente os fen\'otipos intermedi\'arios. 
Com o objetivo de suavizar tal desfavorecimento, adotaremos para o 
segundo intervalo de tempo a nova capacidade de 
sustenta\c{c}\~ao do ambiente, ilustrada na figura \ref{figfuncaoFe} e dada pela seguinte 
equa\c{c}\~ao:  

\begin{equation}\label{eqdcaracoles}
F(k,t)=5\times10^5\times\left\{\begin{array}{ll}
1& ~~~~~ 0\le t\le 4\times10^3, \\ 
1.0-0.8\times exp(-(16-k)^2/64) 
& ~~~~~ 4\times10^3< t\le 5\times10^4,\end{array}\right.
\end{equation}
\bigskip

Em nossas simula\c{c}\~oes  os \aspas{cicl\'{\i}deos} t\^em a morfologia do maxilar 
representada por um par de tiras de bits (fen\'otipo $k$) e a cor \'e tambem representada 
por um par de tiras, independente do primeiro (fen\'otipo $k'$). 

\section{Distribui\c{c}\~ao do tra\cc o sexual numa ecologia fixa}

Como dito no in\'{\i}cio deste cap\'{\i}tulo, estudaremos primeiro o papel da sele\c{c}\~ao 
sexual sozinha, sem qualquer mudan\c{c}a na ecologia. 
A seguir mostramos os resultados das simula\c{c}\~oes  feitas com a mesma 
capacidade de sustenta\c{c}\~ao fixa anterior ($t \le 4 \times 10^3$), mas com indiv\'{\i}duos 
que t\^em agora 
mais uma caracter\'{\i}stica fenot\'{\i}pica, $k'$, ligada \`a sele\c{c}\~ao sexual.  
Lembramos que as simula\c{c}\~oes  come\c{c}am com toda a popula\c{c}\~ao de f\^emeas n\~ao 
seletiva. A tabela 
\ref{tabsexpoli} mostra os par\^ametros utilizados, onde $MS$ \'e a taxa de muta\c{c}\~ao 
no car\'ater da seletividade.

\begin{table}[htbp]
\begin{center}
{\footnotesize
\begin{tabular}{|l||l||l|}
\hline
Descri\c{c}\~ao das grandezas para (macho e f\^emea) & Nome & Valor\\
\hline\hline
Muta\c{c}\~ao (na tira de bits)& $M$ & 1*\\
\hline
Domin\^ancia & $D$ &  3* \\ \hline 
Limite de doen\c{c}as mortais & $L$ & 3* \\ \hline
N\'umero de filhos& $NF$  & 2* \\ \hline 
Idade m\'{\i}nima de reprodu\c{c}\~ao & $R$ &10 \\ \hline \hline
Domin\^ancia - fen\'otipo $k$ & $DF$ & 16* \\ \hline
Domin\^ancia - fen\'otipo $k'$ & $DFS$ & 16* \\ \hline
Muta\c{c}\~ao por bit - fen\'otipo $k$ & $MF$ & 0.1 \\ \hline
Muta\c{c}\~ao por bit - fen\'otipo $k'$ & $MFS$ & 0.5 \\ \hline
Fen\'otipos intermedi\'arios & $k$ e $k'$ & 16*\\ \hline
Muta\c{c}\~ao na seletividade & $MS$ & 0.001 \\ \hline
N\'umero de machos dispon\'{\i}veis & $NM$ & 6\\
\hline
\end{tabular}
\caption{\it Valores dos par\^ametros usados nas simula\coes. Os par\^ametros relativos 
\`a parcela estruturada por idade do genoma est\~ao na parte superior. *Grandezas que 
n\~ao mudam neste cap\'{\i}tulo.}
\label{tabsexpoli}
}
\end{center}
\end{table}

O comportamento da distribui\c{c}\~ao dos fen\'otipos $k$ n\~ao muda em rela\c{c}\~ao 
aos resultados encontrados no cap\'{i}tulo anterior (ver figura \ref{figdbp}, lado esquerdo), 
isto \'e, 
a distribui\c{c}\~ao novamente \'e uma gaussiana centrada em $16$ e independente de $MF$. 
J\'a a distribui\c{c}\~ao dos 
fen\'otipos $k'$ depende dos valores de $MFS$, $MS$ e $NM$, como veremos a seguir. 
\begin{figure}[htbp]
\begin{center}
 \includegraphics[width=6.5cm,angle=270]{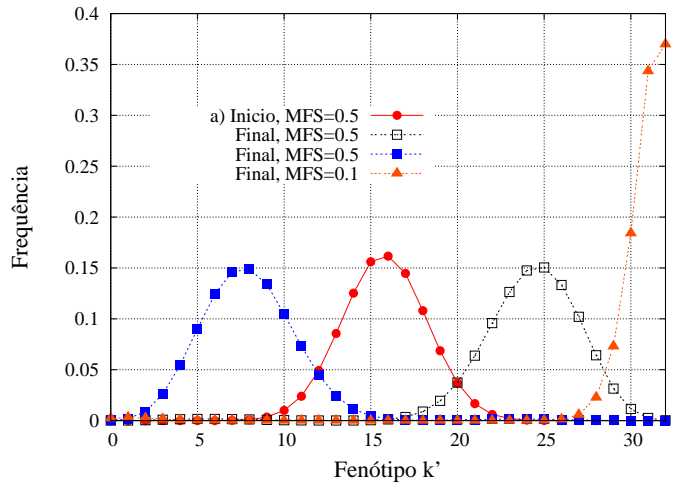}
\end{center}
\caption{\it Distribui\c{c}\~ao do fen\'otipo $k'$ ligado \`a sele\c{c}\~ao sexual. No 
in\'{\i}cio das simula\c{c}\~oes  ($t=200$ passos) a distribui\c{c}\~ao fica centrada em 16 
(c\'{\i}rculos). O valor m\'edio final se desvia de 16 (quadrados), tanto para a direita 
quanto para a esquerda, dependendo da semente aleat\'oria inicial. Quanto menor o valor  
de $MFS$, mais a distribui\c{c}\~ao se aproxima de um dos extremos (tri\^angulos).}
\label{figpolysex}
\end{figure}

No in\'{\i}cio da simula\c{c}\~ao muitas f\^emeas ainda s\~ao n\~ao seletivas e
por esta raz\~ao a distribui\c{c}\~ao fica centrada em 16 - 
figura \ref{figpolysex}, c\'{\i}rculos. \`A medida que as 
f\^emeas se tornam seletivas, uma
das duas carater\'{\i}sticas fenot\'{\i}picas extremas se fixa aleatoriamente - figura 
\ref{figpolysex}, quadrados cheios ou vazios. Este desvio da distribui\c{c}\~ao para um
fen\'otipo extremo acontece tanto com a regra da diferen\c{c}a para o acasalamento 
quanto com a regra com forte dire\cao.~Observe que na regra com forte dire\c{c}\~ao a  
f\^emea seletiva escolhe o macho mais ao extremo de sua pr\'opria cor, mas 
mesmo assim a distribui\c{c}\~ao final n\~ao apresenta uma divis\~ao mas sim, um desvio.
O valor m\'edio da distribui\c{c}\~ao final e o tempo que esta leva 
para ser atingida depende 
dos valores das taxas de muta\c{c}\~ao, $MFS$ e $MS$, e do n\'umero de escolhas 
dispon\'{\i}veis na hora do acasalamento, $NM$. 
Em compara\c{c}\~ao com os valores da tabela \ref{tabsexpoli}, se o valor de $MS$ for 
muito pequeno, o valor m\'edio fica em 16 por um tempo maior; 
se o n\'umero de escolhas $NM$ for muito grande, o valor m\'edio vai para um dos extremos 
(0 ou 32).

\section{Especia\cao}

Como acabamos de ver, numa ecologia (ou capacidade de sustenta\c{c}\~ao) fixa e 
independente do fen\'otipo $k$ e com um grau de competi\c{c}\~ao m\'edio ($X=0,5$), 
a sele\c{c}\~ao sexual n\~ao \'e capaz de levar \`a especia\c{c}\~ao. Introduziremos 
ent\~ao a mudan\c{c}a na ecologia dada pela equa\c{c}\~ao \ref{eqdcaracoles} e observaremos 
como este cen\'ario se modifica.

Antes de apresentar os resultados precisamos definir o que chamaremos de especia\c{c}\~ao, 
neste caso em que os tra\c{c}os s\~ao independentes. Observe-se que esta defini\c{c}\~ao 
\'e bastante problem\'atica para os bi\'ologos \cite{mkn408}. Alguns acreditam 
que para haver especia\c{c}\~ao \'e necess\'aria apenas a sele\c{c}\~ao sexual 
\cite{htakimoto} e neste caso, n\~ao t\^em  d\'uvida quanto \`a especia\c{c}\~ao 
dos cicl\'{\i}deos, pois 
estes apresentam uma clara diverg\^encia na cor. Outros autores, 
entretanto, acreditam que a especia\c{c}\~ao decorre da associa\c{c}\~ao entre uma 
sele\c{c}\~ao ecol\'ogica disruptiva e uma sele\c{c}\~ao sexual \cite{kk,kksexo}. 
Para estes autores, os cicl\'{\i}deos est\~ao ainda em processo de especia\c{c}\~ao, 
que s\'o se concretizar\'a quando os maxilares se dividirem, da mesma forma que a cor.  
N\'os, contudo, temos uma vantagem com rela\c{c}\~ao ao mundo real: podemos medir exatamente 
a densidade de seletivos na popula\c{c}\~ao e saber quando uma parcela da mesma se torna  
reprodutivamente isolada, isto \'e, uma nova esp\'ecie. Assim, sempre que a densidade de 
seletivos final for pr\'oxima de 1 e constante no tempo (sem grandes flutua\c{c}\~oes), diremos 
que houve especia\c{c}\~ao.  

Al\'em das distribui\c{c}\~oes dos fen\'otipos $k$ e $k'$, mostraremos tamb\'em a 
correla\c{c}\~ao, $R(t)$, entre estes dois tra\c{c}os. Ela \'e dada por: 

\[
R(t)=\frac{<kk'>-<k><k'>}{\sigma_k\sigma_{k'}}
\] 
onde $<k'>=\sum_{i=1}^{Pop(t)}k'_i/Pop(t)$ \'e o valor m\'edio da distribui\c{c}\~ao de cores 
da popula\c{c}\~ao total no tempo $t$, $Pop(t)$, e $\sigma_{k'}$ \'e o  
desvio padr\~ao deste valor m\'edio. O mesmo tipo de c\'alculo \'e feito
para a morfologia $k$ do maxilar e para o produto $kk'$.  

\subsection{Resultados similares aos de Kondrashov \cite{kk}}

Na tabela \ref{tabpolims1} mostramos os resultados obtidos em 10 simula\c{c}\~oes distintas, 
usando os valores das taxas de muta\c{c}\~ao $MF$ e $MFS$ tamb\'em mostrados na tabela. 
Em todos os casos a taxa de muta\c{c}\~ao do gene da seletividade vale $MS=0.001$. Os 
n\'umeros entre par\^enteses 
correspondem \`a m\'edia, nas 10 simula\c{c}\~oes, dos valores m\'edios finais da 
correla\c{c}\~ao $R$ e da densidade de 
f\^emeas seletivas $\rho_s$ obtidos em cada simula\c{c}\~ao. Os resultados aqui 
apresentados foram obtidos utilizando-se a regra com 
forte dire\c{c}\~ao para o acasalamento, mas adotando-se a regra da diferen\c{c}a os 
resultados s\~ao os mesmos.
\begin{table}[htbp] 
\begin{center}
\begin{tabular}{c|c|c|ll|ll} 
 \hline 
 \hline 
 $MF$ & $MFS$& {Especia\c{c}\~ao } & \multicolumn{2}{c|}{1Maxilar/1Cor 
 ($R$,\, $\rho_s$)} & \multicolumn{2}{c}{2Maxilares/1Cor ($R$,\, $\rho_s$)}  \\ 
  & &  $R=0.985, \, \, \rho_s=0.997$ &  & \\ \hline 
 0.1  & 1.0   & 9 & 0 &         & 1 & (0.85,\, 0.6) \\
 0.1  & 0.5   & 2 & 1 & (0,\, 0.6) & 7 & (0.85,\, 0.6) \\
 0.1  & 0.1   & 0 & 0 &         & 10 & (0.8,\, 0.6) \\
 0.5  & 0.5   & 0 & 1 & (0,\, 0.6) & 9 & (0.85,\, 0.6)\\
 0.5  & 0.1   & 0 & 0 &         & 10 & (0.5,\, 0.5) \\
 1.0  & 0.1   & 0 & 0 &         & 10 & (0,\, 0.5)\\
\hline
\hline
 \end{tabular} 
 \caption{\it Resultados, no final das simula\c{c}\~oes, para $MS=0.001$.  
Em todos os casos de especia\c{c}\~ao as distribui\c{c}\~oes  finais do tra\cc os  
$k$ e $k'$ s\~ao bimodais e podem ser vistas na figura 
\ref{figestable} (a) e (b). O estado 1Maxilar/1Cor indica que no final da 
simula\c{c}\~ao foram encontradas distribui\c{c}\~oes  unimodais tanto para 
$k$ como para $k'$, centradas nos extremos. 2Maxilares/1Cor \'e o caso em que  
a distribui\c{c}\~ao de $k'$ \'e unimodal e centrada num extremo e a  
distribui\c{c}\~ao de $k$ \'e bimodal. 
Os n\'umeros entre par\^enteses s\~ao a m\'edia dos valores m\'edios finais de   
$R$ e $\rho_s$ obtidos nas 10 simula\coes.}  
\label{tabpolims1} 
\end{center}
\end{table}

\begin{figure}[htbp]
\begin{center}
\includegraphics[width=5.5cm,angle=270]{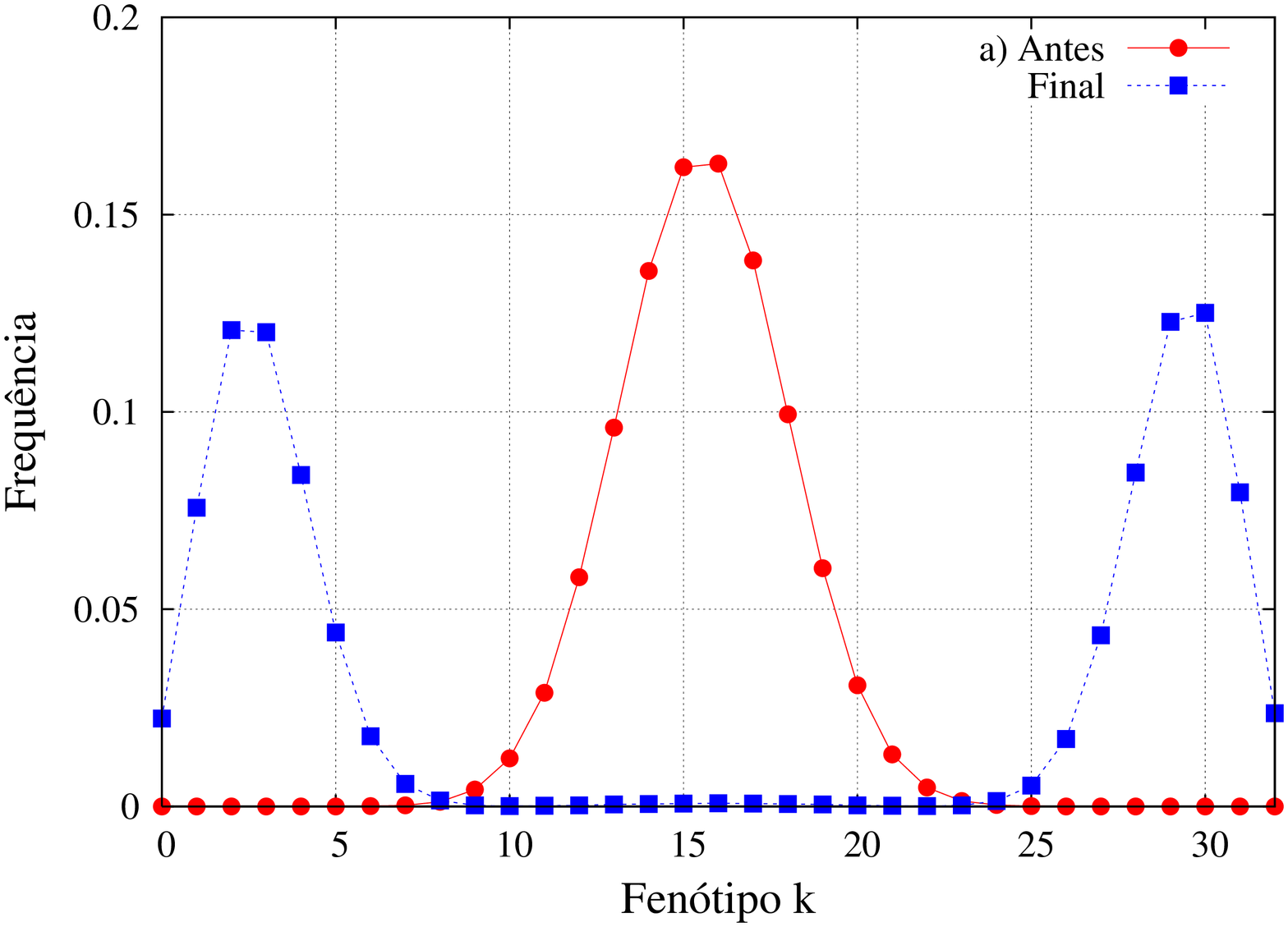}
\includegraphics[width=5.5cm,angle=270]{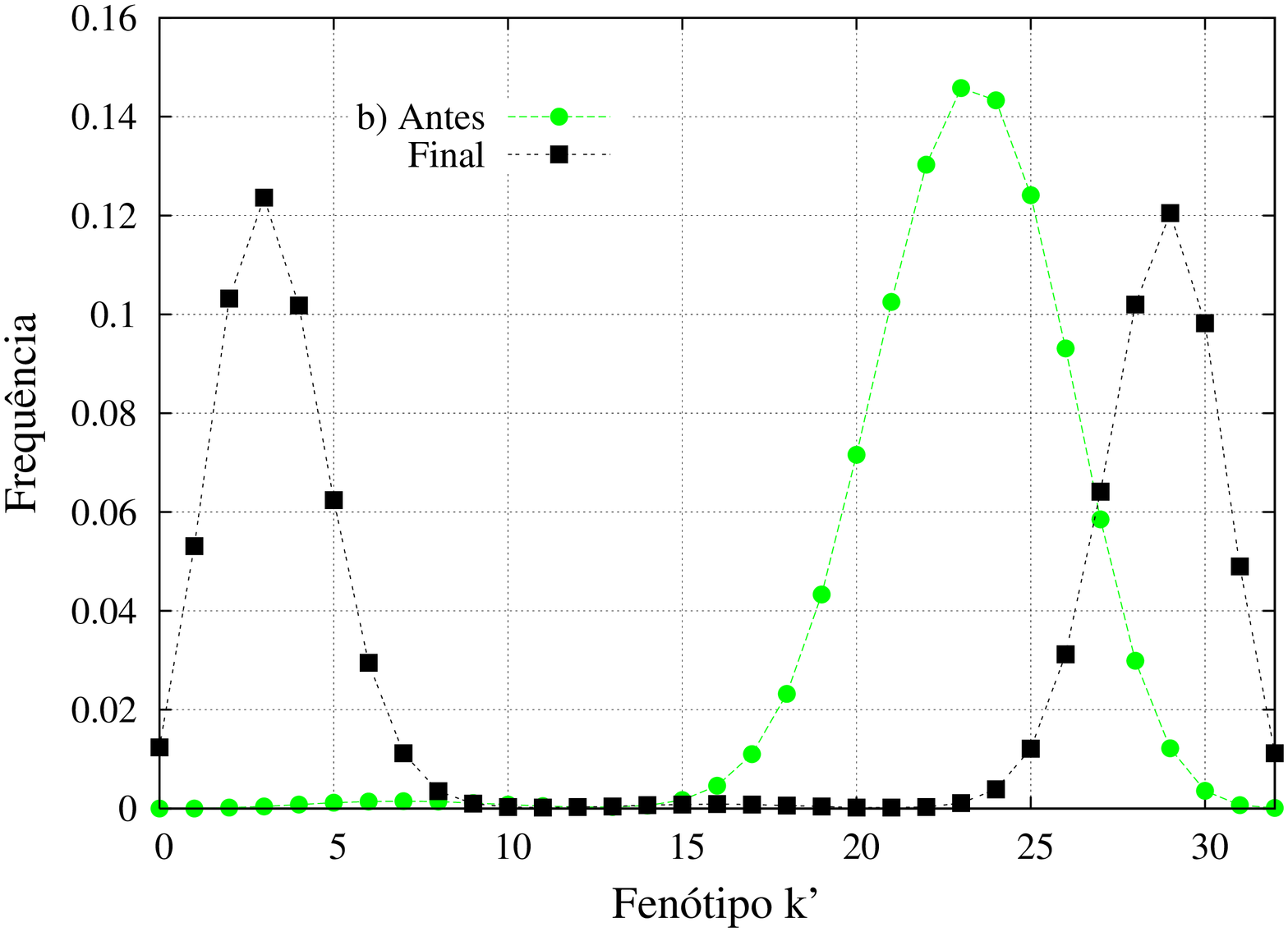}
\includegraphics[width=5.5cm,angle=270]{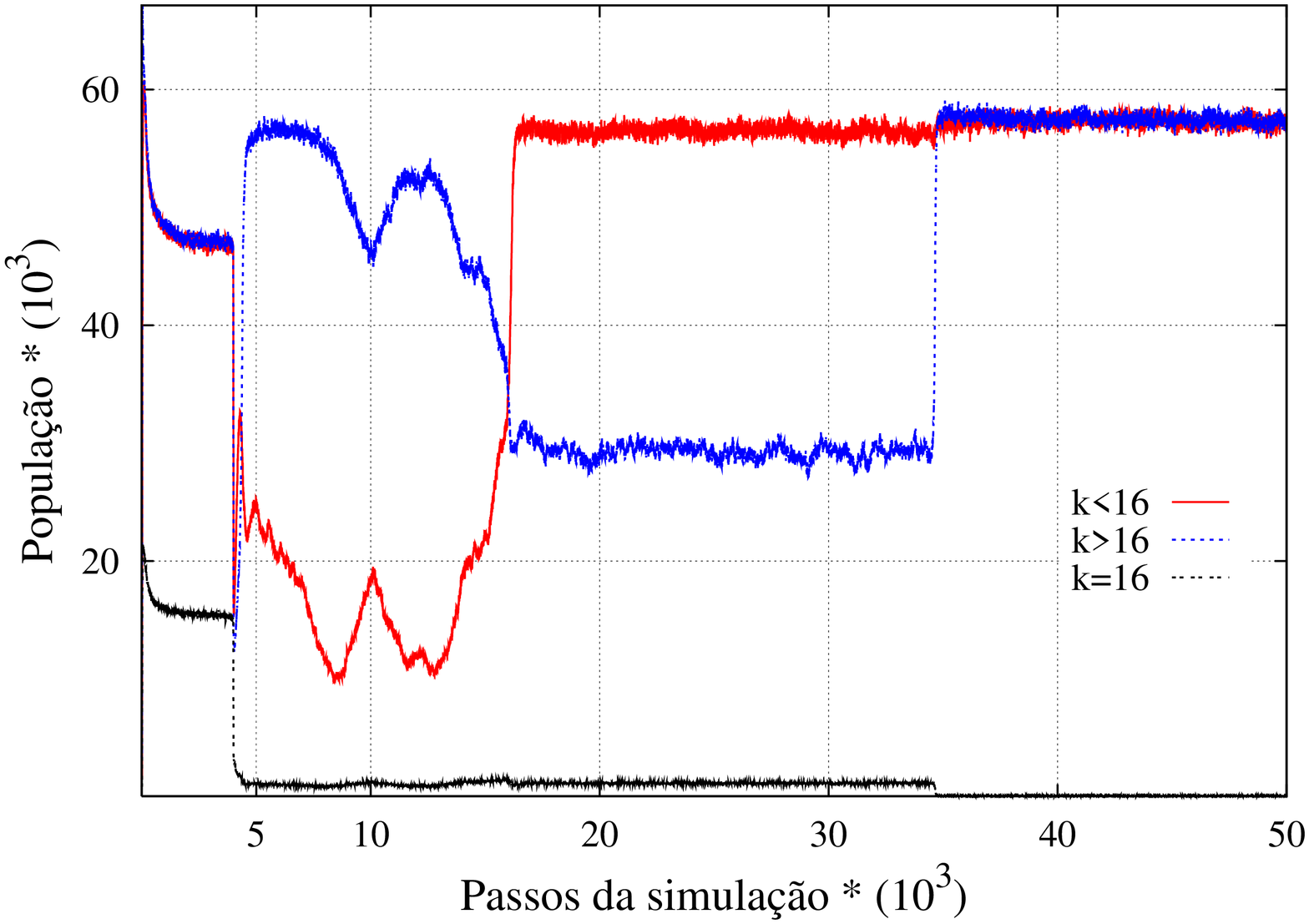}
\includegraphics[width=5.5cm,angle=270]{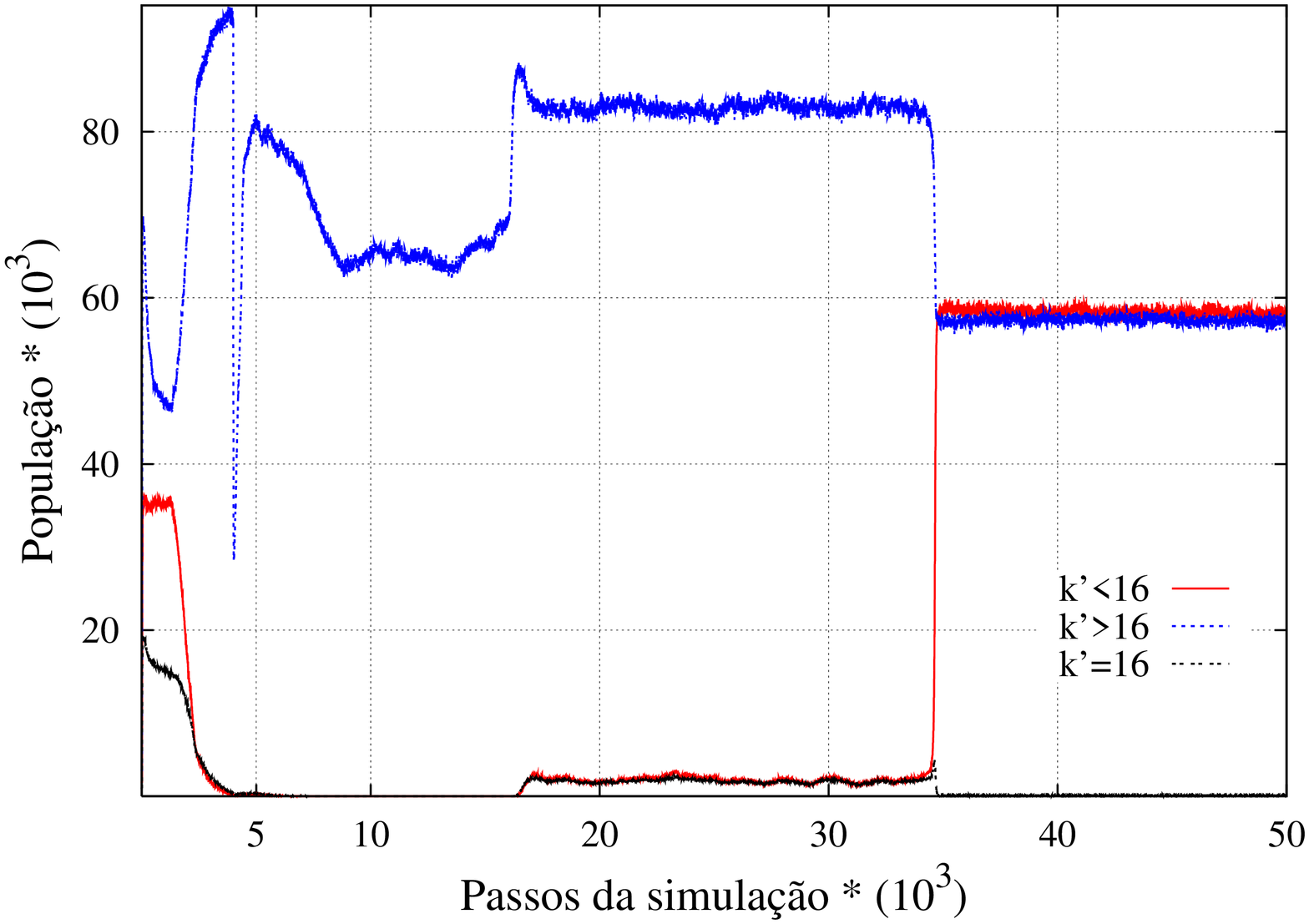}
\includegraphics[width=5.5cm,angle=270]{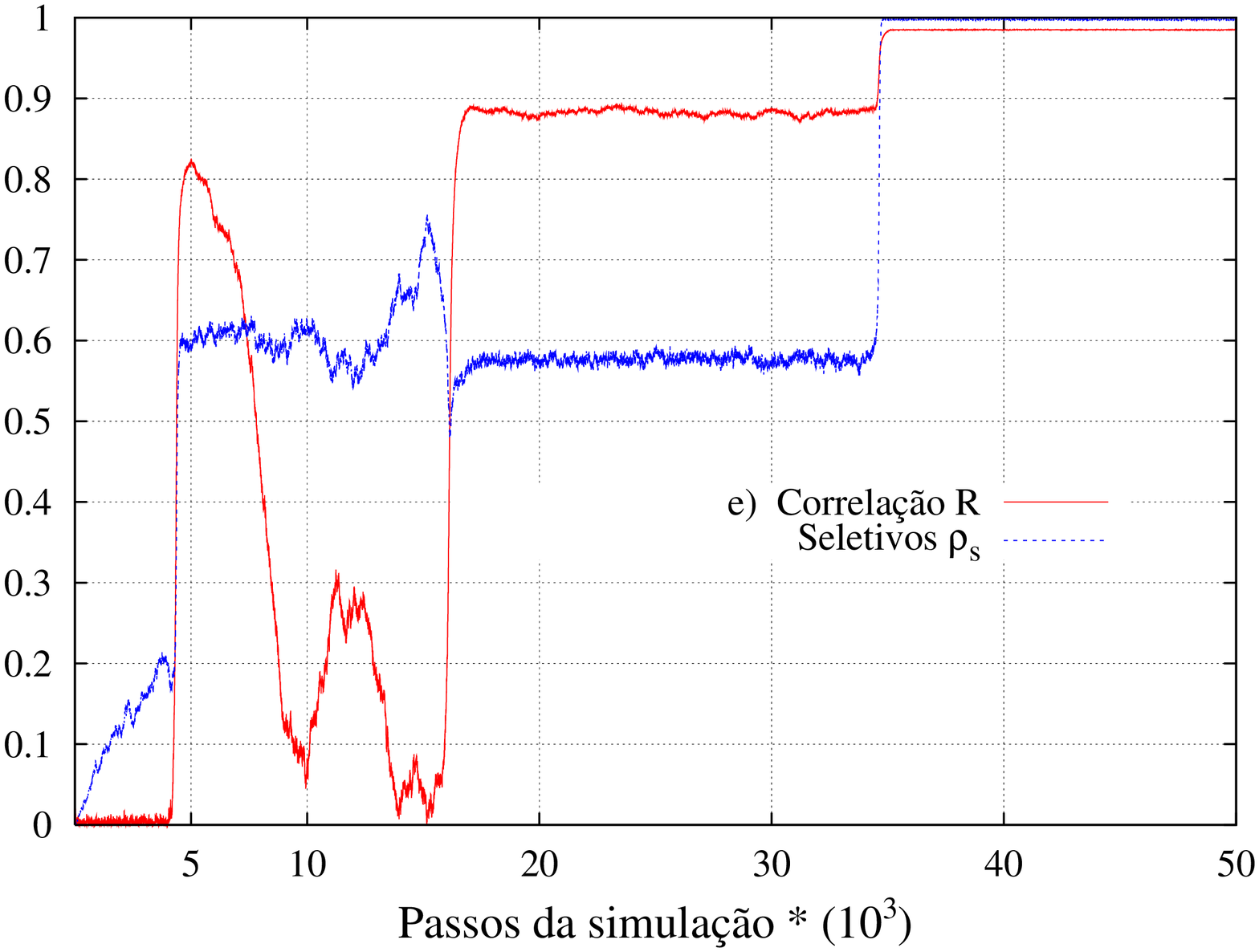}
\includegraphics[width=5.5cm,angle=270]{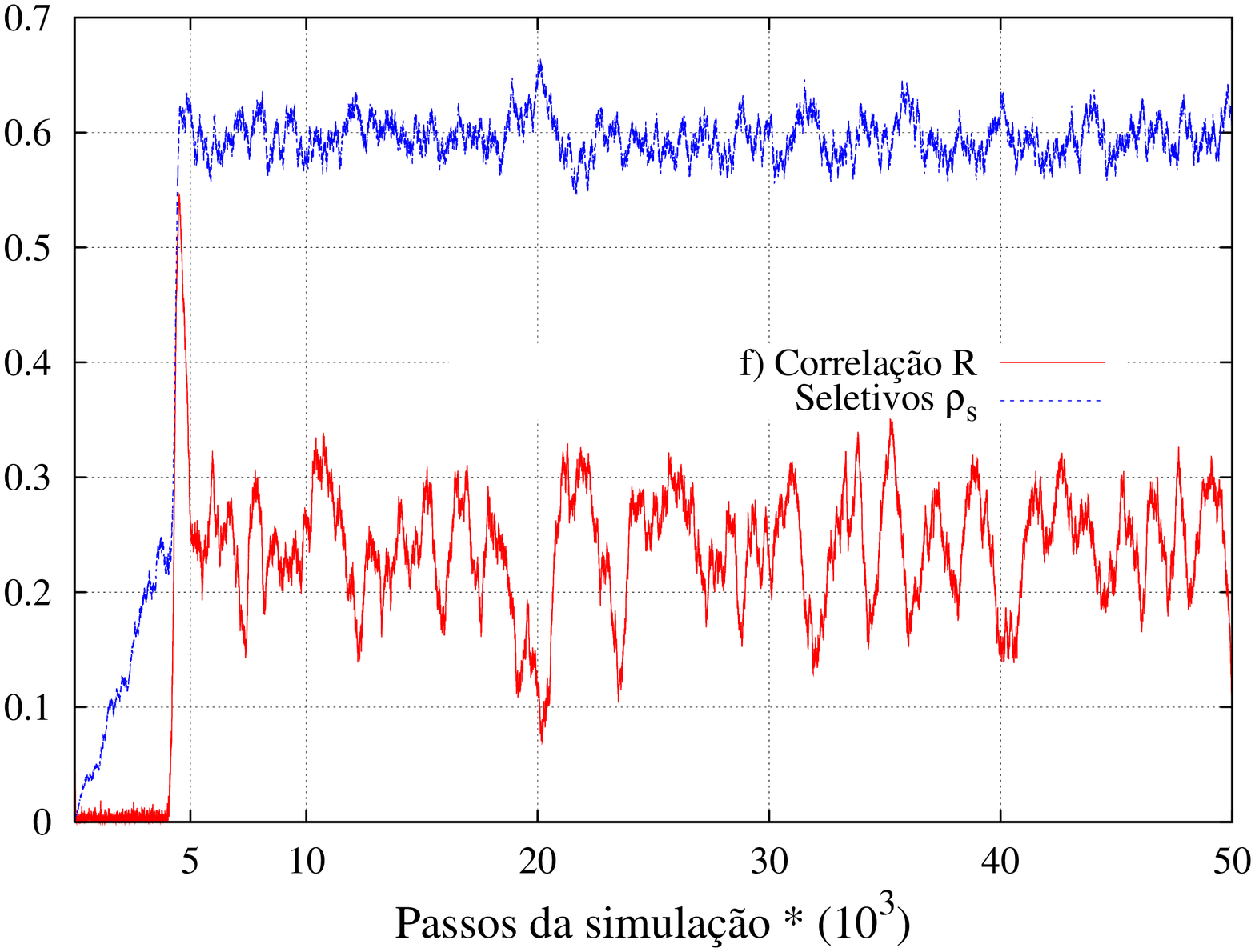}
\end{center}
\caption{\it Especia\c{c}\~ao com $MS=0.001$, $MF=0.1$ e $MFS=0.5$; a mudan\c{c}a na ecologia 
se d\'a no passo $t=4 \times 10^3$. (a) Distribui\c{c}\~ao do tra\cc o ecol\'ogico $k$. 
(b) Distribui\c{c}\~ao do tra\cc o sexual $k'$. Os c\'{\i}rculos correspondem  
ao intervalo de tempo inicial ($t\le 4\times10^3$) e os quadrados ao final da simula\cao.~
(c) Evolu\c{c}\~ao temporal das popula\c{c}\~oes  com fen\'otipos $k$ e   
(d) evolu\c{c}\~ao das popula\c{c}\~oes  com fen\'otipos $k'$, durante o
processo de especia\cao.~e) Varia\c{c}\~ao temporal do valor absoluto da correla\c{c}\~ao $R$ 
(linha s\'olida) e da densidade de seletivos $\rho_s$ (linha pontilhada).
(f) Varia\c{c}\~ao temporal da correla\c{c}\~ao e da densidade de seletivos num caso onde 
n\~ao houve especia\c{c}\~ao (2Maxilares/1Cor).}
\label{figestable}
\end{figure} 

Os casos que chamamos de especia\c{c}\~ao correspondem 
\`aqueles em que as distribui\c{c}\~oes finais dos dois fen\'otipos se tornam bimodais 
est\'aveis 
e que, como dito anteriormente, apresentam uma {\it densidade de f\^emeas seletivas pr\'oxima 
de 1 e est\'avel}. Os casos $1Maxilar/1Cor$ correspondem \`aqueles em que 
ambas as distribui\c{c}\~oes se mantiveram unimodais depois da mudan\c{c}a na ecologia. Nos 
casos assinalados como $2Maxilares/1Cor$ apenas a distribui\c{c}\~ao do fen\'otipo $k$ se 
dividiu, tendo a distribui\c{c}\~ao de $k'$ se mantido unimodal.  

Foi escolhido um caso da tabela \ref{tabpolims1} onde ocorreu 
especia\cao, para mostrar as 
ditribui\c{c}\~oes  estacion\'arias dos tra\cc os ecol\'ogico e sexual,  
a evolu\c{c}\~ao temporal das popula\c{c}\~oes de cada fen\'otipo e os comportamentos da 
densidade de seletivos e da correla\c{c}\~ao durante o processo de especia\c{c}\~ao - 
figura \ref{figestable}, (a) $\rightarrow$ (e). Nos outros casos da tabela o comportamento 
das correspondentes grandezas s\~ao os mesmos. 
No caso (f) da figura \ref{figestable} a especia\c{c}\~ao 
n\~ao ocorreu e pode-se comparar as flutua\c{c}\~oes, tanto em $R$ como em $\rho_s$, com 
aquelas mostradas em (e), onde houve especia\c{c}\~ao.

Observando a tabela \ref{tabpolims1}, nota-se que a probabilidade de especia\c{c}\~ao 
diminui \`a medida em que $MF$ aumenta. 
A especia\c{c}\~ao \'e mais prov\'avel quando  
$MF$ \'e pequeno porque a distribui\c{c}\~ao bimodal dos fen\'otipos $k$  
apresenta os dois modos mais separados do  
que para valores maiores de $MF$ (ver tabela \ref{tabpolyresul}).  
Uma maior separa\c{c}\~ao destes dois modos provoca um aumento nas  
popula\c{c}\~oes  dos extremos (que s\~ao as seletivas) e uma maior elimina\c{c}\~ao da popula\c{c}\~ao com 
fen\'otipo intermedi\'ario (que acasala aleatoriamente).

Uma vez fixado um valor de $MF$, a probabilidade de especia\c{c}\~ao  
diminui \`a medida em que $MFS$ diminui. 
Valores pequenos de $MFS$ difilcultam a especia\c{c}\~ao porque fazem com que a 
distribui\c{c}\~ao de
cores (fen\'otipo $k'$), antes da mudan\c{c}a na ecologia, se desloque do centro para 
um valor pr\'oximo a um dos extremos, como j\'a mencionado anteriormente por ocasi\~ao da 
fig.\ref{figpolysex} e como tamb\'em pode ser visto na figura \ref{figestable}-(b). 
Observando a fig.\ref{figestable}-(d) v\^e-se que de fato uma das cores, no caso $k'< 16$, 
praticamente 
desaparece da popula\c{c}\~ao no momento da mudan\c{c}a na ecologia, para reaparecer cerca  
de 30 mil passos depois e se igualar \`a popula\c{c}\~ao que havia permanecido. Pela  
fig.\ref{figestable}-(c), vemos que o mesmo n\~ao ocorre com a distribui\c{c}\~ao dos 
maxilares, isto \'e, assim que se d\'a a mudan\c{c}a na ecologia a popula\c{c}\~ao com 
um certo tipo de maxilar (no caso $k < 16$) diminui, mas rapidamente volta a crescer e 
estabelece-se um polimorfismo no que diz respeito \'a morfologia dos maxilares. Contudo, 
o processo de especia\c{c}\~ao s\'o se concretiza quando a popula\c{c}\~ao com $k' < 16$ 
reaparece, quando ent\~ao a correla\c{c}\~ao entre os dois tra\c{c}os e a densidade de 
f\^emeas seletivas v\~ao para 1. Conv\'em ressaltar que apresentamos tanto na tabela 
\ref{tabpolims1} quanto na 
fig.\ref{figestable}-(e) o valor absoluto da correla\c{c}\~ao, mas que esta tanto pode ser 
positiva quanto negativa. Ela \'e positiva quando os extremos das popula\c{c}\~oes de 
maxilar e cor coincidem, isto \'e, quando todos os indiv\'iduos com $k > 16$ t\^em tamb\'em 
$k' > 16$ e todos com $k < 16$ t\^em $k' < 16$. A correla\c{c}\~ao \'e negativa quando 
os extremos s\~ao opostos ($k > 16$ com $k'< 16$ e $k < 16$ com $k'> 16$), mas em ambos 
os casos toda a popula\c{c}\~ao com um tipo de maxilar tem uma dada cor e toda a 
popula\c{c}\~ao com o outro tipo de maxilar tem uma mesma cor, diferente da primeira.         
 
Tanto a evolu\c{c}\~ao temporal dos fen\'otipos $k$ e $k'$ durante o processo de 
especia\c{c}\~ao como o comportamento da correla\c{c}\~ao que se estabelece entre eles 
s\~ao bastante semelhantes aos obtidos em \cite{kk}, apesar dos modelos serem distintos. 
Contudo, n\~ao parecem descrever corretamente o que ocorre com os cicl\'{\i}deos, j\'a 
que nestes observa-se uma clara divis\~ao na cor mas n\~ao na morfologia dos maxilares 
e ainda, as correla\c{c}\~oes entre os dois tra\c{c}os s\~ao pequenas se comparadas tanto com 
as obtidas em \cite{kk} como com as nossas.  
 
Os processos de especia\c{c}\~ao com $MFS=1.0$ s\~ao mais r\'apidos do que 
com $MFS=0.5$, mas todos apresentam as mesmas carater\'{\i}sticas dadas pela figura 
\ref{figestable}, (a) $\rightarrow$ (e). 

\subsection{Resultados mais adequados aos cicl\'{\i}deos}

Nesta subse\c{c}\~ao adotaremos exatamente o mesmo procedimento da subse\c{c}\~ao anterior, 
mas agora para uma taxa de muta\c{c}\~ao do gene da seletividade $MS=0.0001$, isto \'e, dez 
vezes menor.

\begin{table}[htbp] 
\begin{center}
\begin{tabular}{c|c|c|ll|ll} 
 \hline 
 \hline 
  $MF$ & $MFS$& {Especia\cao} & 
 \multicolumn{2}{c|}{1Maxilar/1Cor ($R$,\, $\rho_s$)} & \multicolumn{2}{c}{2Maxilares/1Cor 
($R$,\, $\rho_s$)}  \\ 
  & &  $R=0.974, \, \, \rho_s=1.000$ &  & \\ \hline 
 0.1  & 1.0   & 8 & 0 &           &2& (0.9,\, 0.55)  \\
 0.1  & 0.5   & 6 & 3 & (0,\, 0.65) &1&  (0.5,\, 0.62) \\
 0.1  & 0.1   & 1 & 0 &          &9& (0.3,\, 0.6)\\
 0.5  & 0.5   & 3 & 4 & (0,\, 0.7)  &3& (0.4,\, 0.65)\\
 0.5  & 0.1   & 1 & 0 &   &9& (0,\, 0.5)\\
 1.0  & 0.1   & 0 & 0 &          &10& (0,\, 0.5)\\
\hline
\hline
 \end{tabular} 
 \caption{\it Resultados, no final das simula\coes,~para $MS=0.0001$.
  Nos casos de especia\c{c}\~ao as distribui\c{c}\~oes  finais dos tra\cc os  
  $k$ e $k'$ s\~ao bimodais e podem ser vistas na figura \ref{figinestable}-(a) e (b).} 
 \label{tabpolims2} 
 \end{center}
 \end{table}

\begin{figure}
\begin{center}
\includegraphics[width=5.5cm,angle=270]{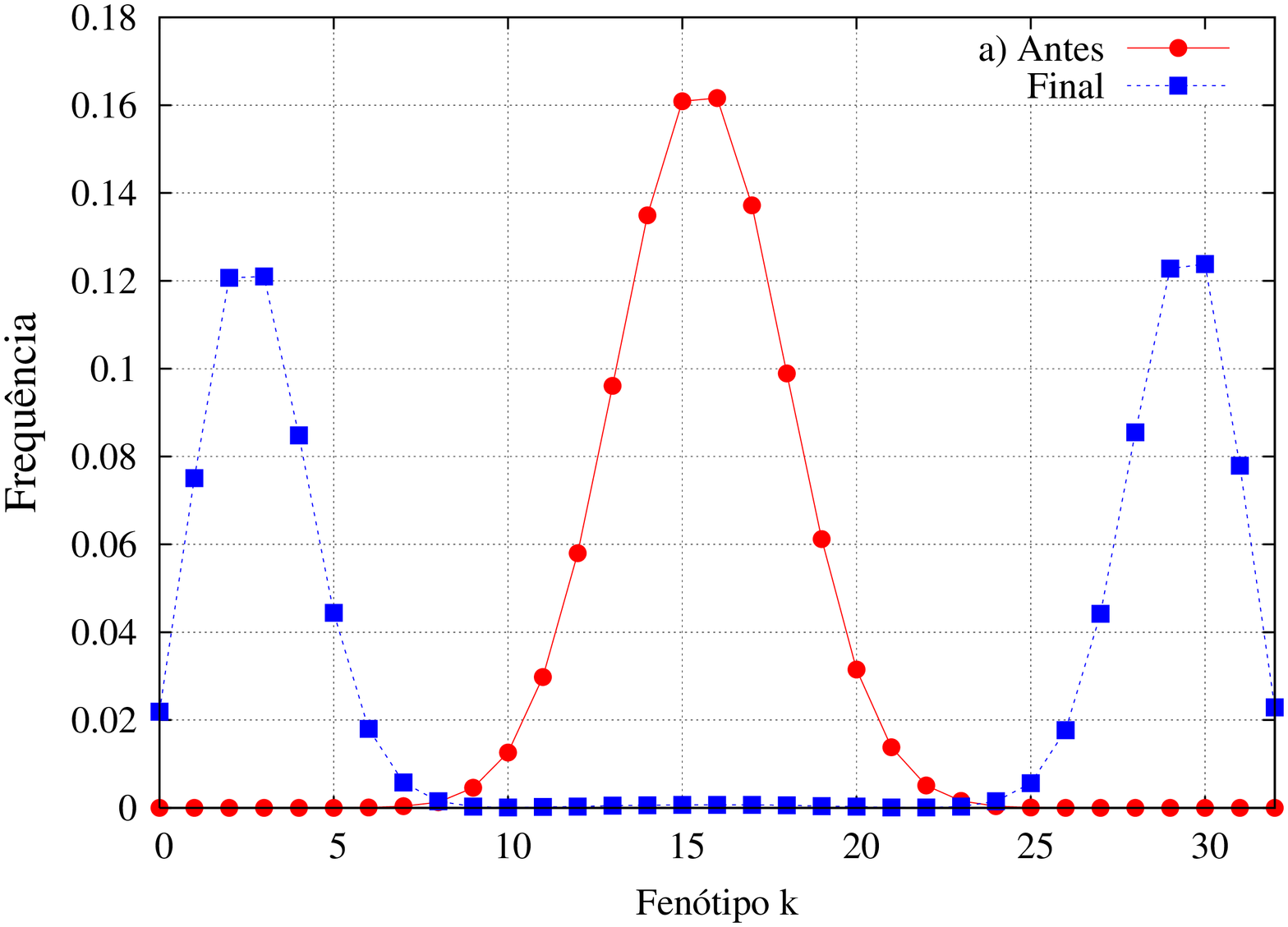}
\includegraphics[width=5.5cm,angle=270]{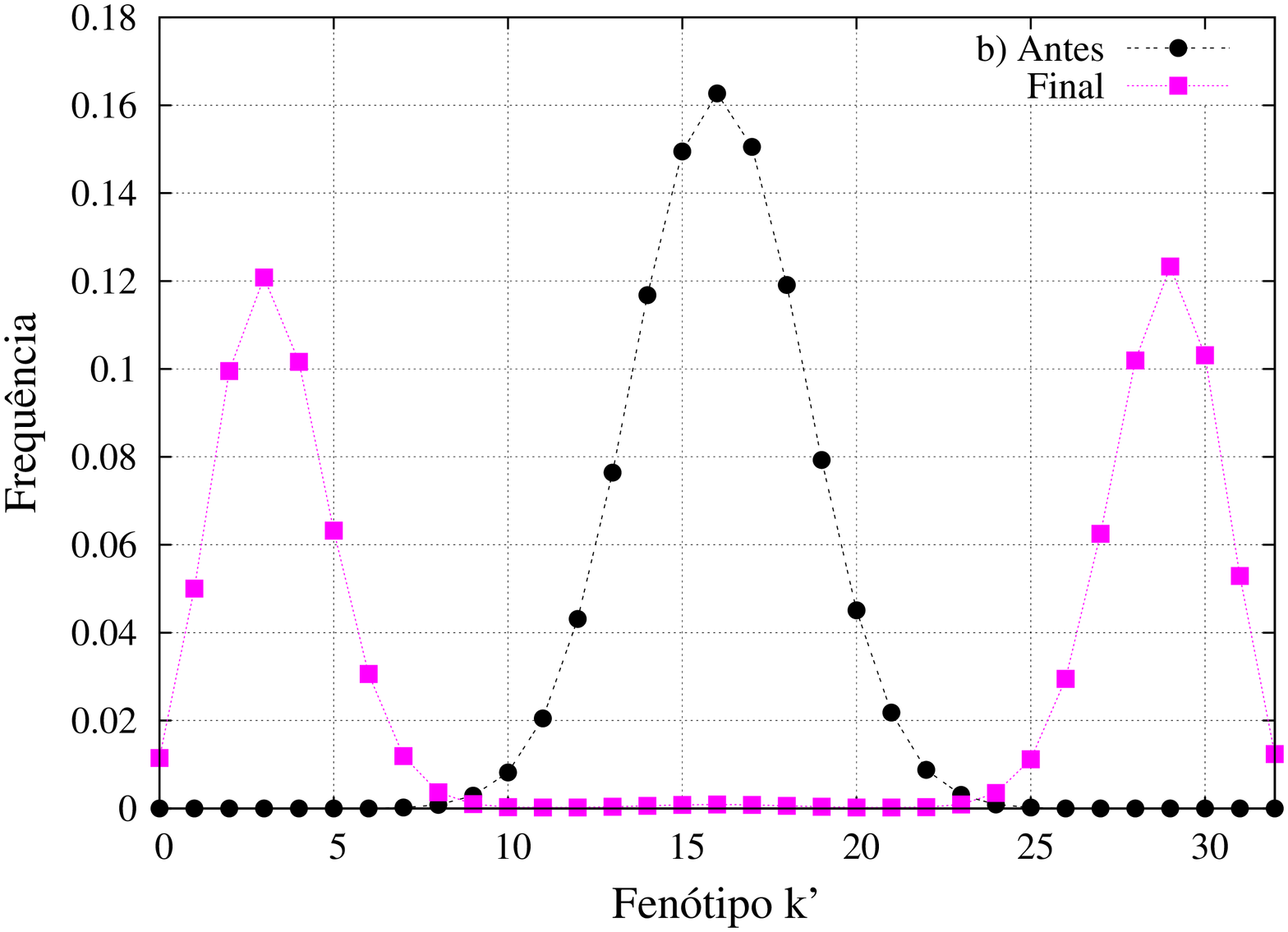}
\includegraphics[width=5.5cm,angle=270]{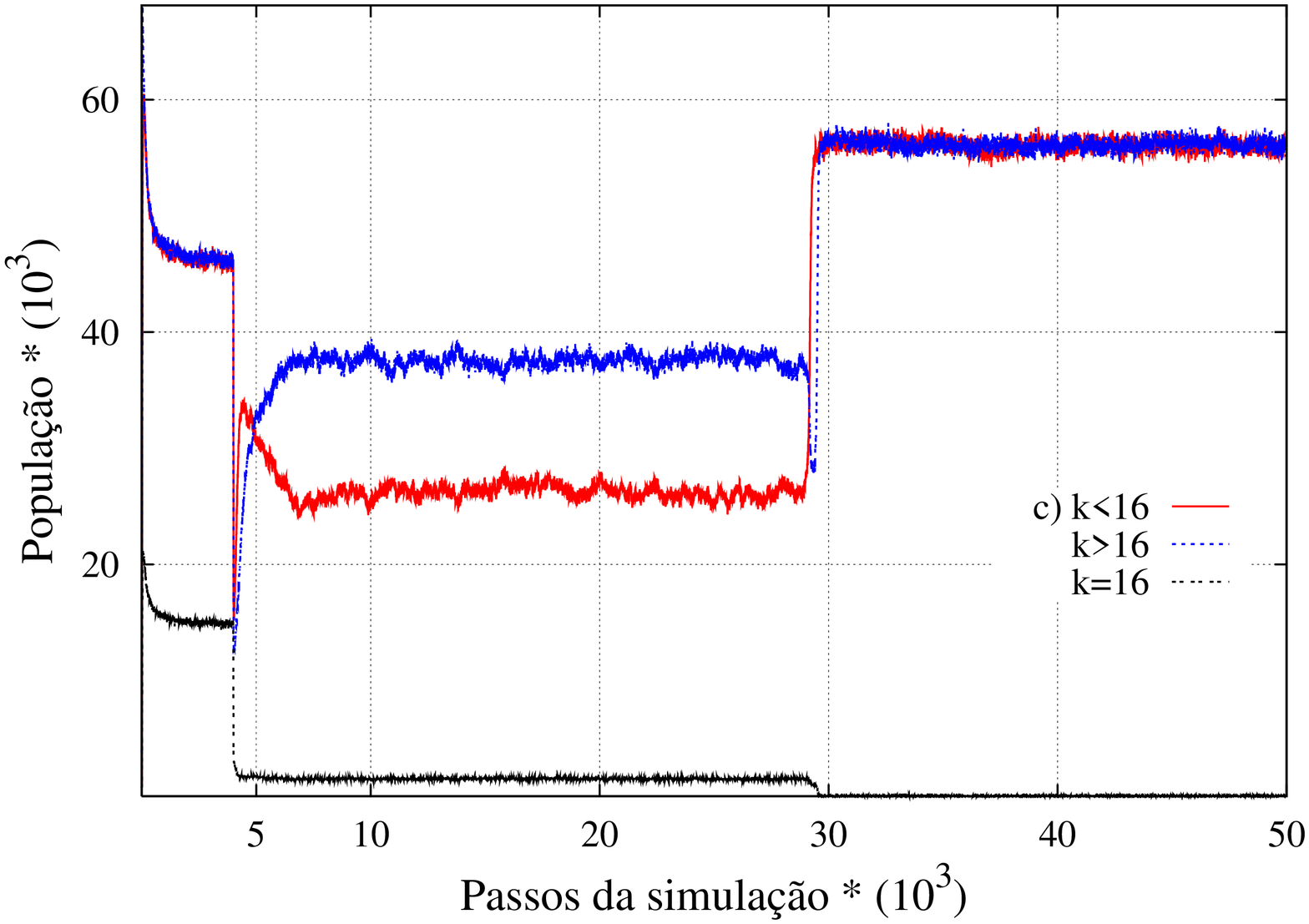}
\includegraphics[width=5.5cm,angle=270]{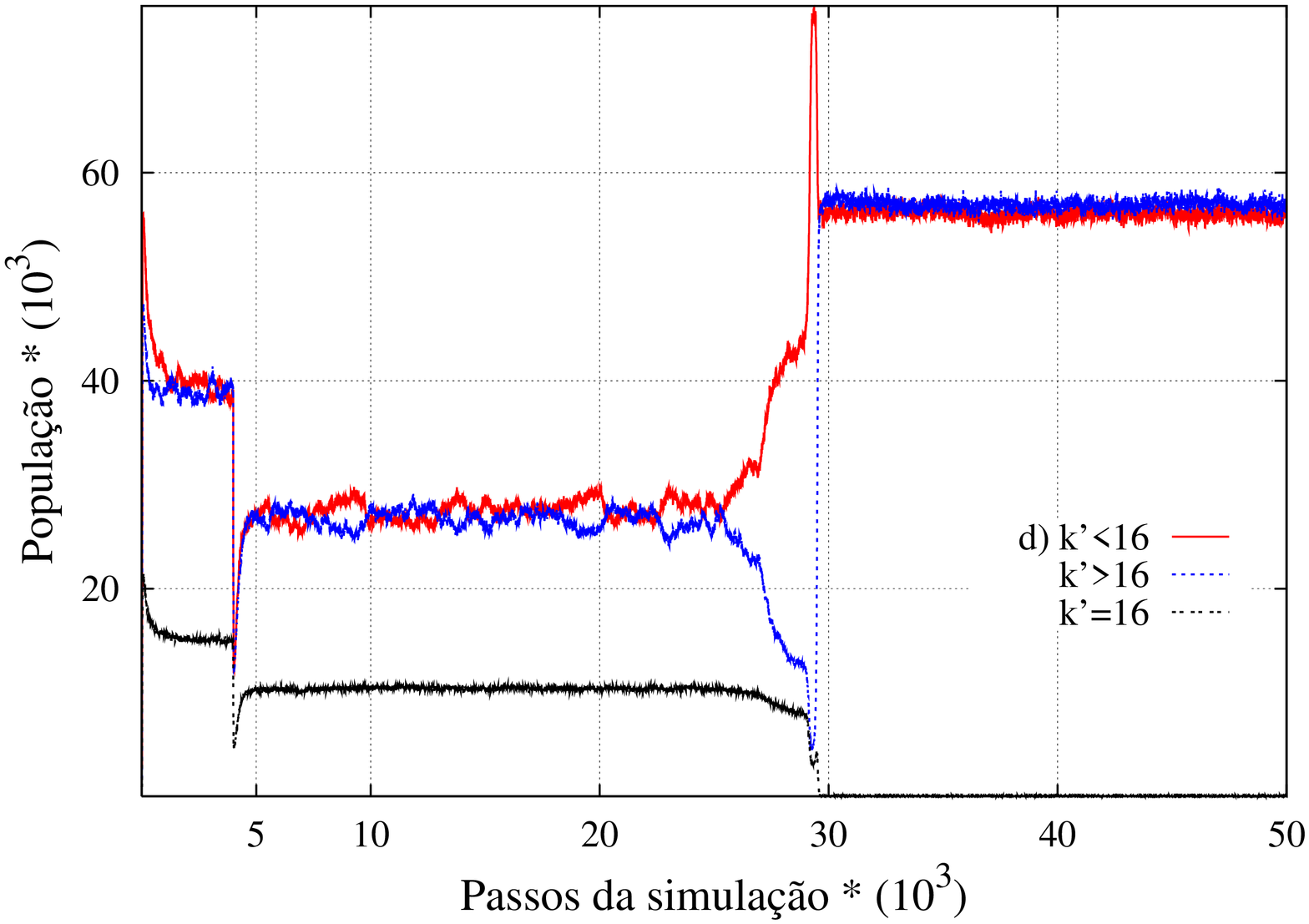}
\includegraphics[width=5.5cm,angle=270]{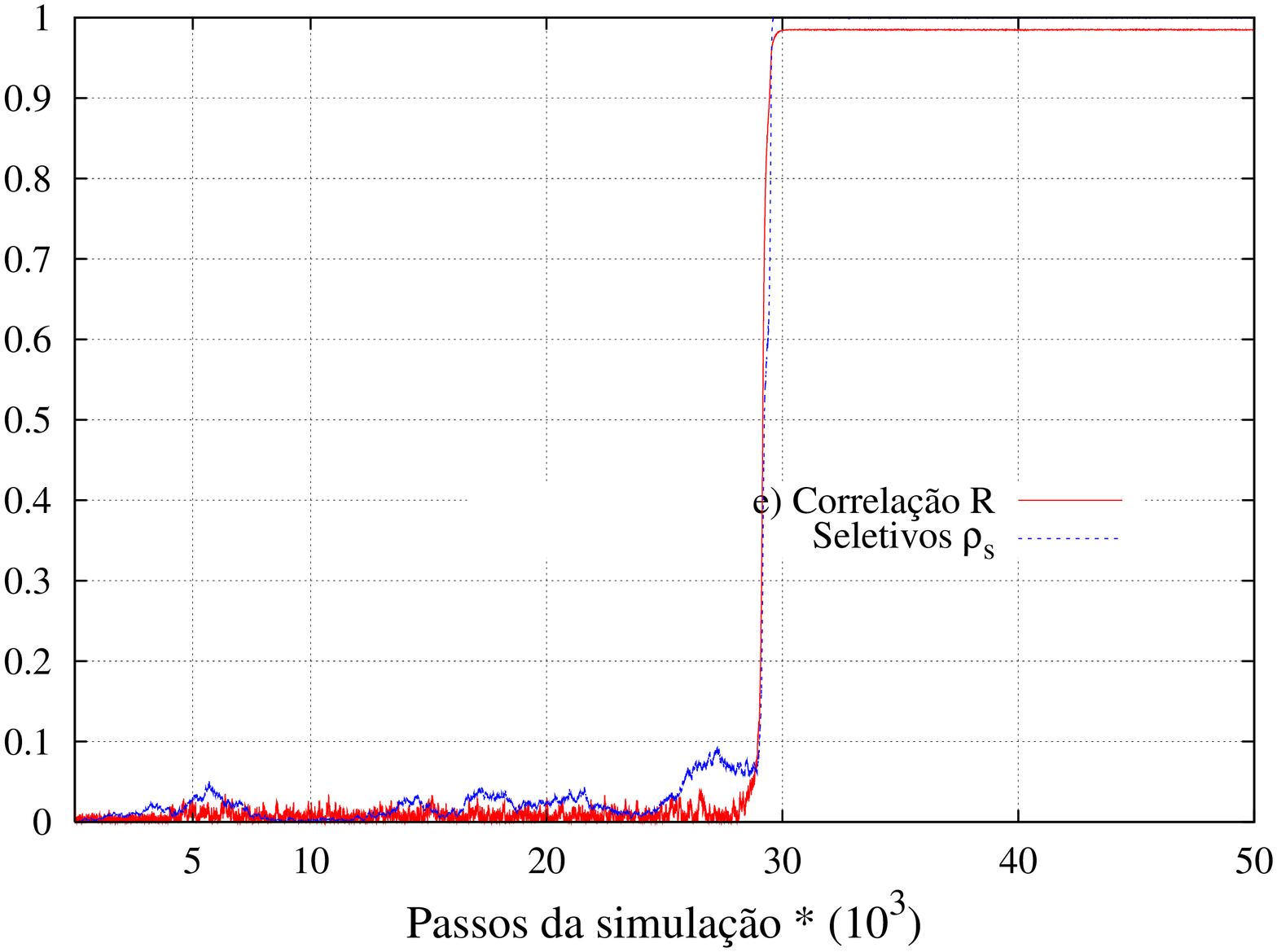}
\includegraphics[width=5.5cm,angle=270]{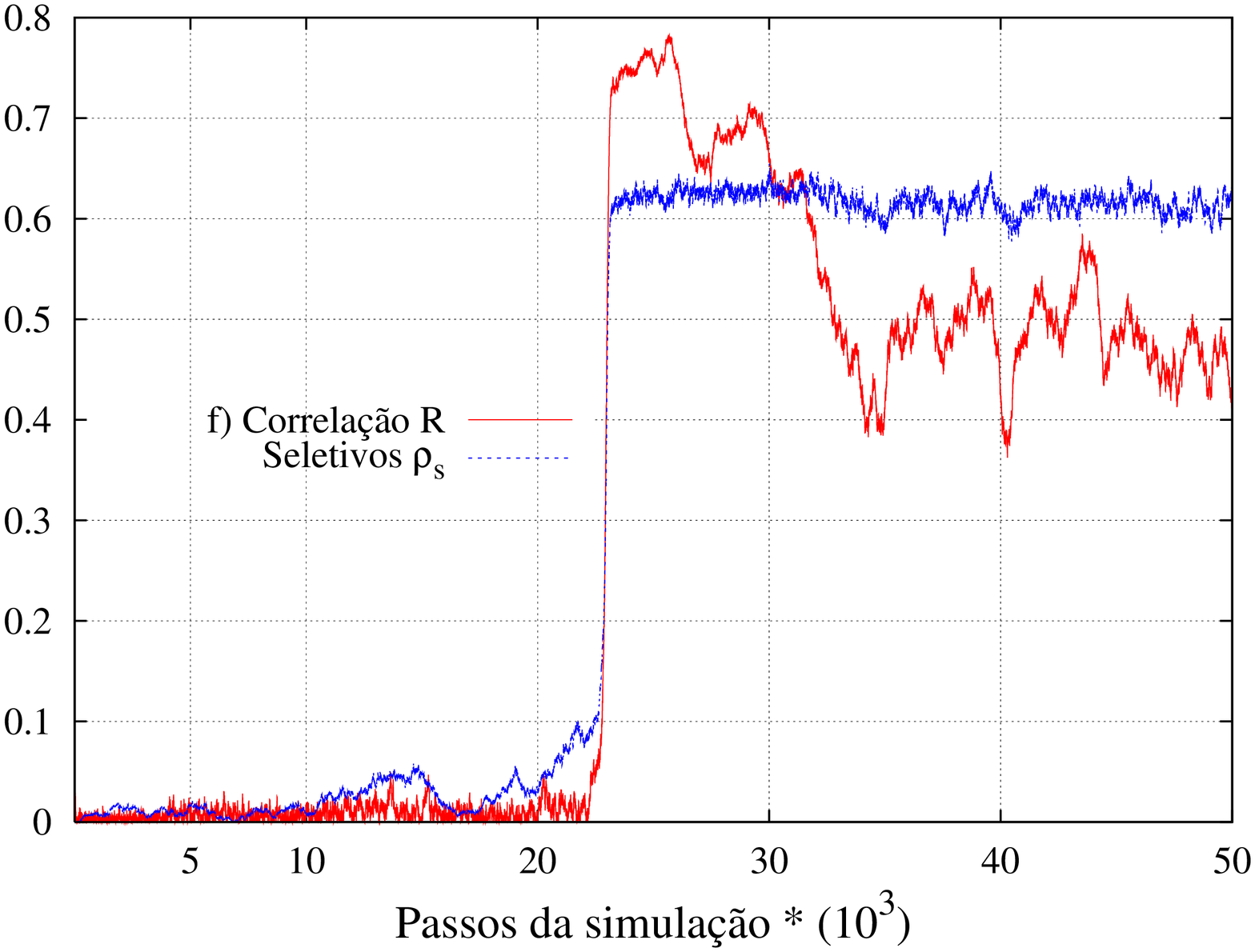}
\end{center}
\caption{\it Especia\c{c}\~ao para $MS=0.0001$, $MF=0.1$ e $MFS=0.5$. (a) Distribui\c{c}\~ao dos 
fen\'otipos $k$ - tra\cc o ecol\'ogico. (b) Distribui\c{c}\~ao dos fen\'otipos $k'$ - tra\cc o 
sexual. (c) Comportamento da popula\c{c}\~ao com fen\'otipo $k$ e 
(d) com fen\'otipo $k'$, durante o
processo de especia\cao.~(e) Varia\c{c}\~ao temporal do valor absoluto da correla\c{c}\~ao $R$ 
(linha s\'olida) e da densidade de seletivos $\rho_s$ (linha pontilhada). 
(f) Varia\c{c}\~ao temporal da correla\c{c}\~ao e da densidade de seletivos num caso onde 
n\~ao ocorreu especia\cao (2Maxilares/1Cor).}
\label{figinestable}
\end{figure} 
 
Comparando os resultados desta subse\c{c}\~ao (ver tabela \ref{tabpolims2} e figura  
\ref{figinestable}) com os da subse\c{c}\~ao anterior, pode-se observar que   
a diminui\c{c}\~ao de $MS$ leva aos seguintes efeitos:   
\bigskip

\noindent 1) O n\'umero de casos em que ocorre especia\c{c}\~ao aumenta (comparar tabelas). 

\noindent 2) A distribui\c{c}\~ao de cores permanece centrada no meio ($k'=16$) at\'e o 
momento em que se d\'a a mudan\c{c}a na ecologia (fig.\ref{figinestable}-(b)). 

Os \'{\i}tens 1) e 2) corroboram a conclus\~ao de que quanto mais centrada estiver a 
distribui\c{c}\~ao de cores antes da ecologia mudar, maior a probabilidade da popula\c{c}\~ao 
especiar. 

\noindent 3) A correla\c{c}\~ao entre os tra\cc os ecol\'ogico e sexual diminui ligeiramente 
(ver tabelas \ref{tabpolims1} e \ref{tabpolims2}).  

\noindent 4) A divis\~ao na distribui\c{c}\~ao dos maxilares ocorre ao mesmo tempo que 
a da cor e n\~ao antes, como no caso anterior.

Os \'{\i}tens 3) e 4) parecem indicar que valores menores de $MF$ s\~ao mais apropriados 
para descrever os cicl\'{\i}deos. Veremos nas conclus\~oes deste cap\'{\i}tulo que 
\'e poss\'{\i}vel melhorar um pouco mais estes resultados.  
 
\begin{figure}[htbp]
\begin{center}
\includegraphics[width=6.5cm,angle=270]{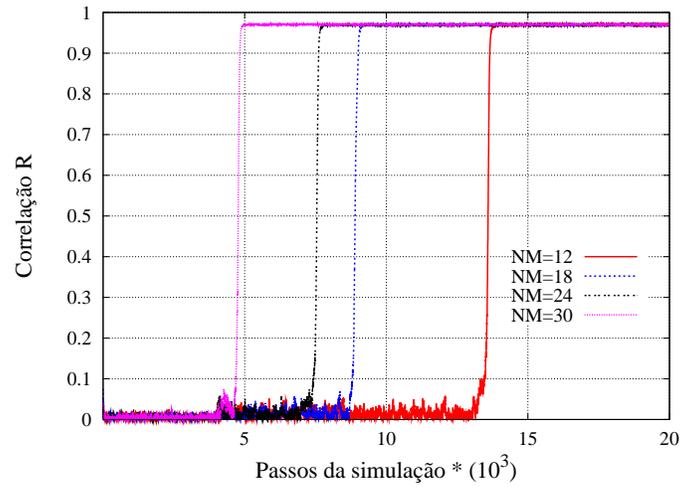}
\end{center}
\caption{\it Correla\c{c}\~ao em fun\c{c}\~ao do tempo. O tempo que a densidade de seletivos 
leva para se aproximar de 1 
depende do n\'umero de escolhas $NM$. Quanto mais exigente \'e a femea, mais r\'apida 
\'e a especia\cao.}
\label{figveloms3}
\end{figure}

Nas tabelas \ref{tabpolims1} e \ref{tabpolims2} n\~ao 
foi encontrado nenhum caso de especia\c{c}\~ao para $MF=1.0$ e $MFS=0.1$, mas 
estes podem ser obtidos utilizando-se valores altos do n\'umero $NM$ de machos que a 
f\^emea t\^em \`a disposi\c{c}\~ao para escolher na hora do acasalamento. Quanto maior for 
$NM$, mais pr\'oximo do momento em que a ecologia se divide se d\'a o crescimento s\'ubito 
da densidade de seletivos e consequentemente, mais r\'apida \'e a especia\c{c}\~ao, como 
mostra a figura \ref{figveloms3}.
  
\begin{figure}
\begin{center}
\includegraphics[width=5.5cm,angle=270]{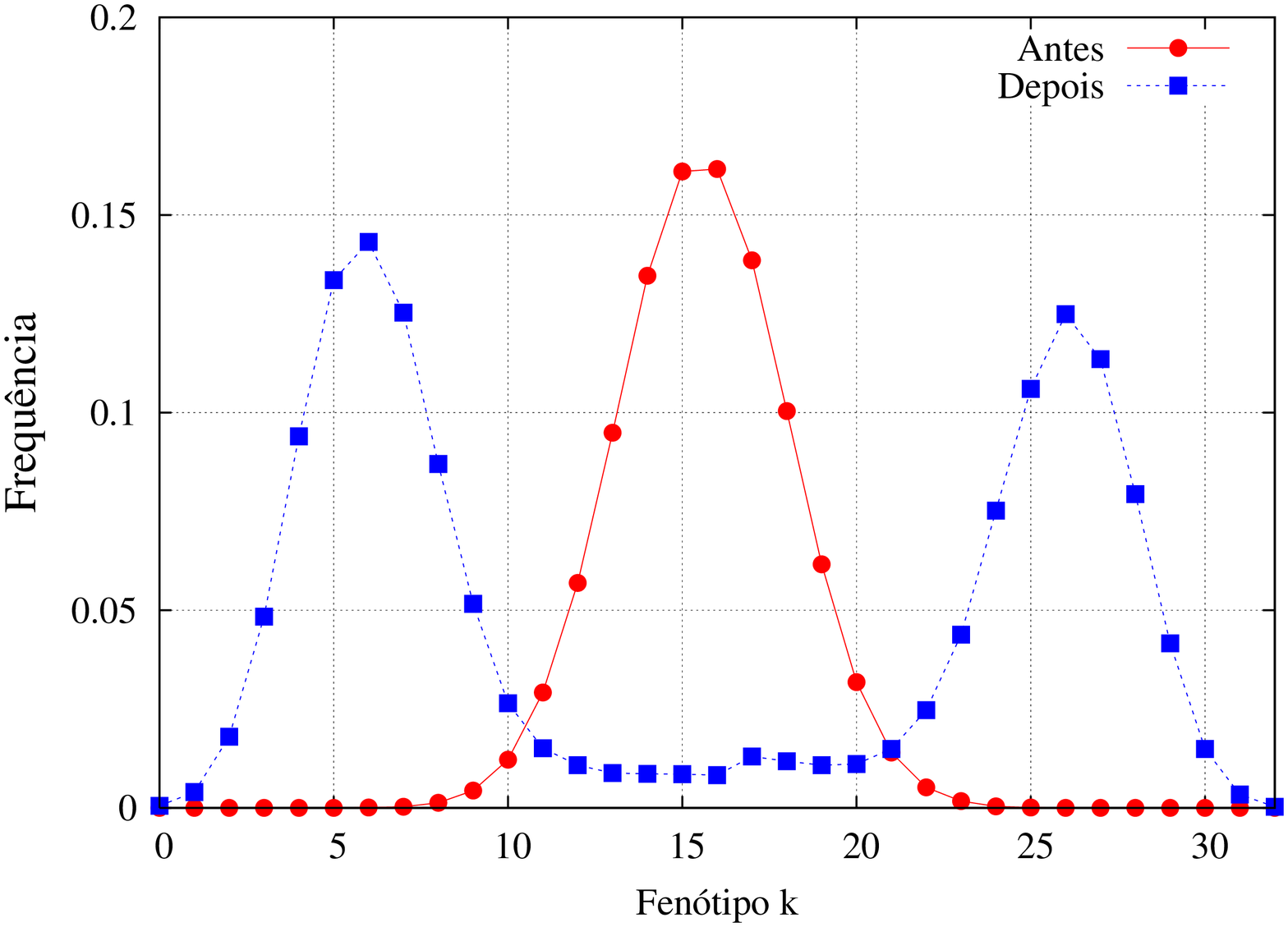}
\includegraphics[width=5.5cm,angle=270]{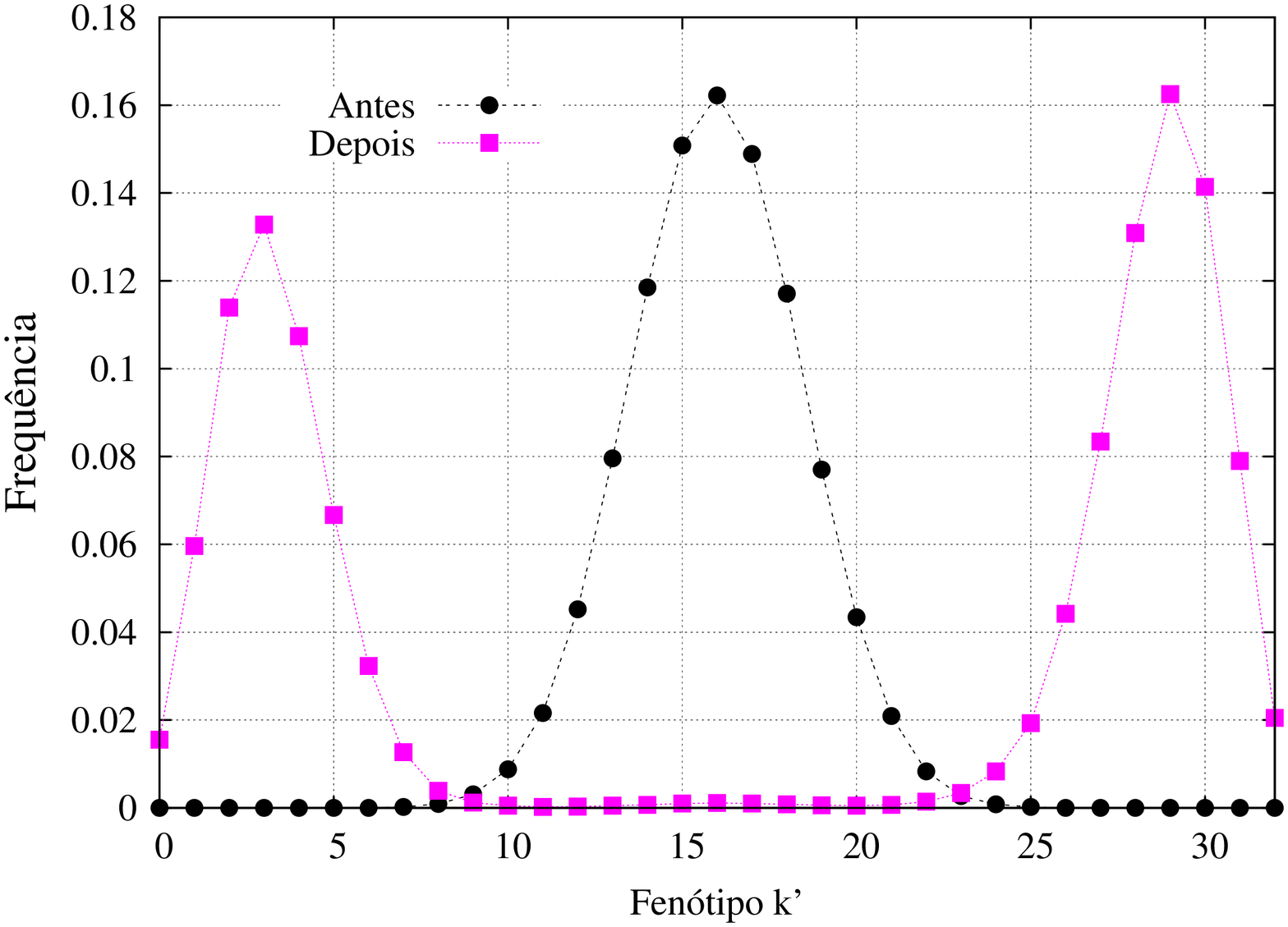}
\end{center}
\caption{\it Especia\c{c}\~ao para $MS=0.0001$, $MF=0.5$ e $MFS=0.5$. Lado esquerdo: 
distribui\c{c}\~ao dos 
fen\'otipos $k$ - tra\cc o ecol\'ogico. Lado direito: distribui\c{c}\~ao dos fen\'otipos 
$k'$ - tra\cc o sexual.} 
\label{figconclusao}
\end{figure} 

\section{Perspectivas}

Como pode ser notado, neste cap\'{\i}tulo n\~ao foi dito se os estados 
1Maxilar/1cor e 2Maxilares/1cor s\~ao est\'aveis, pois aumentando-se 
o tempo de simula\c{c}\~ao para $160\times10^5$ passos as distribui\c{c}\~oes  dos dois 
tra\cc os n\~ao mudam, mas a densidade de seletivos flutua entre  
0.1 e 1.0, em intervalos variados com e sem sicronia, dependendo da 
semente aleat\'oria inicial. 
Gostar\'{\i}amos de estudar estes estados, analisando as s\'eries temporais
da densidade de seletivos e das popula\c{c}\~oes  com m\'etodos como o TISEAN
(Nonlinear Time Series Analysis) \cite{tisean}. O objetivo \'e conhecer o tipo
de estabilidade dos casos de especia\c{c}\~ao das tabelas e o tipo de 
instabilidade dos estados 1Maxilar/1Cor e 2Maxilares/1Cor. 

Gostar\'{\i}amos ainda de continuar procurando, na regi\~ao de $MS$ pequeno, uma  
combina\c{c}\~ao de par\^ametros que resulte numa correla\c{c}\~ao mais baixa e numa 
divis\~ao da cor anterior \`a dos maxilares, como ocorre nos cicl\'{\i}deos.
Na figura \ref{figconclusao}, por exemplo, mudamos apenas o valor de $MF=0.1$ para $MF=0.5$; 
com isto, j\'a obtivemos uma popula\c{c}\~ao de fen\'otipos intermedi\'arios $k$ 
(maxilares intermedi\'arios) que n\~ao desaparece completamente, embora as cores j\'a 
estejam totalmente separadas. Contudo, a correla\c{c}\~ao (n\~ao mostrada) continua 
pr\'oxima de 1.

\chapter{Transi\c{c}\~ao de fase na especia\c{c}\~ao simp\'atrica}

Neste cap\'itulo utilizaremos um modelo no qual  
o tra\cc o fenot\'{\i}pico relacionado \`a sele\c{c}\~ao ecol\'ogica \'e o mesmo  
utilizado para a sele\c{c}\~ao sexual e que comanda a competi\c{c}\~ao entre os indiv\'{\i}duos. 
A motiva\c{c}\~ao biol\'ogica deste modelo se relaciona com o aspecto 
da irreversibilidade do processo de especia\c{c}\~ao em fun\c{c}\~ao do 
isolamento reprodutivo.

Inicialmente mostraremos o comportamento
da distribui\c{c}\~ao dos fen\'otipos para diferentes graus de 
competi\c{c}\~ao $X$. Veremos que para valores altos de $X$, isto \'e, quando os 
fen\'otipos intermedi\'arios est\~ao submetidos a uma forte competi\c{c}\~ao, \'e  
poss\'{\i}vel obter especia\c{c}\~ao sem qualquer mudan\cc a na ecologia.
A seguir, estudando o valor m\'edio da densidade de seletivos em fun\c{c}\~ao do  
grau de competi\cao,~neste ambiente ecol\'ogico que independe 
do tempo e do fen\'otipo, observaremos um comportamento 
caracter\'{\i}stico de uma transi\c{c}\~ao de fase onde a densidade de seletivos corresponde 
ao par\^ametro de ordem e o grau de competi\c{c}\~ao, ao par\^ametro de controle.  

\section{Motiva\c{c}\~ao Biol\'ogica}
A larva da mosca da ma\cc\~a, {\it Rhagoletis pomonella}, \'e um dos 
exemplos de uma poss\'{\i}vel especia\c{c}\~ao simp\'atrica em curso \cite{kjj407}.
A {\it Rhagoletis} acasala preferencialmente perto da fruta que \'e sua planta
hospedeira; diferen\cc as nas prefer\^encias por determinado hospedeiro podem
resultar num completo isolamento pr\'e-acasalamento entre indiv\'{\i}duos 
da mesma esp\'ecie. A ma\cc\~a Crataegus spp. \'e o hospedeiro nativo 
da {\it Rhagoletis pomonella} na Am\'erica do Norte, mas nos anos 80, 
uma popula\c{c}\~ao nova surgiu na ma\c{c}\~a dom\'estica, Malus pumila. 
 Os cromossomas desta nova popula\c{c}\~ao foram mapeados e
comparados aos da primeira esp\'ecie. Os dados indicam uma invers\~ao do
polimorfismo, ou seja, h\'a uma possibilidade das duas popula\c{c}\~oes estarem voltando a 
fundir-se numa \'unica \cite{jjkjnj163}. 
Este comportamento nas larvas da mosca levam a uma
quest\~ao muito interessante a respeito da reversibilidade ou irreversibilidade do processo 
de especia\cao.~Este tipo de quest\~ao se enquadra perfeitamente  
em fen\^omenos do tipo transi\c{c}\~ao de fase.

\section{Especia\c{c}\~ao independente da ecologia}
\begin{table}[htbp]
\begin{center}
{\footnotesize
\begin{tabular}{|l||c||c|}
\hline
Descri\c{c}\~ao das grandezas (macho e f\^emea) &Vari\'avel & Valor\\
\hline\hline
 Muta\c{c}\~ao (por tira de bits) & $M$ & 1* \\ \hline
 Domin\^ancia&$D$ & 3*\\ \hline  
Limite de doen\cc as& $L$ & 3* \\ \hline
Idade m\'{\i}nima de reprodu\c{c}\~ao & $R$ &10* \\ \hline 
N\'umero de filhos& $NF$  & 2*\\ \hline \hline
Domin\^ancia & $DF$ & 16 \\ \hline
Muta\c{c}\~ao do fen\'otipo por bit& $MF$ & 0.01*\\ \hline
Muta\c{c}\~ao na seletividade& $MS$ &  0.001*  \\ \hline
Fen\'otipos extremos & $n_1$ & 13\\ \hline
Fen\'otipos extremos & $n_2$ & 19\\ \hline
N\'umero de machos dispon\'{\i}veis & $NM$ & 50\\ \hline
\end{tabular}
\caption{Par\^ametros do modelo, relativos tanto aos    
genomas cronol\'ogicos (parte superior da tabela) quanto aos genomas sem estrutura 
de idade (parte inferior da tabela). Tempo de simula\c{c}\~ao de $4\times10^4$ passos.
*Grandezas que n\~ao mudam neste cap\'{\i}tulo.}
\label{tabgenes}
}
\end{center}
\end{table}
Nesta se\c{c}\~ao ser\~ao mostradas  as distribui\c{c}\~oes dos fen\'otipos $k$ 
da popula\c{c}\~ao para diferentes graus de competi\c{c}\~ao $X$, num ambiente ecol\'ogico
onde a capacidade de sustenta\c{c}\~ao {\it n\~ao muda no tempo nem depende do 
valor de $k$}, isto \'e, $F(k,t)\equiv F=2\times10^5$. Observe que neste caso 
os ingredientes de 
nosso modelo s\~ao os mesmos do modelo de campo m\'edio de Dieckmann e Dobeli \cite{dd}, que 
como mencionado no cap\'{\i}tulo anterior (se\c{c}\~ao motiva\c{c}\~ao biol\'ogica), \'e 
considerado mais realista que o modelo tamb\'em do tipo campo m\'edio de Kondrashov e 
Kondrashov \cite{kk}.   
Desde o in\'{\i}cio da simula\c{c}\~ao os indiv\'{\i}duos est\~ao sujeitos \`as   
competi\c{c}\~oes por alimento e acasalamento,~em fun\c{c}\~ao de suas carater\'{\i}sticas 
fenot\'{\i}picas, $k$.
A simula\c{c}\~ao come\cc a com todos os indiv\'{\i}duos n\~ao seletivos, mas a popula\c{c}\~ao 
de f\^emeas pode se tornar seletiva e neste caso acasalar de acordo com a regra com forte 
dire\cao.~\'E importante lembrar  
que em nossas simula\c{c}\~oes dividimos a popula\c{c}\~ao em
tr\^es grupos, de acordo com os valores de $k$. O grupo $P_1$ consiste 
dos indiv\'{\i}duos com $0\le k<n_1$; o grupo do meio, $P_m$, corresponde
aos fen\'otipos intermedi\'arios onde $n_1\le k\le n_2$ e o terceiro grupo, 
$P_2$, corresponde aos fen\'otipos com $n_2<k\le32$. 
O tipo de competi\c{c}\~ao entre os indiv\'{\i}duos de cada grupo \'e dado pela 
equa\c{c}\~ao \ref{eqcompe}. Os valores utilizados 
para $n_1$, $n_2$ e demais par\^ametros podem ser vistos na tabela \ref{tabgenes}. 
Observe que agora os fen\'otipos intermedi\'arios correspondem a uma faixa de valores, ao  
inv\'es de apenas a $k=16$.

\subsection{Resultados} 
\begin{figure}[htbp]
\begin{center}
\includegraphics[width=5.5cm,angle=270]{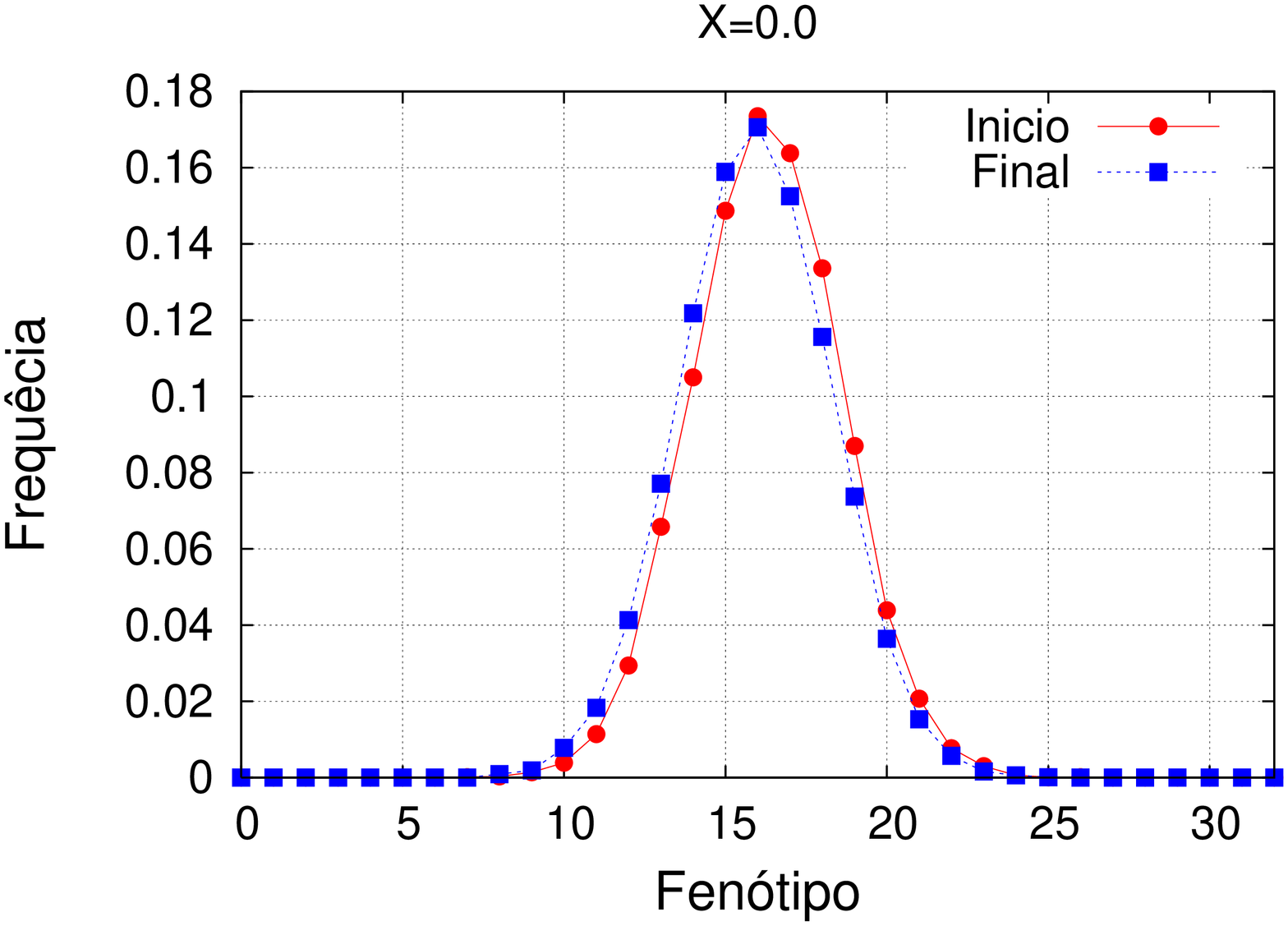}
\includegraphics[width=5.5cm,angle=270]{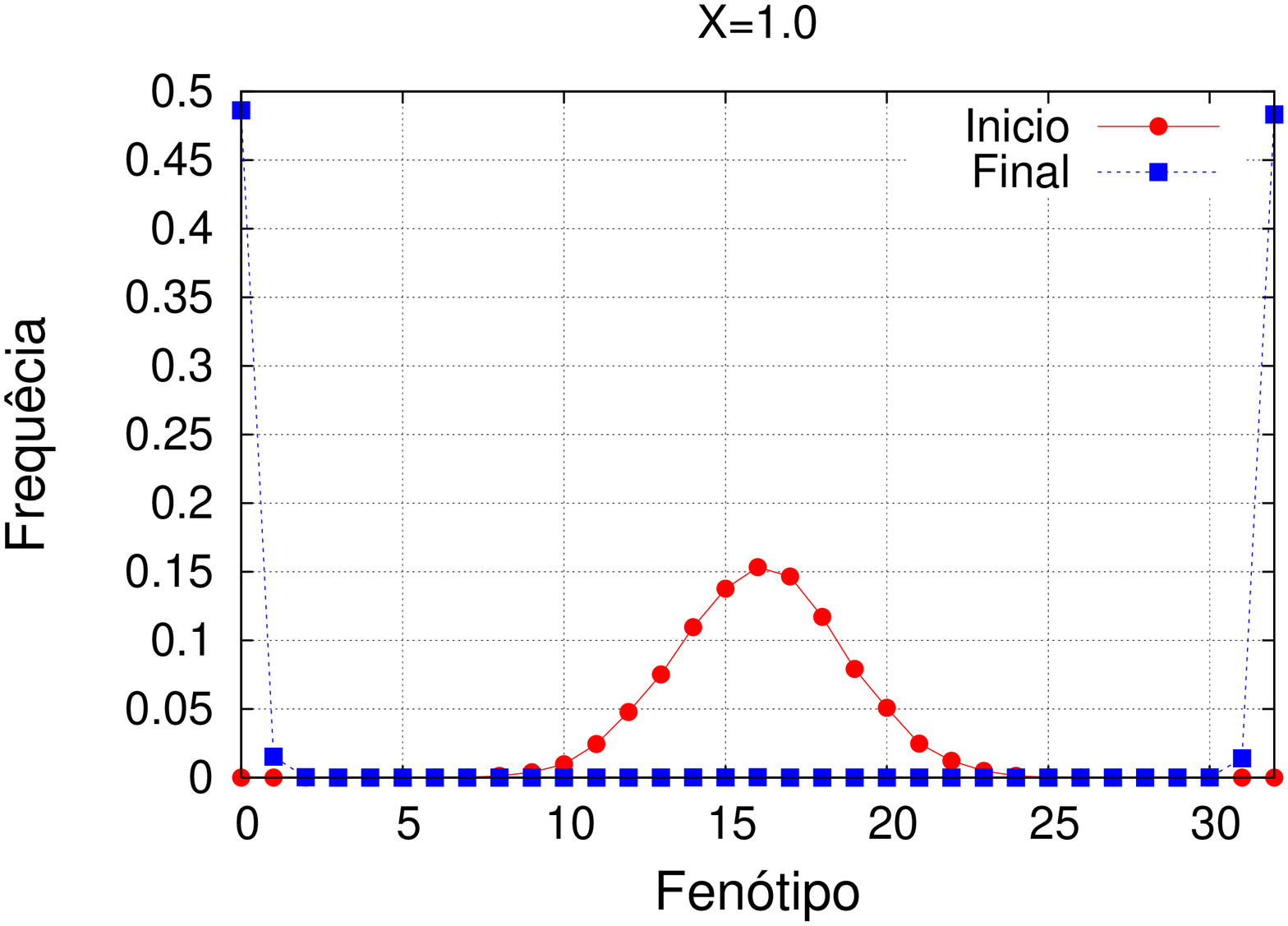}
\end{center}
\caption{\it Distribui\c{c}\~ao dos fen\'otipos nos primeiros 200 passos
de simula\c{c}\~ao e distribui\c{c}\~ao est\'avel no final da 
simula\cao,~ap\'os $4\times10^4$ passos.
Lado esquerdo: Competi\c{c}\~ao favor\'avel para a popula\c{c}\~ao com 
fen\'otipo intermedi\'ario ($X=0.0$). Lado direito: Competi\c{c}\~ao altamente desfavor\'avel para a 
popula\c{c}\~ao com fen\'otipo intermedi\'ario ($X=1.0$).} 
\label{figemx0}
\end{figure} 

\begin{figure}[htbp] 
\begin{center}
\includegraphics[width=5.5cm,angle=270]{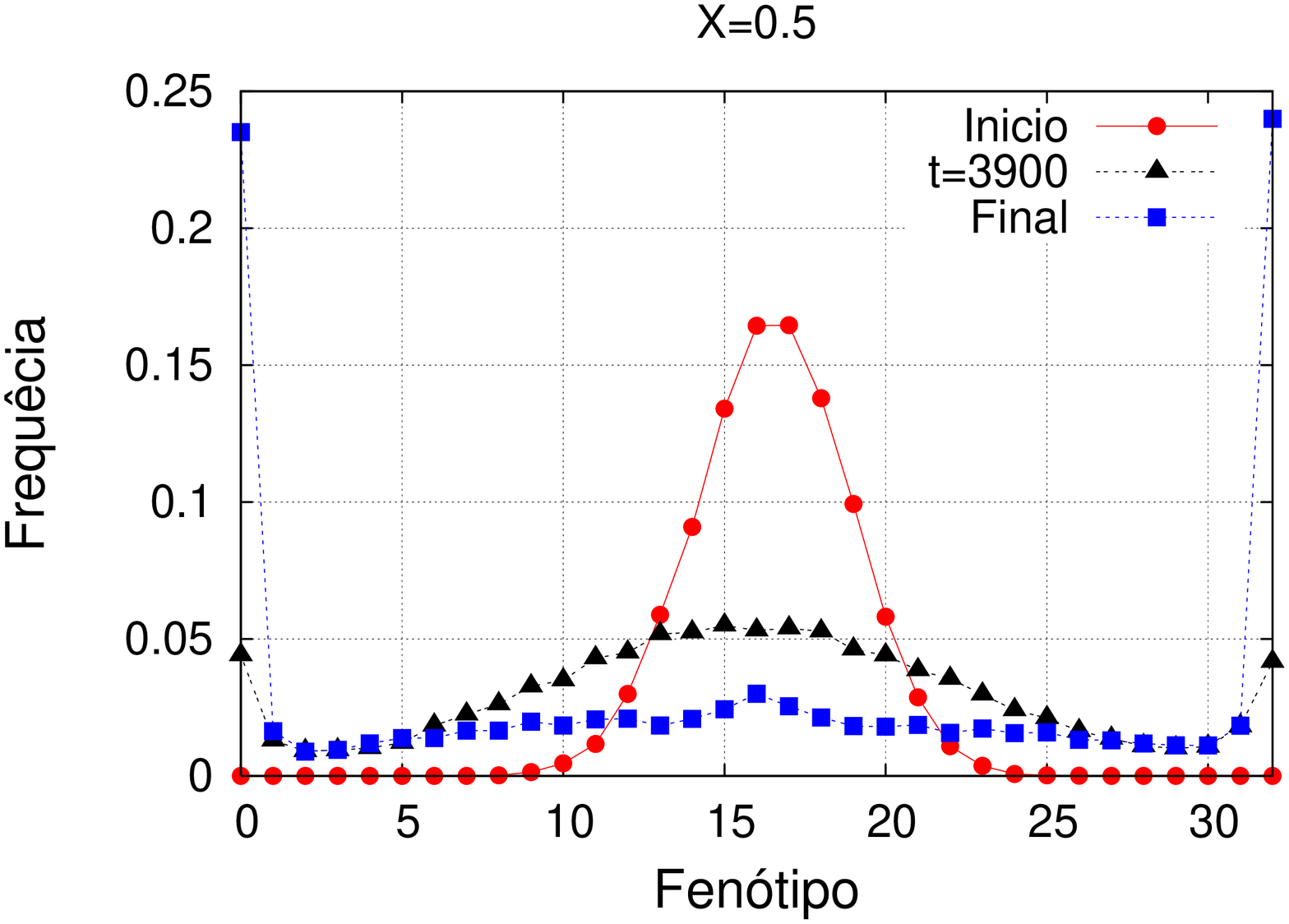}
\includegraphics[width=5.5cm,angle=270]{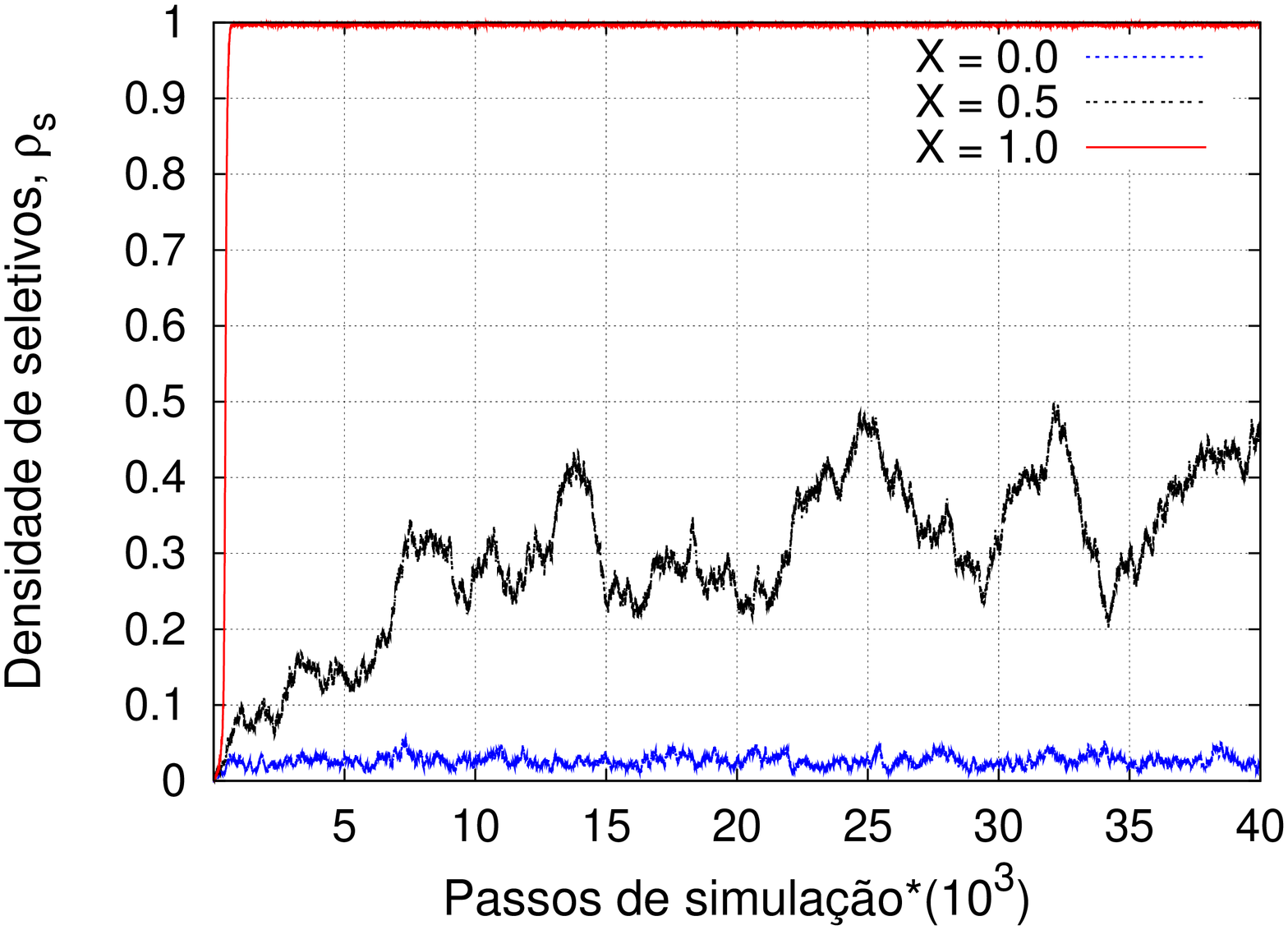}
\end{center}
\caption{\it Lado esquerdo: Distribui\c{c}\~ao dos fen\'otipos 
da popula\c{c}\~ao em tr\^es diferentes instantes: ap\'os 200 passos (c\'{\i}rculos vermelhos), 
ap\'os 3900 passos (tri\^angulos pretos) e no final 
da simula\cao,~ap\'os $4\times10^4$ passos (quadrados azuis). Nos tr\^es casos $X=0.5$. 
Lado direito: Compara\c{c}\~ao da densidade de seletivos em fun\c{c}\~ao do n\'umero de passos 
de simula\cao,~para diferentes graus de competi\c{c}\~ao $X$.} 
\label{figemx1}
\end{figure}

Quando o par\^ametro de competi\c{c}\~ao $X=0.0$, os indiv\'{\i}duos com fen\'otipos intermedi\'arios 
s\~ao favorecidos pois n\~ao competem com os grupos fenot\'{\i}picos
extremos (equa\c{c}\~ao \ref{eqcompe}). Neste caso a distribui\c{c}\~ao de fen\'otipos corresponde 
a uma gaussiana est\'avel e independente da semente aleat\'oria inicial e que pode ser vista 
na figura \ref{figemx0}, lado esquerdo.
O comportamento da densidade 
de seletivos em fun\c{c}\~ao do tempo mostra que a popula\c{c}\~ao atinge um estado 
estacion\'ario onde n\~ao existem f\^emeas que reproduzam seletivamente,  
ou seja, a popula\c{c}\~ao n\~ao especia - 
ver figura \ref{figemx1}, lado direito, curva inferior (azul).

Contrariamente, quando introduzimos um forte grau de competi\c{c}\~ao $X=1.0$ 
a popula\c{c}\~ao com fen\'otipo intermedi\'ario \'e severamente desfavorecida, pois compete 
com as duas popula\c{c}\~oes inteiras de fen\'otipos extremos. Novamente os resultados 
independem da semente aleat\'oria inicial e a distribui\c{c}\~ao final dos 
fen\'otipos da popula\c{c}\~ao \'e est\'avel. Esta distribui\c{c}\~ao est\'avel mostra que
a maioria da popula\c{c}\~ao tem fen\'otipo $k=0$ ou $k=32$, ou seja, os fen\'otipos 
intermedi\'arios  desaparecem como pode ser visto na figura \ref{figemx0}, lado direito. 
O comportamento da densidade de seletivos em fun\c{c}\~ao do tempo
mostra que a popula\c{c}\~ao atinge novamente um estado estacion\'ario, mas que neste caso 
consiste apenas de f\^emeas seletivas - ver figura 
\ref{figemx1}, lado direito, curva superior (vermelha). Vemos portanto que com $X=1.0$  
a popula\c{c}\~ao especia sem necessidade de uma mudan\cc a ecol\'ogica, devido
ao forte grau de competi\cao.

As distribui\c{c}\~oes dos fen\'otipos para um grau de competi\c{c}\~ao intermedi\'ario 
$X=0.5$ podem ser vistas na figura \ref{figemx1}, lado esquerdo, em tr\^es diferentes 
instantes. A correspondente densidade de seletivos em fun\c{c}\~ao do tempo apresenta grandes 
flutua\c{c}\~oes, como mostra a figura \ref{figemx1}, lado direito, curva do meio (preta).

\section{Transi\c{c}\~ao de fase} 
Como acabamos de ver, para $X=0.0$ 
a popula\c{c}\~ao n\~ao especia, isto \'e, a densidade de seletivos \'e est\'avel e pr\'oxima 
de zero.
Para um grau de competi\c{c}\~ao alto, $X=1.0$, a popula\c{c}\~ao especia com uma densidade 
de seletivos est\'avel e pr\'oxima de 1. Contudo, para $X=0.5$ a densidade de seletivos 
flutua muito, 
o que sugere a exist\^encia de uma transi\c{c}\~ao de fase entre um estado onde toda a 
popula\c{c}\~ao permanece n\~ao seletiva e outro onde todas as f\^emeas se tornam seletivas 
e ocorre a especia\c{c}\~ao. Nesta se\c{c}\~ao estudaremos mais detalhadamente como se comporta a 
densidade de seletivos em fun\c{c}\~ao $X$, a fim de verificar se de fato \'e poss\'{\i}vel 
afirmar a exist\^encia e caracterizar esta transi\c{c}\~ao de fase.

\begin{figure}[htbp]
\begin{center}
\includegraphics[width=5.5cm,angle=270]{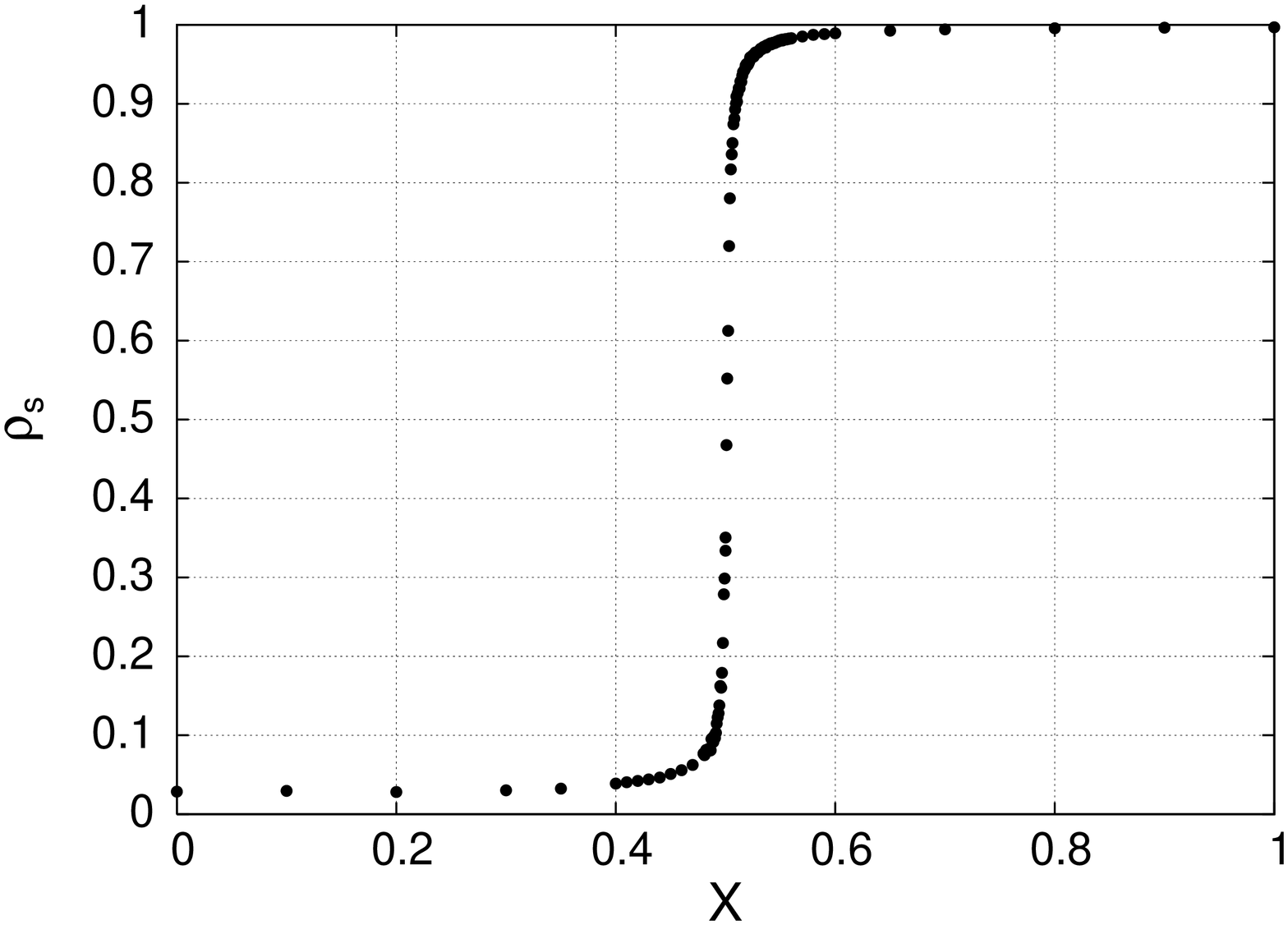}
\includegraphics[width=5.5cm,angle=270]{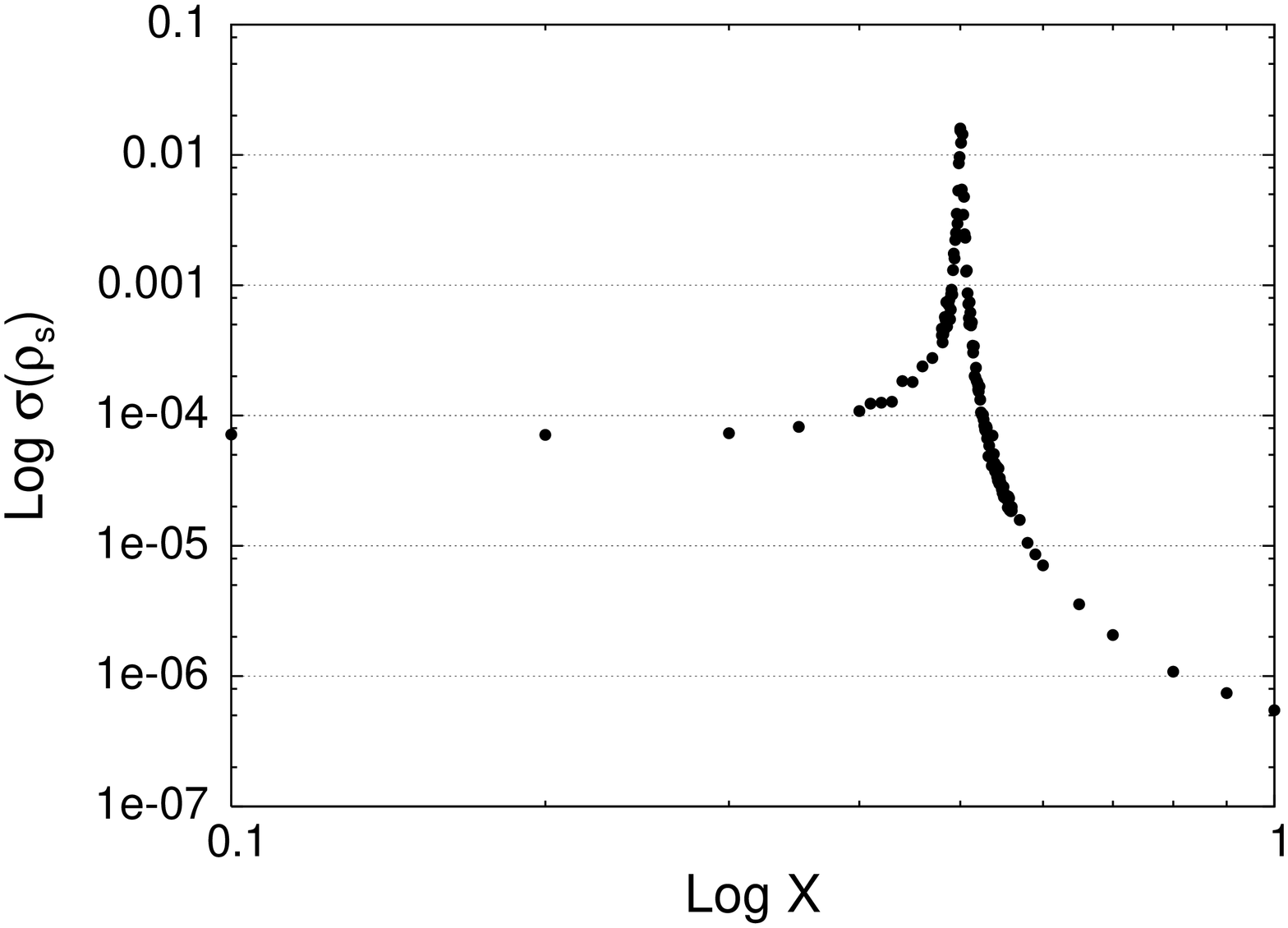}
\end{center}
\caption{\it Lado esquerdo: Comportamento da 
densidade m\'edia de seletivos, obtida com 10 simula\c{c}\~oes distintas, em fun\c{c}\~ao do 
grau de competi\c{c}\~ao $X$. Lado direito: Comportamento do desvio padr\~ao
da densidade m\'edia de seletivos.} 
\label{figtran}
\end{figure}
Para cada valor de competi\c{c}\~ao $X$ foram feitas 10 
simula\c{c}\~oes mudando-se apenas a semente aleat\'oria; 
em cada simula\c{c}\~ao foi calculado o valor m\'edio da densidade de seletivos, 
$\rho_s$, nos \'ultimos $10^4$ passos. Calculamos ent\~ao a m\'edia destes
10 valores m\'edios, cujo comportamento em fun\c{c}\~ao do grau de competi\c{c}\~ao pode ser visto 
na figura \ref{figtran}, lado esquerdo.
No lado direito da figura pode-se observar o comportamento do 
desvio padr\~ao desta m\'edia em fun\c{c}\~ao de $X$.

\begin{figure}[htbp]
\begin{center}
\includegraphics[width=5.5cm,angle=270]{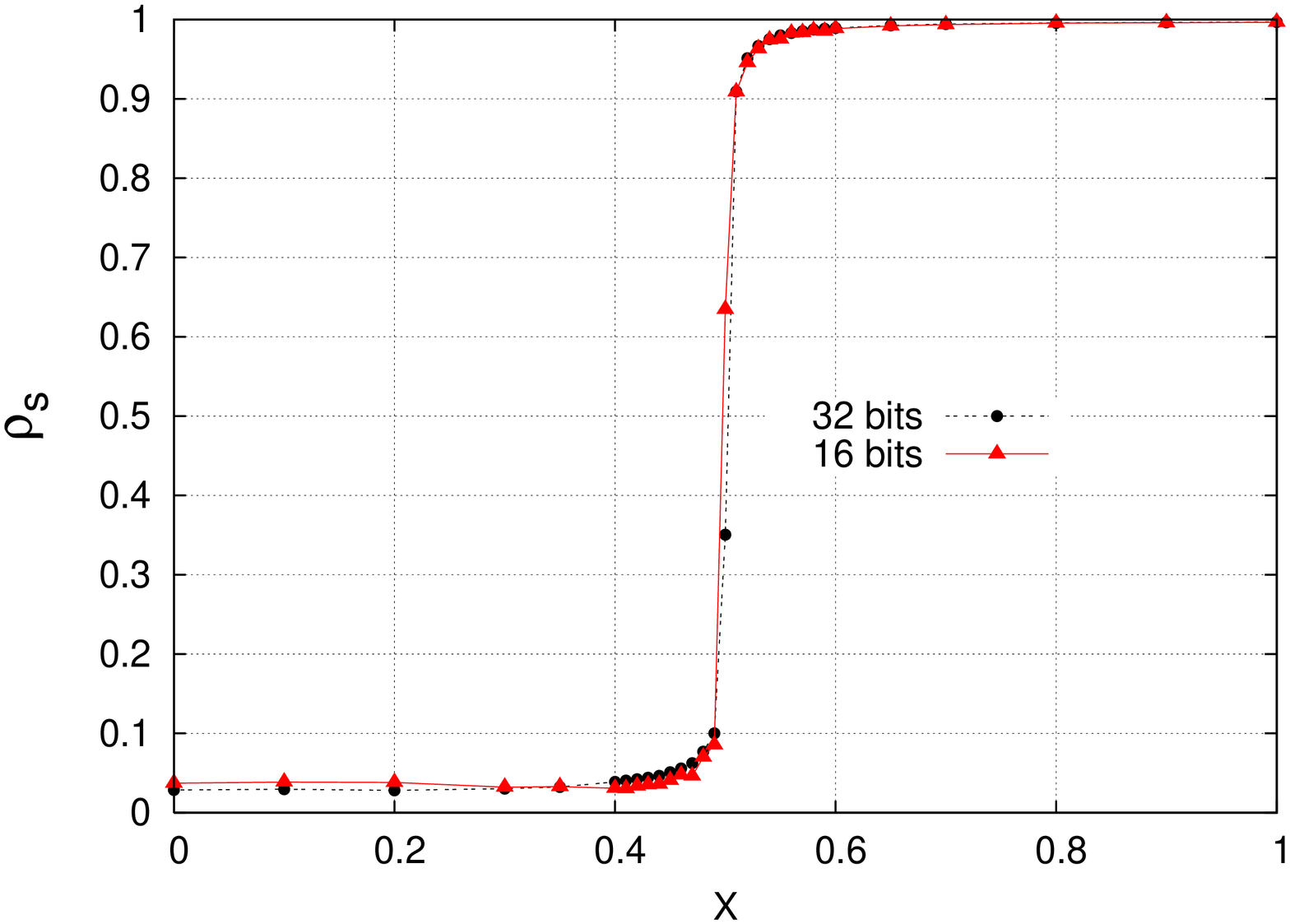}
\includegraphics[width=5.5cm,angle=270]{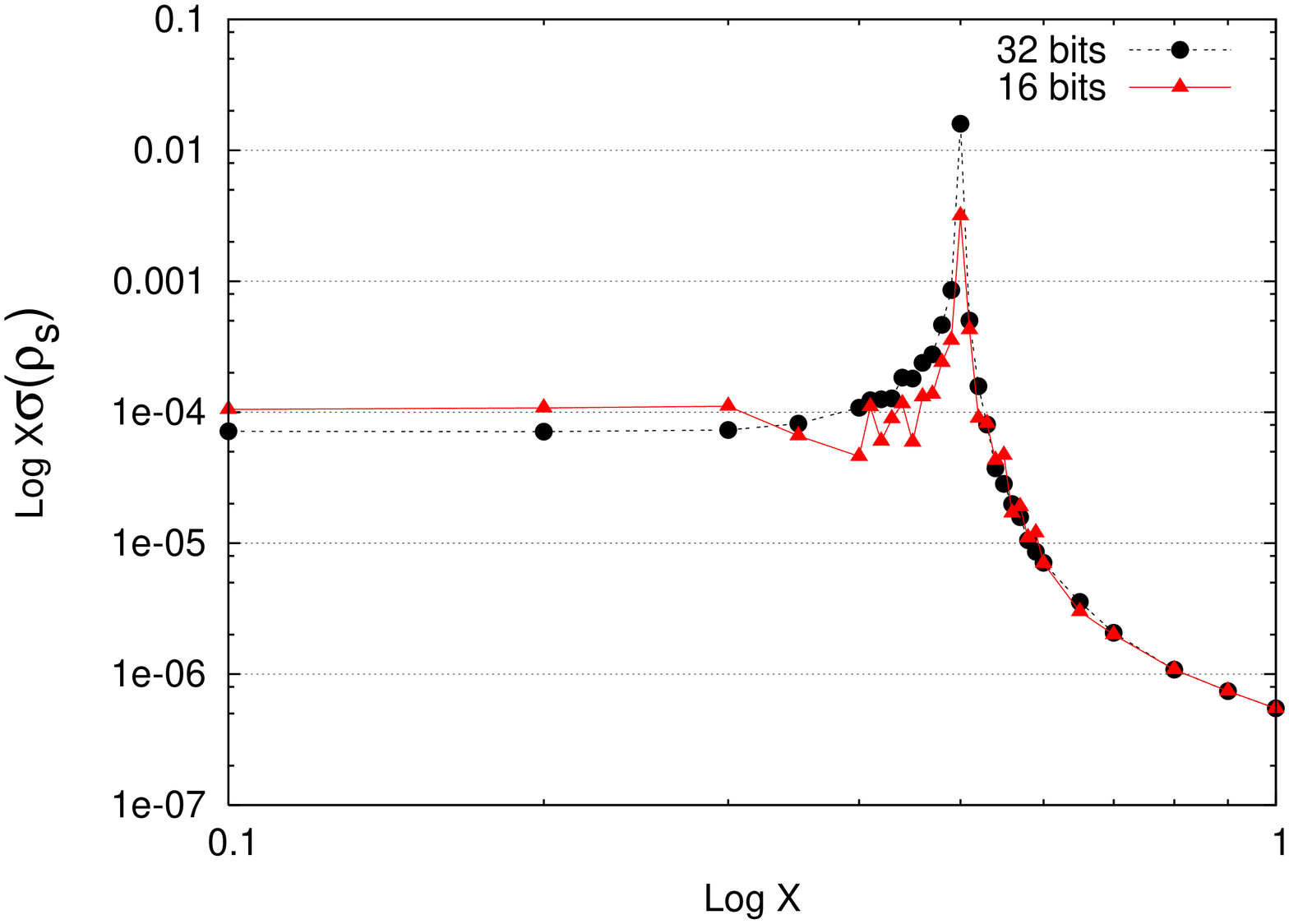}
\end{center}
\caption{\it Comparando, para diferentes tamanhos de tiras de bits, 
os comportamentos da densidade de seletivos m\'edia (lado esquerdo) e de seu desvio padr\~ao 
(lado direito), em fun\c{c}\~ao do grau de competi\c{c}\~ao $X$.}
\label{figsistema}
\end{figure}

Em todas as simula\c{c}\~oes anteriores a carater\'{\i}stica fenot\'{\i}pica de cada indiv\'{\i}duo
era dada por um par de tiras de 32 bits cada. Reduzindo o tamanho das tiras de 32 para 16 bits,
a domin\^ancia fenot\'{\i}pica de $DF=16$ para $DF=8$ e 
a defini\c{c}\~ao dos fen\'otipos $k$ intermedi\'arios de $n_1=13$ e $n_2=19$ para $n_1=7$ e 
$n_2=9$, observamos que os comportamentos de $\rho_s$ e $\sigma(\rho_s)$ variam ligeiramente  
no lado $X<0.5$, como mostra a figura \ref{figsistema}.

Associando a densidade de seletivos ao par\^ametro de ordem e o grau de 
competi\c{c}\~ao $X$ ao par\^ametro de controle, vemos que de fato a figura \ref{figtran} 
indica a exist\^encia de uma transi\c{c}\~ao de fase em torno de $X_c=0.5$, entre um estado 
onde \'e livre a troca de genes entre os indiv\'{\i}duos ($\rho_s=0$) e um outro estado 
no qual coexistem duas popula\c{c}\~oes que n\~ao acasalam entre si ($\rho_s=1$). 
Se esta \'e de fato uma transi\c{c}\~ao de fase, ela se d\'a fora do equil\'{\i}brio \cite{dickman}, 
uma vez que os indiv\'{\i}duos da popula\c{c}\~ao do modelo de Penna original 
vivem num estado meta-est\'avel.

A densidade de seletivos se comporta como o par\^ametro de ordem 
pois muda de um valor zero para um diferente de zero a 
partir de $X=X_c$ e sua derivada, o desvio padr\~ao de seu valor m\'edio, mostra um 
comportamento singular em $X_c=0.5$. Esta singularidade no 
desvio padr\~ao aumenta com o tamanho das tiras de bits usadas 
para caracterizar o fen\'otipo (ver figura \ref{figsistema}, lado direito), indicando  
que o comprimento das tiras de bits faz o papel do tamanho do sistema numa  
transi\c{c}\~ao de fase usual. 
 
O grau de competi\c{c}\~ao cr\'{\i}tico $X=0.5$ pode ser determinado anal\'{\i}ticamente, como 
veremos a seguir. 
Observamos pelas simula\c{c}\~oes que as popula\c{c}\~oes de fen\'otipos extremos t\^em o mesmo 
tamanho, $P_1(k,t)\approx P_2(k,t)$, como era de se esperar, devido 
\`a simetria com que foi definida a competi\c{c}\~ao intra-espec\'{\i}fica 
dada pela equa\c{c}\~ao \ref{eqcompe}. Assim, colocando $P_1=P_2=P_e$ nesta equa\c{c}\~ao 
e chamando $V_1=V_2=V_e$, obtemos que: 
\begin{eqnarray}
V_e(k,t)=(P_e + P_m)/F, \nonumber \\
V_m(k,t) = (P_m + X * 2P_e)/F. \label{eqcompe2}
\end{eqnarray}

Pela equa\c{c}\~ao acima vemos que se $X=0$, $V_m < V_e$ e portanto a probabilidade de 
sobreviv\^encia dos fen\'otipos intermedi\'arios \'e maior que a dos fen\'otipos extremos. 
J\'a para $X=1$, ocorre o oposto.  

O valor de $X$ para o qual a probabilidade de sobreviv\^encia ou, equivalentemente, a 
competi\c{c}\~ao entre todos os indiv\'{\i}duos \'e a mesma, pode ser 
imediatamente obtido impondo-se que $V_m(k,t)=V_e(k,t)$ na equa\c{c}\~ao \ref{eqcompe2} 
e vale $X_c=0.5$.

\begin{figure}[htbp]
\begin{center}
\includegraphics[width=5.5cm,angle=270]{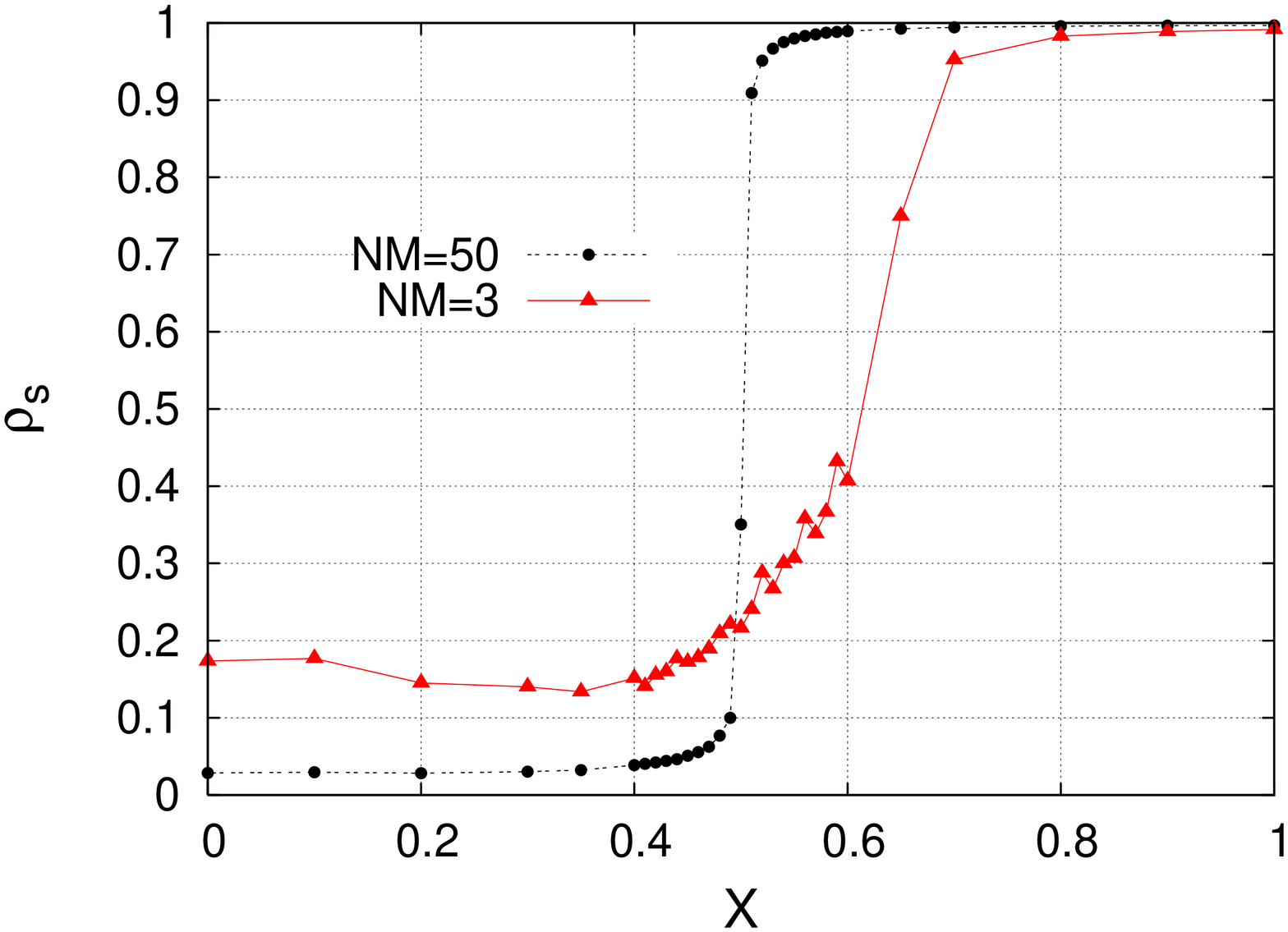}
\includegraphics[width=5.5cm,angle=270]{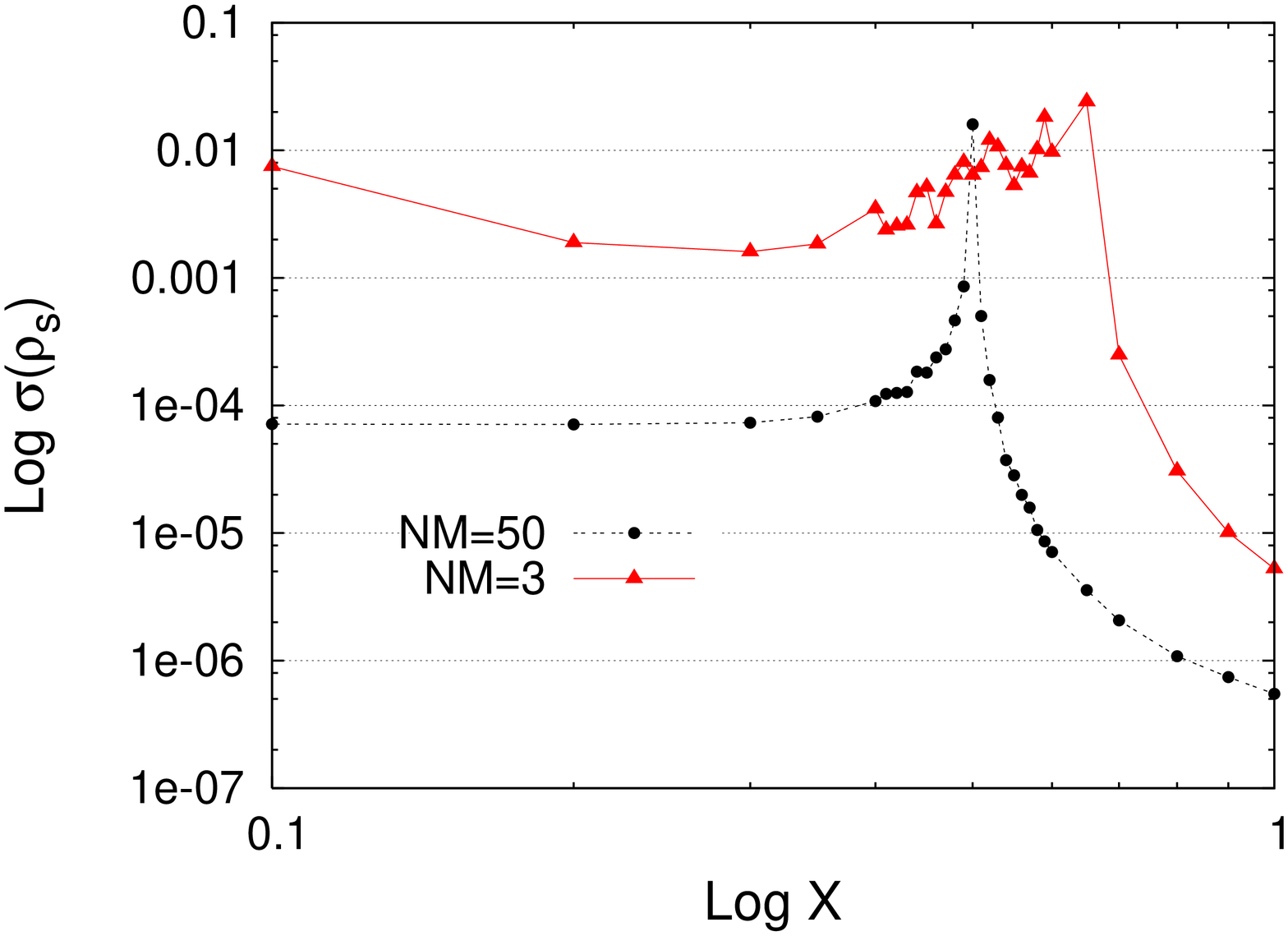}
\end{center}
\caption{\it
Lado esquerdo: Densidade m\'edia de seletivos em fun\c{c}\~ao de $X$ para diferentes 
n\'umeros de escolhas no acasalamento, $NM$. 
Lado direito: Desvio padr\~ao da densidade m\'edia de seletivos em fun\c{c}\~ao de $X$ para 
diferentes valores de $NM$.} 
\label{figcampo}
\end{figure}   

Diminuindo o n\'umero de machos que a f\^emea tem para escolher ao acasalar 
de $NM=50$ para $NM=3$, verificamos que o comportamento da densidade m\'edia de seletivos 
e de seu desvio padr\~ao em fun\c{c}\~ao de $X$ variam, conforme mostra a  
figura \ref{figcampo}. 

Observe que para valores pequenos de $NM$ a transi\c{c}\~ao de fase \'e destruida. Desta forma, 
podemos associar o papel do n\'umero de machos dispon\'{\i}veis para o acasalamento ao inverso 
do campo numa transi\c{c}\~ao de fase usual.

\begin{figure}[htbp]
\begin{center}
\includegraphics[width=5.5cm,angle=270]{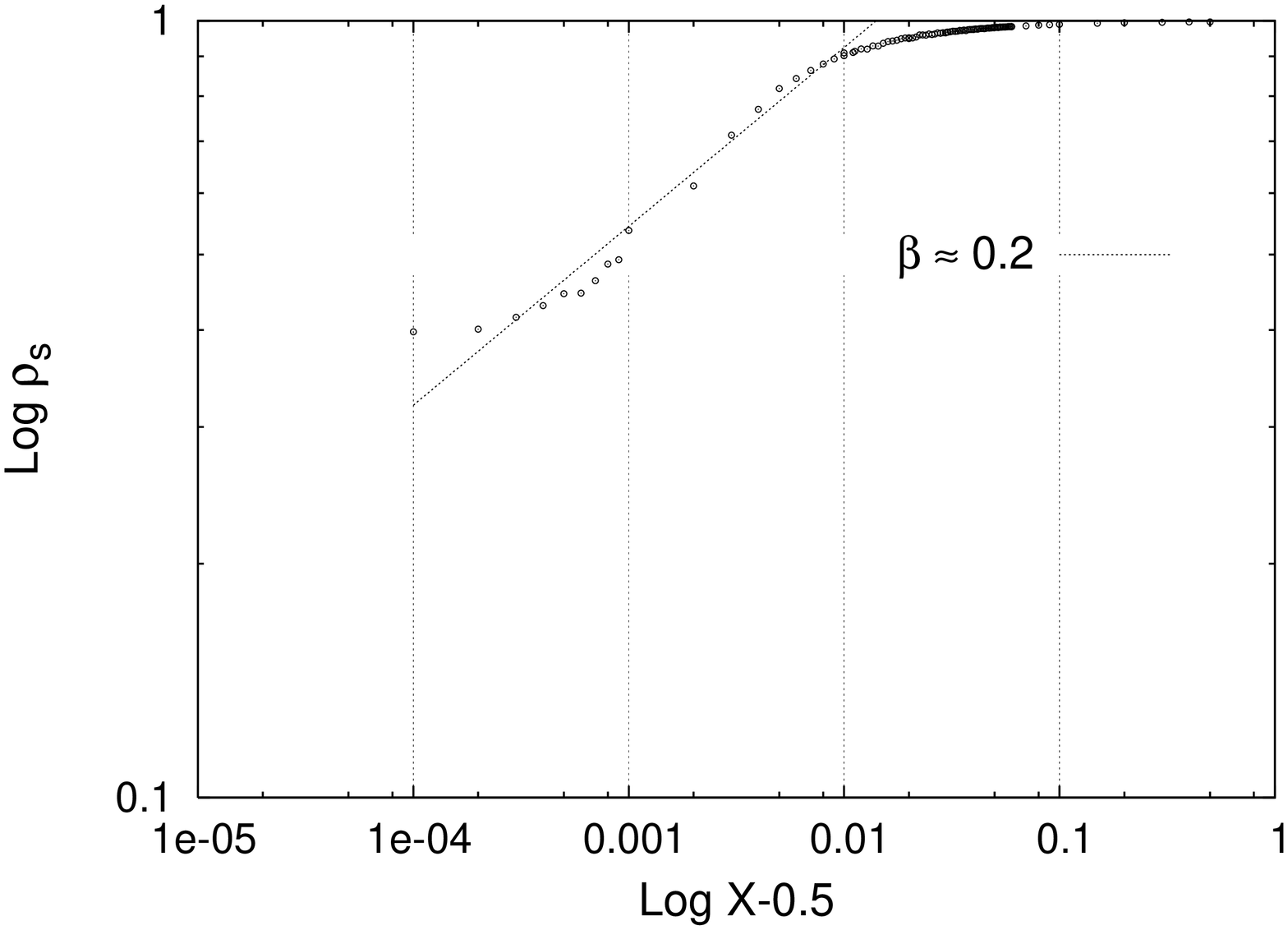}
\includegraphics[width=5.5cm,angle=270]{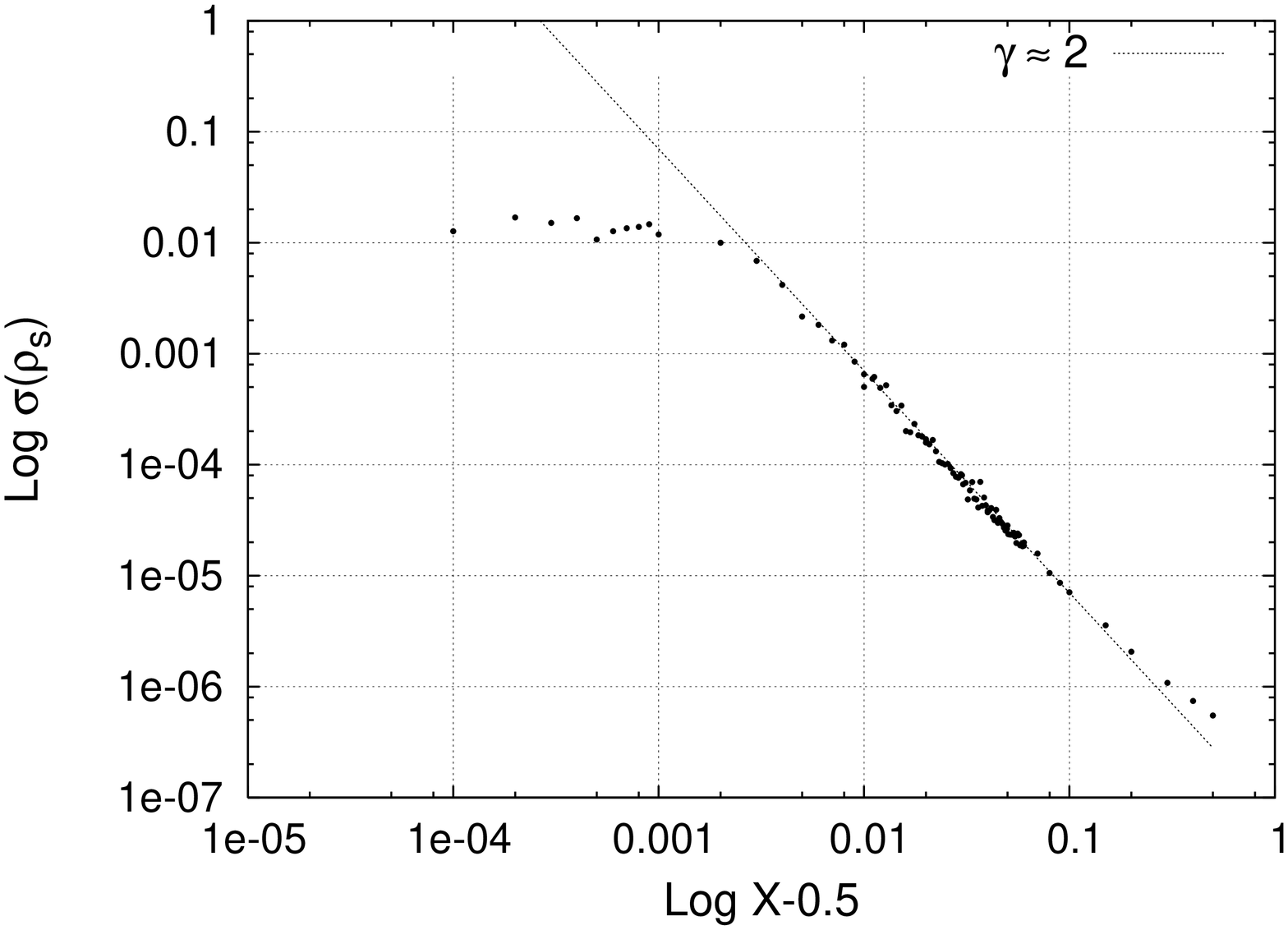}
\end{center}
\caption{\it A inclina\c{c}\~ao das retas foram obtidas
a partir de 100 simula\c{c}\~oes distintas, ao inv\'es de 10. 
Lado esquerdo: Comportamento, perto do ponto cr\'{\i}tico, da densidade m\'edia de 
seletivos. Lado direito: Comportamento, perto do ponto cr\'{\i}tico,  
do desvio padr\~ao da densidade m\'edia de seletivos.}
\label{figexpoentes}
\end{figure}    

Com o objetivo de caracterizar esta aparente transi\c{c}\~ao de fase 
medimos, para 100 simula\c{c}\~oes distintas no intervalo $X-X_c=X-0.5\in [0.001,0.01]$, a 
densidade m\'edia de seletivos e seu desvio padr\~ao. As leis de pot\^encia encontradas podem 
ser vistas na figura \ref{figexpoentes}. A primeira, do lado esquerdo da figura, 
relaciona-se    
ao comportamento do par\^ametro de ordem perto do ponto cr\'{\i}tico, ou 
seja, $\rho_s(X\approx X_c) \sim |X-X_c|^{\beta}$. A segunda, do lado direito, 
est\'a relacionada com o desvio padr\~ao tamb\'em pr\'oximo ao ponto cr\'{\i}tico, isto \'e,  
$\sigma_{\rho_s}(X\approx X_c)\sim |X-X_c|^{-\gamma}$.  

Sabemos que os resultados aqui apresentados n\~ao s\~ao conclusivos no que diz respeito  
\`a ordem da transi\c{c}\~ao. Para concluir se esta transi\c{c}\~ao \'e de primeira ou de 
segunda ordem \'e necess\'ario, por exemplo, repetir as simula\c{c}\~oes que deram origem 
\`a figura \ref{figsistema} para tamanhos de tiras de bits maiores e verificar o que 
ocorre com o pico da suscetibilidade (desvio padr\~ao da densidade m\'edia de seletivos). 
Se este pico divergir, a transi\c{c}\~ao \'e de segunda ordem, se n\~ao, \'e de primeira. 
Esta perspectiva futura de trabalho nos parece de fundamental import\^ancia, pois sinalizar\'a 
se o processo de especia\c{c}\~ao pode ou n\~ao ser revers\'{\i}vel, apesar do isolamento 
reprodutivo inerente ao mesmo. Contudo, tal perspectiva demandar\'a um tempo razoavelmente  
grande, visto que exigir\'a modifica\c{c}\~oes bastante complexas nos programas. 

Note que nos dois cap\'{\i}tulos anteriores t\'{\i}nhamos $n_1=n_2=16$ e neste 
cap\'itulo tomamos $n_1=13$ e $n_2=19$. A raz\~ao de utilizarmos este intervalo foi 
porque percebemos que com estes valores os pontos das curvas presentes na figura 
\ref{figexpoentes} apresentavam uma menor dispers\~ao. Observamos ainda que se neste 
modelo forem usados valores altos de $MF$ ($MF \approx 0.1$ ou maior) 
a especia\c{c}\~ao obtida para $X>0.5$ \'e perdida. 

\begin{figure}[htbp]
\begin{center}
\includegraphics[width=6.5cm,angle=270]{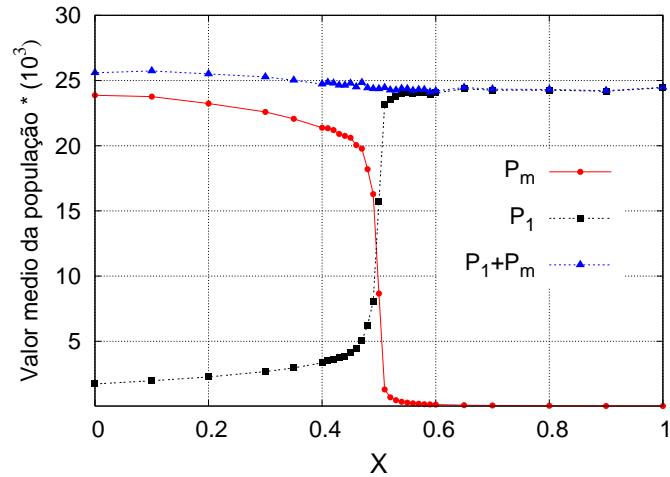}
\end{center}
\caption{\it Numa transi\c{c}\~ao de fase muitas grandezas sofrem mudan\cc as 
de comportamento pr\'oximas ao ponto cr\'{\i}tico, como ocorre por exemplo com a popula\c{c}\~ao de  
fen\'otipos intermedi\'arios (c\'{\i}rculos vermelhos) e com qualquer uma das popula\c{c}\~oes de 
fen\'otipos extremos (quadrados pretos). Contudo, a soma destas duas popula\c{c}\~oes, 
$P_m + P_1$ ou $P_m + P_2$, permanece aproximadamente constante (tri\^angulos azuis).} 
\label{figconstante}
\end{figure}

Numa transi\c{c}\~ao de fase no equil\'{\i}brio \cite{reichl}, pr\'oximo ao 
ponto cr\'{\i}tico, acontecem mudan\cc as dr\'asticas no comportamento  
da maioria das grandezas caracter\'{\i}sticas do sistema, como na entropia, 
na entalpia, etc.
Aqui tamb\'em foram encontradas mudan\cc as no comportamento de outras grandezas al\'em da 
densidade de seletivos,
como por exemplo, na popula\c{c}\~ao total, na popula\c{c}\~ao com 
fen\'otipo intermedi\'ario, nas popula\c{c}\~oes com fen\'otipos extremos, etc. 
Mas existe uma grandeza que se mant\'em aproximadamente constante durante 
todas as simula\c{c}\~oes que \'e $P_e+P_m$, onde $P_m$ \'e a popula\c{c}\~ao com 
fen\'otipos intermedi\'arios e $P_e$ \'e qualquer uma das popula\c{c}\~oes com fen\'otipos
extremos, como mostra a figura \ref{figconstante}. 
Esta constante foi confirmada pela vers\~ao preliminar do modelo anal\'{\i}tico 
que vem sendo desenvolvido h\'a bem pouco tempo principalmente pelo 
estudante de doutorado Veit Schw\"ammle.  
 
Nesta vers\~ao preliminar utilizamos as seguintes 
equa\c{c}\~oes diferenciais, do tipo presa-predador de Lotka-Volterra:  
\begin{eqnarray}\label{eqlv}
 \dot{P_1}&=&aP_1+bP_m-c(P_1+P_m)P_1\\ \nonumber
\dot{P_m}&=&aP_m+bP_1-c(2XP_1+P_m)P_m
\end{eqnarray}

O primero termo destas equa\c{c}\~oes representa o crescimento das popula\c{c}\~oes 
em fun\c{c}\~ao dos nascimentos. Se s\'o houvesse este termo, o crescimento seria  
expoencial ($a>0$).
O segundo termo representa o resultado das muta\c{c}\~oes e 
cruzamentos que fazem com que indiv\'{\i}duos de uma popula\c{c}\~ao gerem filhos 
com fen\'otipos correspondentes ao da outra popula\c{c}\~ao. Os \'ultimos termos destas 
equa\c{c}\~oes correspondem aos fatores de Verhulst 
(equa\c{c}\~ao \ref{eqcompe}) das popula\c{c}\~oes de acordo com seus respectivos fen\'otipos. 
Este modelo, apesar de muito simples, vem confirmando v\'arios resultados obtidos 
com as simula\c{c}\~oes e ser\'a um dos temas da tese do Veit. 

\section{Perspectivas}

Conforme j\'a mencionado, foi demonstrada qualitativamente a exist\^encia de uma 
transi\c{c}\~ao de fase no processo de especia\c{c}\~ao simp\'atrica,  
mas n\~ao foi encontrada a ordem desta transi\cao. 
Pretendemos nos aprofundar nesta quest\~ao, n\~ao com o intuito de comparar 
nossos resultados com dados reais, mas com o objetivo de estudar a 
reversibilidade ou n\~ao deste processo. Lembramos que o n\'umero de machos
dispon\'{\i}veis, ligado ao isolamento reprodutivo, faz o papel do campo 
na transi\c{c}\~ao de fase encontrada.
  
No que diz respeito \`a especia\c{c}\~ao da 
mosca da ma\c{c}\~a, tal estudo n\~ao tem apenas uma import\^ancia te\'orica, mas tamb\'em 
econ\^omica. Por exemplo, a fruticultura brasileira tem preju\'{\i}zos 
da ordem de R\$ 360 milh\~oes anuais (US\$ 120 milh\~oes) por causa da 
mosca-da-fruta, que s\'o se reproduz uma vez.
O presidente da Biof\'abrica 
Moscamed Brasil disse que \cite{noticias} a partir de setembro do pr\'oximo ano  
o Brasil produzir\'a moscas est\'ereis, que ser\~ao utilizadas no combate \`a mosca-da-fruta.

Pretendemos tamb\'em utilizar este modelo onde a capacidade de sustenta\c{c}\~ao \'e 
fixa e independente de $k$, para estudar as distribui\c{c}\~oes dos fen\'otipos para 
diferentes graus de competi\c{c}\~ao, mas agora considerando dois tra\c{c}os fenot\'{\i}picos 
independentes.

\chapter{Conclus\~oes} 

Comparar os modelos que descrevem o processo de especia\c{c}\~ao simp\'atrica
no geral \'e amb\'{\i}guo, pois como se viu nos cap\'{\i}tulos 2 e 3, o modelo depende
da esp\'ecie que est\'a em processo de especia\c{c}\~ao e quais s\~ao as 
condi\c{c}\~oes  ecol\'ogicas pelas quais passa a dita esp\'ecie.
Por exemplo, dizer que o modelo com dois tra\c{c}os fenot\'{\i}picos \'e 
mais realista que o modelo com apenas um tra\c{c}o \'e verdade para o caso dos 
cicl\'{\i}deos, 
pois eles apresentam duas carater\'{\i}sticas fenot\'{\i}picas claramente distintas.  
Contudo, isto j\'a n\~ao \'e verdade para os tentilh\~oes, pois  
a carater\'{\i}stica fenot\'{\i}pica ligada \`a competi\c{c}\~ao \'e a mesma ligada 
\`a sele\c{c}\~ao sexual, uma vez que o som, atrav\'es do qual a f\^emea escolhe o parceiro, 
est\'a diretamente relacionado \`a morfologia do bico. 

Uma das perguntas mais t\'{\i}picas em ci\^encia \'e se o modelo utilizado descreveu
ou n\~ao alguma carater\'{\i}stica especial real do fen\^omeno que foi estudado. 
No caso da especia\c{c}\~ao dos tentilh\~oes, a bibliografia descreve   
com riqueza diversas carater\'{\i}sticas qualitativas
destes p\'assaros, mas ainda n\~ao existem dados quantitativos suficientes. 
A seguir ressaltaremos os resultados qualitativos 
das simula\c{c}\~oes  que coincidem com aqueles observados nos tentilh\~oes.
O resultado qualitativo que mais chama a aten\c{c}\~ao \'e a r\'apida adapta\c{c}\~ao da 
morfologia do bico a uma mudan\c{c}a clim\'atica abrupta (ver figura \ref{figdbp} 
\aspas{inset}) e a exist\^encia de um comportamento oscilat\'orio no aparecimento 
das popula\c{c}\~oes  com bicos extremos (ver figuras \ref{figdoscila} 
e \ref{figpoposcila}, lado direito).
Note que na figura \ref{figsong} da introdu\cao,~as oscila\c{c}\~oes  no tamanho dos bicos dos
tentilh\~oes est\~ao diretamente relacionadas \`as oscila\c{c}\~oes  clim\'aticas. Por outro lado,  
nossos resultados sugerem que s\'o uma mudan\c{c}a abrupta da ecologia  
produz oscila\c{c}\~oes com per\'{\i}odos de uma gera\cao.~Por esta
raz\~ao pretendemos continuar procurando na bibliografia dos tentilh\~oes 
dados sobre estas oscila\c{c}\~oes, com o  
objetivo de introduzir no modelo oscila\c{c}\~oes   peri\'odicas da ecologia  
e verificar se as oscila\c{c}\~oes  das popula\c{c}\~oes  extremas seguem as da ecologia,  
ou se o per\'{\i}odo de uma gera\c{c}\~ao encontrado aqui se mant\'em.

No caso dos cicl\'{\i}deos existem dados gen\'eticos e uma 
descri\c{c}\~ao detalhada da forma de acasalamento e alimenta\c{c}\~ao destes peixes, como
foi visto na introdu\c{c}\~ao e no cap\'{\i}tulo 3.
A princ\'{\i}pio foi dif\'{\i}cil entender o processo de especia\c{c}\~ao dos cicl\'{\i}deos, 
primeiro porque na bibliografia o n\'umero de palavras t\'ecnicas \'e muito grande e  
segundo porque as simula\c{c}\~oes  com o modelo de dois tra\c{c}os demandam   
muito trabalho, no sentido de que existem v\'arios par\^ametros para 
\aspas{sintonizar} a fim de se obter a especia\cao. 
Como foi visto nas tabelas, \ref{tabpolims1} e \ref{tabpolims2}, a 
obten\c{c}\~ao da especia\c{c}\~ao n\~ao s\'o depende dos par\^ametros mas
tamb\'em da semente aleat\'oria com que \'e iniciada a simula\cao.~Finalmente 
este tema ficou claro, pois agora percebemos que para reproduzir a especia\c{c}\~ao dos cicl\'{\i}deos
\'e preciso obter primeiro uma separa\c{c}\~ao na morfologia da cor e 
posteriormente, uma separa\c{c}\~ao na morfologia dos maxilares. N\'os obtivemos 
dois resultados diferentes: num deles a separa\c{c}\~ao dos maxilares 
acontece primeiro, da mesma forma que em \cite{kk}, e no outro os maxilares e 
as cores se separaram ao mesmo tempo. Pretendemos continuar
procurando os efeitos \aspas{microsc\'opicos} que produzam  
a separa\c{c}\~ao nas cores antes da dos maxilares. Para isto pretendemos estudar a 
estabilidade das distribui\c{c}\~oes  dos fen\'otipos, cor e maxilar,
como foi exposto nas perspectivas do cap\'{\i}tulo 3 e suavizar a mudan\c{c}a 
brusca na ecologia, j\'a que as mudan\c{c}as clim\'aticas detectadas nos lagos onde 
foram estudados estes peixes n\~ao parecem ser muito relevantes.

O resultado encontrado no cap\'{\i}tulo 4, acerca do comportamento da densidade
de seletivos em fun\c{c}\~ao da competi\c{c}\~ao, \'e bastante interessante 
do ponto de vista da compreens\~ao dos modelos para especia\c{c}\~ao simp\'atrica 
em geral. Por exemplo, existe uma grande diferen\c{c}a entre os modelos \cite{kk} e \cite{dd}, 
j\'a que o modelo \cite{kk} precisa de uma mudan\c{c}a ecol\'ogica 
radical enquanto o modelo \cite{dd} n\~ao utiliza mudan\c{c}a alguma 
e os dois obt\^em especia\c{c}\~ao simp\'atrica. Agora que conhecemos o 
comportamento tipo transi\c{c}\~ao de fase na especia\c{c}\~ao,  
compreendemos que o modelo \cite{kk} precisa da mudan\c{c}a na ecologia 
devido \`a fraca competi\c{c}\~ao \`a qual os fen\'otipos intermedi\'arios est\~ao sujeitos,  
o que equivale \`a regi\~ao $X<0.5$ na figura 
\ref{figtran}, lado esquerdo. J\'a o modelo \cite{dd}, que utiliza uma 
forte competi\cao,~equivale \`a regi\~ao $X>0.5$ da mesma figura,  
n\~ao precisando de qualquer mudan\c{c}a radical na ecologia.  

O comportamento do tipo transi\c{c}\~ao de fase na 
especia\c{c}\~ao simp\'atrica tamb\'em nos fez perceber o tipo de estudo 
que pode ser realizado na mosca da ma\c{c}a, como foi explicado
no cap\'{\i}tulo 4. Al\'em do ponto de vista econ\^omico, o tema 
da irreversibilidade na especia\c{c}\~ao simp\'atrica \'e tamb\'em  
interessante do ponto de vista te\'orico, 
pois a identifica\c{c}\~ao da irreversibilidade ou reversibilidade do
processo d\'a a oportunidade de se saber se o sistema
est\'a dissipando energia e que grandeza est\'a fazendo o papel da
energia neste modelo.


\addcontentsline{toc}{chapter}{Bibliografia}

\begin{thebibliography}{100}
\bibitem{schro} Erwin Schr$\ddot{o}$edinger (1977), {\it O que \'e vida?}, Unesp.
\bibitem{librogene} Monroe W. Strickberger (1988), Gen\'etica, 
terceira edi\cao~Omega Barcelona, 738.
\bibitem{biohoje} S\'ergio Linhares e Fernando Gewandsznajder (1998), 
{\it Biologia Hoje}, vol. 3, editora \'atica.
\bibitem{richard} Richard E. Leakey (1982), Autor do texto introdut\'orio
de {\it A origem das esp\'ecies de Darwin}, publicado pela editora 
Universidade de Bras\ia lia/Melhoramentos, SP.
\bibitem{theorygp}Jonathan Roughgarden, {\it Theory of Population Genetics and 
Evolutionary Ecology, an introduction}, Prentice Hall 1996.
\bibitem{robert} Robert Foley (1993), {\it Apenas mais uma esp\'ecie \'unica}, Edusp.
\bibitem{bgs214} Peter T. Boag, Peter R. Grant (1981), Intense natural selection
in a population of Darwin's Finches (Geospizinae) in the Galapagos, {\it Science},
{\bf 214}, 
82-84.
\bibitem{sgs227} Schluter D., Price T. D. and Grant P. R.(1985), Ecological
character displacement in Darwin's Finches, {\it Science}, {\bf 227}, 1056-1059.
\bibitem{gge50} B. Rosemary Grant and Peter R. Grant (1996), Cultural
inheritance of song 
and its role in the evolution of Darwin's finches, {\it Evolution}, {\bf 50(6)},
2471-2487.
\bibitem{editor} Tom Tregenza and Roger K. Butlin (1999), Speciation 
without isolation, {\it Nature}, {\bf 400}, 311-312.
\bibitem{wilson} Anthony B. Wilson, Katharina Noack-Kunnmann and Axel Meyer (2000),
Incipient speciation in sympatric Nicaraguan crater lake cichlid fishes: sexual selection
versus ecological diversification, 
{\it Proc. R. Soc. Lond. B}, {\bf 267}, 2133-2141.
\bibitem{penna} T.J.P. Penna (1995), A Bit-String model for Biological Aging, 
{\it J. Stat. Phys.} {\bf 78}, 1629.
\bibitem{spa257} S. Moss de Oliveira (1998), A small review of the Penna model
for biological ageing, {\it Physica A}, {\bf 257}, 465-469.
\bibitem{livrosu} S. Moss de Oliveira, P.M.C. de Oliveira and D. Stauffer (1999), 
{\it Evolution, Money, War and Computers}, Teubner, Leipzig and Stuttgart, 
ISBN 3-519-00279-5.
\bibitem{teseadriana} Adriana Racco (1999),  Modelo de tira de bits:
Le\ia~unit\'aria de mortalid\'ade, estrat\'egias de reprodu\cao~e
regulamenta\cao~de pesca, Tese de Mestrado UFF.
\bibitem{asppa253} R.M.C. de Almeida, S. Moss de Oliveira, T.J.P. Penna (1998),
Theoretical approach to biological aging, {\it Physica A}, {\bf 253}, 366-378.
\bibitem{prlpenna1}J. B. Coe, Y.Mao and M. E. Cates (2002),
Solvable senescence model showing a mortality plateau,
{\it Physical Review Letters} {\bf 89}, 28, 288103.
\bibitem{pamilo}P. Pamilo, M. Nei and W.H. Li (1987), Genet. Res. Camb. 49, 135.
\bibitem{jsmctb119}J.S. Sa Martins, S. Cebrat (2000), 
Random Deaths in a Computational model for age-structured populations, 
{\it Theory in Biosciences} {\bf 119}, 156-165. 
\bibitem{medeiros} J.S. S\'a Martins, S. Moss de Oliveira and G.A. de Medeiros
(2001), 
Simulated ecology-driven sympatric speciation, Phys. Rev. E {\bf 64}, 021906.
\bibitem{toshitony} K. Luz-Burgoa, Tony Dell and Toshinori Okuyama (2004), 
Student papers Complex System Summer School Santa Fe, New Mexico USA, 
June 6-July 2, artigo para revista cient\ia fica de circula\cao~internacional em prepara\cao.
\bibitem{pn409} Podos J. (2001), Correlated evolution of morphology and vocal
signal 
structure in Darwin's finches, {\it Nature}, {\bf 409} 185-188.
\bibitem{pnbio54} Podos J. and Nowicki S. (2004), Beaks, adaptation and vocal
evolution 
in Darwin's finches, {\it Bioscience} {\bf 54} (6) 501-510.
\bibitem{psr207} Podos J., Southall J. A., Rossi-Santos M. R. (2004), Vocal
mechanics in 
Darwin's finches: correlation of beak gape and song frequency, 
{\it Journal of experimental biology} {\bf 207} (4): 607-619.
\bibitem{mkn408} Mark Kirkpatric (2000), Fish found in flagrante delicto, {\it Nature}, 
{\bf 408}, 298.
\bibitem{kk} Alexey S. Kondrashov and Fyodor A. Kondrashov (1999), Interactions among 
quantitative traits in the course of sympatric speciation, {\it Nature}, {\bf 400}, 351-354.
\bibitem{dd} Ulf Dieckmann and Michael Doebeli (1999), On the origin of species 
by sympatric speciation, {\it Nature}, {\bf 400}, 354-357.
\bibitem{bjp} K. Luz-Burgoa, S. Moss de Oliveira, J. S. S\'a Martins, D.
Stauffer and 
A.O. Souza (2003), Computer simulation of sympatric speciation with Penna Ageing
Model, 
{\it Brazilian Journal Physics}, {\bf 33} 3, 623-627. 
\bibitem{cise}S. Moss de Oliveira, J.S. S\'a Martins, P.M.C. de
Oliveira, 
K. Luz-Burgoa, A. Ticona and T.J.P. Penna (2004), The Penna model for biological
aging and 
speciation, {\it Computing in Science and Engineering}, IEEE CS and the AIP.
\bibitem{turelli}Michael Turelli, Nicholas H. Barton and Jerry A. Coyne (2001),
Theory and speciation, {\it TRENDS in Ecology $\&$ Evolution}, {\bf 16} No.7, 330-343.
\bibitem{htakimoto} M. Higashi, G. Takimoto $\&$ N. Yamamura (1999) Sympatric 
Speciation by sexual selection, {\it Nature}, {\bf 402}, 523-526.
\bibitem{kksexo} Matthew E. Arnegard and Alexey S. Kondrashov (2004),
Sympatric Speciation by sexual selection alone is unlikely, {\it Evolution}, 
{\bf 58} 2, 222-237.
\bibitem{tisean}http://ls11-www.cs.uni-dortmund.de/people/hermes/NLDdocs/docs/
\bibitem{kjj407} Kenneth E. Filchak, Joseph B. Roethele and Jefrey L. Feder
(2000), Natural selection and sympatric divergence in the apple maggot
{\it Rhagoletis pomonella}, {\it Nature}, {\bf 407}, 739-742.
\bibitem{jjkjnj163} Jefrey L. Feder, Joseph B. Roethele, Kenneth Filchak, Julie
Niedbalski and Jeanne Romero-Severson (2003), Evidence for inversion Polymorphism
Related to Sympatric Host Race Formation in the apple Maggot Fly, {\it
Rhagoletis pomonella} ,{\it Genetics}, {\bf 163}, 939-953.
\bibitem{dickman} Joaqu\ia n Marro and Ronald Dickman (1999), {\it Nonequilibrium
Phase Transitions in Lattice Models}, Cambridge University press.
\bibitem{reichl} L. E. Reichl (1998), {\it A modern course in Statistical 
Physics}, 2nd Edition John Wiley $\&$ Sons, INC. 
\bibitem{noticias} http://www.faemg.org.br/$fruti_-noticias.asp$,
Controle biol\'ogico para a mosca da fruta.
\end{thebibliography}
\end{document}